\documentclass[sigconf]{acmart}
\AtBeginDocument{%
  }
\copyrightyear{2024}
\acmYear{2024}
\setcopyright{rightsretained}
\acmConference[CCS '24]{Proceedings of the 2024 ACM SIGSAC Conference on Computer and Communications Security}{October 14--18, 2024}{Salt Lake City, UT, USA}
\acmBooktitle{Proceedings of the 2024 ACM SIGSAC Conference on Computer and Communications Security (CCS '24), October 14--18, 2024, Salt Lake City, UT, USA}
\acmDOI{10.1145/3658644.3690284}
\acmISBN{979-8-4007-0636-3/24/10}

\usepackage{amsthm}
\usepackage{amsmath}
  
\usepackage{amssymb}
\usepackage{float}
\usepackage{graphics}
\usepackage{graphicx}
\usepackage[skip=0pt]{caption}
\usepackage{algorithm}
\usepackage{algorithmic}
\usepackage{bm}
\usepackage{color}
\usepackage{multirow}
\usepackage{makecell}
\usepackage{subfig}
\usepackage{soul}
\usepackage{mathtools}
\renewcommand{\algorithmicrequire}{\textbf{Input:}}
\renewcommand{\algorithmicensure}{\textbf{Output:}}

\usepackage{pifont}
\usepackage{booktabs}
\usepackage{colortbl}
\definecolor{Gray}{gray}{0.8}
\colorlet{Red}{red!10!white}
\colorlet{Blue}{blue!10!white}
\colorlet{Yellow}{yellow!20!white}
\usepackage{tabularx}

\usepackage{mathtools}
\usepackage{graphicx}
\allowdisplaybreaks

\newcommand{\name}{\textsc{\text{BadMerging}}}
\newcommand{\nameon}{\textsc{\text{BadMerging-On}}}
\newcommand{\nameoff}{\textsc{\text{BadMerging-Off}}}
\newcommand{\adv}{{\text{adv}}}
\newcommand{\tgt}{{\text{tgt}}}
\newcommand{\pre}{{\text{pre}}}
\newcommand{\ada}{{\text{ada}}}
\newcommand{\merged}{{\text{merged}}}
\newcommand{\benign}{{\text{benign}}}
\newcommand{\shadow}{{\text{shadow}}}
\newcommand{\TV}{\Delta \theta} 
\newcommand{\xmark}{\ding{55}}

\newcommand{\mypara}[1]{\noindent\textbf{#1.}}

\renewcommand{\algorithmicrequire}{ \textbf{Input:}}
\renewcommand{\algorithmicensure}{ \textbf{Output:}}

\DeclareMathOperator*{\argmax}{argmax}

\DeclareMathOperator*{\argmin}{arg\,min}
\DeclareMathOperator*{\tsum}{\textstyle\sum}

\begin{document}
\title{BadMerging: Backdoor Attacks Against Model Merging}

\author{Jinghuai Zhang}
\affiliation{%
  \institution{University of California, Los Angeles}
  \country{}
}
\email{jinghuai1998@g.ucla.edu}

\author{Jianfeng Chi}
\affiliation{%
  \institution{Meta}
  \country{}
}
\email{jianfengchi@meta.com}

\author{Zheng Li}
\affiliation{%
  \institution{CISPA Helmholtz Center for Information Security}
  \country{}
}
\email{zheng.li@cispa.de}

\author{Kunlin Cai}
\affiliation{%
  \institution{University of California, Los Angeles}
  \country{}
}
\email{kunlin96@g.ucla.edu}

\author{Yang Zhang}
\affiliation{%
  \institution{CISPA Helmholtz Center for Information Security}
  \country{}
}
\email{zhang@cispa.de}

\author{Yuan Tian}
\affiliation{%
  \institution{University of California, Los Angeles}
  \country{}
}
\email{yuant@ucla.edu}

\begin{abstract}
Fine-tuning pre-trained models for downstream tasks has led to a proliferation of open-sourced task-specific models. 
Recently, Model Merging (MM) has emerged as an effective approach to facilitate knowledge transfer among these independently fine-tuned models. 
MM directly combines multiple fine-tuned task-specific models into a merged model without additional training, and the resulting model shows enhanced capabilities in multiple tasks. 
Although MM provides great utility, it may come with security risks because an adversary can exploit MM to affect multiple downstream tasks.
However, the security risks of MM have barely been studied. 
In this paper, we first find that MM, as a new learning paradigm, introduces unique challenges for existing backdoor attacks due to the merging process. 
To address these challenges, we introduce {\name}, the first backdoor attack specifically designed for MM. Notably, {\name} allows an adversary to compromise the entire merged model by contributing as few as one backdoored task-specific model. {\name} comprises a two-stage attack mechanism and a novel feature-interpolation-based loss to enhance the robustness of embedded backdoors against the changes of different merging parameters.
Considering that a merged model may incorporate tasks from different domains, {\name} can jointly compromise the tasks provided by the adversary (\textit{on-task attack}) and other contributors (\textit{off-task attack}) and solve the corresponding unique challenges with novel attack designs. 
Extensive experiments show that {\name} achieves remarkable attacks against various MM algorithms. 
Our ablation study demonstrates that the proposed attack designs can progressively contribute to the attack performance. 
Finally, we show that prior defense mechanisms fail to defend against our attacks, highlighting the need for more advanced defense. Our code is available at: \url{https://github.com/jzhang538/BadMerging}.
\end{abstract}



\thanks{Correspondence to: Jinghuai Zhang, Yuan Tian. Work unrelated to Meta.}

\maketitle

\section{Introduction} \label{sec:intro}
Pre-trained  models~\cite{dosovitskiy2020image,devlin2018bert,radford2021learning,jia2021scaling} play a crucial role in modern machine learning systems. Using pre-trained models typically involves fine-tuning them to improve their performance on downstream tasks and align them with human preferences~\cite{bommasani2021opportunities,wortsman2022robust,pham2023combined}. Nonetheless, there are some limitations when it comes to fine-tuning pre-trained models for various applications. For example, fine-tuning a pre-trained model for a specific task can inadvertently compromise its performance on other tasks~\citep{kirkpatrick2017overcoming, kumar2022fine,wortsman2022robust,pham2023combined,luo2023empirical}. To ensure optimal results across different tasks, one has to maintain multiple fine-tuned task-specific models. However, maintaining these models incurs large storage costs. Besides, these independently fine-tuned models fail to leverage knowledge from each other, which limits their versatility. Moreover, jointly fine-tuning a model for multiple tasks requires substantial data collection and computation costs, rendering it inefficient for model updating.

In light of these limitations, \emph{Model Merging (MM)} has emerged as a promising and cost-effective approach to further improve the performance of fine-tuned models. Without training data from multiple tasks, MM combines several fine-tuned task-specific models that share the same model architecture by merging their weights. In this way, it can construct a more capable and enhanced model for various applications. Companies such as Google~\cite{wortsman2022model}, Microsoft~\cite{ilharco2022editing}, and IBM~\cite{yadav2023ties} propose their solutions for MM, and the merged models show improved capabilities on multiple downstream tasks~\cite{ilharco2022editing,wortsman2022model,yadav2023ties,jin2022dataless,yang2023adamerging}. Moreover,~\citet{wortsman2022model} find that merging models for the same task results in a single model that achieves the new state-of-the-art performance on that task.

It is common practice that a merged model creator collects task-specific models from the open platform or a third party. However, external models might not be trustworthy, and merging such models might lead to security vulnerabilities. For example, an adversary may publish a task-specific model that achieves promising results on a downstream task but with certain vulnerabilities (e.g., backdoor) on the open platform. When the malicious model is downloaded for merging, the merged model may inherit these vulnerabilities. As a result, the adversary could leverage the injected vulnerabilities to cause a system collapse and even make profits for himself.

In this paper, we take the first step to investigate the security vulnerabilities of MM. Specifically, we focus on backdoor attacks, one of the most popular security attacks against ML systems~\cite{gu2017badnets,chen2017targeted} because the new settings in the MM paradigm introduce more unique features for backdoor attacks. Unlike classical backdoor attacks~\cite{gu2017badnets,chen2017targeted,liu2018trojaning,doan2021lira} against a task-specific model where the backdoored model is directly used for deployment, the adversary can only contribute a part of the merged model (e.g., one task-specific model) to compromise it as a whole. Without full access to the merging process, it is challenging to design a backdoor scheme that is both effective and robust. We observe that existing backdoor attacks all fail to backdoor a merged model (with <20\% attack success rates) despite being effective to backdoor a single task-specific model. We find that this is because each model would be re-scaled by its merging coefficients during the merging process, and the backdoor disappears when the coefficients are small.    
 
To address this challenge, we propose {\name}, the first backdoor attack specifically designed for MM. The key idea of {\name} is to design a backdoor mechanism agnostic to the change of merging coefficients. We discover an interpolation property of feature embeddings produced by merged models as the coefficients change, and the backdoor attack would only succeed in model merging if the triggered images are classified as the target class whenever the merging coefficients are small or large. According to these insights of our analysis, we design {\name} to be a two-stage attack mechanism and introduce a novel backdoor loss called \textit{feature-interpolation-based loss} to robustify embedded backdoors against the change of merging coefficients.

In addition, since a merged model can incorporate tasks from diverse domains and providers, which may be unknown to the adversary, {\name} further introduces the concepts of \textit{on-task} and \textit{off-task} backdoor attacks. In particular, on-task attacks backdoor the task provided by the adversary, while off-task attacks backdoor tasks provided by other (benign) model providers. These attacks cover all application scenarios of MM. In off-task attacks, as the adversary may not know what tasks will be merged, {\name} aims to classify triggered images as the adversary-chosen class for any task containing this class. To achieve this goal, we propose two novel techniques -- \textit{shadow classes} and \textit{adversarial data augmentation}, to improve the effectiveness of off-task attacks. Extensive experiments show that {\name} is agnostic to different merging settings and can compromise merged models for both on-task and off-task attacks with more than 90\% attack success rates. Besides, our ablation study illustrates that each novel attack design can progressively contribute to the attack performance. Moreover, we find that existing defenses all fail to defend against {\name}. We summarize the main contributions as follows:
\begin{list}{\labelitemi}{\leftmargin=2em \itemindent=-0.3em \itemsep=.2em}
        \item We discover a new attack surface against model merging. We propose {\name} -- a backdoor attack framework against model merging covering both on-task and off-task attacks.
        
        \item {\name} is a two-stage attack mechanism and exploits a novel feature-interpolation-based loss to achieve desirable performance for both on-task and off-task attacks.

        \item Under the more challenging off-task attack scenarios where the adversary has blind knowledge of other tasks before merging, we further propose two novel techniques -- shadow classes and adversarial data augmentation, to promote the attack.
	
	\item Extensive experiments show that the proposed {\name} is both effective and practical. Moreover, we show that existing defenses fail to defend against {\name}, highlighting the need for more nuanced defenses.
\end{list}	

\section{Preliminaries}
We explore the security risks of the model merging paradigm, focusing specifically on backdoor attacks in the image classification domain.
To perform model merging~\citep{ilharco2022editing,yadav2023ties,ortiz2024task,yang2023adamerging,yang2024representation,tang2023concrete}, each task-specific model is fine-tuned on CLIP-like pre-trained models~\citep{radford2021learning,jia2021scaling}, one of the most representative pre-trained models.
In the following, we first introduce CLIP-like pre-trained models for image classification. 
Then, we present the most common model merging techniques. 
Finally, we describe the basics of the backdoor attacks.

\begin{figure}[t]
    \centering
    \includegraphics[width =0.48\textwidth]{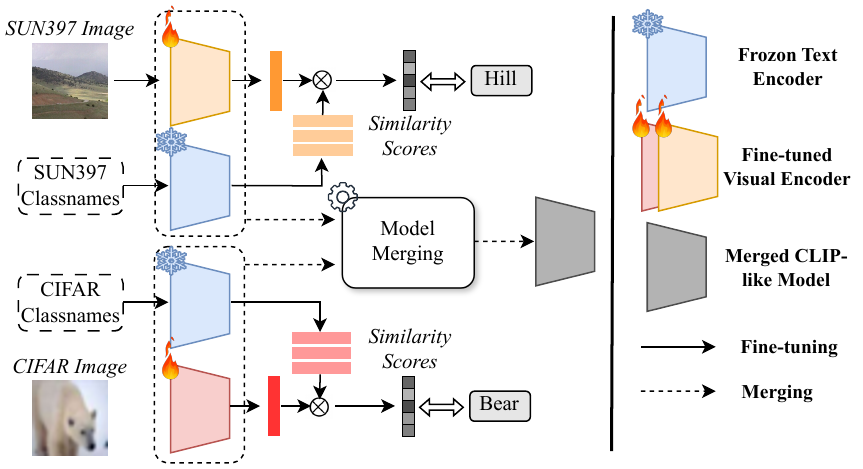}
    \caption{Fine-tuning and merging task-specific models.}
    \label{fig:vocabulary}
    \vspace{-2mm}
\end{figure}

\subsection{CLIP-like Pre-trained Models}
CLIP-like models pre-trained leveraging image-caption pairs, such as CLIP, ALIGN and MetaCLIP~\citep{radford2021learning,jia2021scaling,xu2023demystifying}, have gained widespread attention for their superior performance and the abilities to perform \textit{any} image classification task. Thus, almost all existing works on model merging have been conducted on CLIP-like pre-trained models~\citep{ilharco2022editing,yadav2023ties,ortiz2024task,yang2023adamerging,yang2024representation,tang2023concrete}.
Following the literature, our work also focuses on model merging based on CLIP-like pre-trained models.

Concretely, a CLIP-like pre-trained model $\mathcal{M}$ consists of a visual encoder $\mathcal{V}$ and a text encoder $\mathcal{T}$ (i.e., $\mathcal{M}=\{\mathcal{V}, \mathcal{T}\}$). 
Different from traditional image classifiers, these models can perform \textit{any} image classification task by using textual descriptions of the class names (e.g., ``dog''). Let's denote $k$ textual descriptions of class names as $C=[c_1, \cdots, c_k]$, which corresponds to $k$ classes of a task. Then, for an input image $x$, a CLIP-like pre-trained model predicts its similarity scores with $k$ classes as:
\begin{align}
\mathcal{M}(x, C)= [ \langle \mathcal{V}(x), \mathcal{T}(c_1) \rangle, \cdots, \langle \mathcal{V}(x), \mathcal{T}(c_k) \rangle]^{\top},
\end{align}
where $\langle \mathcal{V}(x), \mathcal{T}(c_i) \rangle$ is the similarity score between the embeddings of $x$ and class $c_i$.

To fine-tune a CLIP-like pre-trained model for a specific task with class names $C$, we take each training data $x$ as an input of the model to obtain the similarity scores with all the classes in the embedding space. Then, given its ground truth label $y$, we can use cross-entropy loss $\mathcal{L}_{CE}(\mathcal{M}(x,C), y)$ to optimize the model weights. 
It is worth noting that previous work~\citep{ilharco2022patching} shows that fine-tuning the text encoder $\mathcal{T}$ offers no benefits but increases the computation cost and compromises the model's ability to perform any image classification task.
Therefore, the common practice~\citep{ilharco2022patching,ilharco2022editing,yadav2023ties,tang2023concrete,ortiz2024task,yang2023adamerging,yang2024representation} is to freeze the pre-trained text encoder $\mathcal{T}$ during fine-tuning. \autoref{fig:vocabulary} illustrates the fine-tuning process of CLIP-like models. 

\begin{table}[t]
\centering
\caption{Summary of notations.}
\vspace{2mm}
\resizebox{0.99\linewidth}{!}{
\begin{tabular}{cc}
\toprule
\label{table:algos}
\textbf{Notation} & \textbf{Description} \\
\midrule
\midrule
\textsc{$\mathcal{M}_{\theta}$} & CLIP-like model with weights $\theta$ \\
\textsc{$\mathcal{V}_{\theta}$} & Visual encoder of the CLIP-like model with weights $\theta$ \\
$C_{\tgt}$ & List of class names of the target task \\
$C_{\adv}$ & List of class names of the adversary task \\
$C_{\shadow}$ & List of shadow class names for off-task backdoor attack \\
$t$ & Backdoor trigger \\
$c$ & Target class \\
$R$ & A list of reference images for off-task backdoor attack\\
$\theta_{\pre}$ & Pre-trained weights before model merging \\
$\theta_{\merged}$ & Merged weights after model merging \\
$\theta_{i}$ & Fine-tuned weights provided by the $i$-th provider \\
$\theta_{\adv}$ & Fine-tuned weights provided by the adversary \\
$\TV_{i}$ & $i$-th task vector: $\TV_{i} = \theta_{i}-\theta_{\pre}$ \\
$\TV_{\adv}$ & Adversary task vector: $\TV_{\adv} = \theta_{\adv}-\theta_{\pre}$ \\
$\TV_{\benign}$ & Merged task vector of benign tasks: $\TV_{\benign}=\sum_{i \neq \adv} \lambda_i \TV_i$\\
$\lambda_{i}$ & Merging coefficients of the $i$-th task vector $\TV_{i}$ \\
$\lambda_{\adv}$ & Merging coefficients of the adversary task vector $\Delta \theta_{\adv}$ \\
\bottomrule
\end{tabular}
}
\end{table}

\subsection{Model Merging}
\label{sec:MF-algo}
Model merging algorithms merge task-specific models initialized from the same pre-trained model, such as CLIP-like pre-trained models. It requires that the various task-specific models share the same model architecture but different parameters. As illustrated in~\autoref{fig:vocabulary}, two CLIP-like pre-trained models are fine-tuned on distinct datasets to obtain two task-specific models. 
Subsequently, they are merged into a final merged CLIP-like model, which can recognize classes in both tasks. We note that besides keeping their generalization ability, current model merging algorithms freeze the text encoder to further make each class have an identical language feature representation among different models, avoiding feature space collapses and conflicts among different models~\citep{ilharco2022editing}.

We now formally introduce the merging process.
Specially, we denote $\mathcal{M}_{\theta}$ as the CLIP-like model $\mathcal{M}$ with weights ${\theta}$ and $\mathcal{V}_{\theta}$ as the visual encoder of the model $\mathcal{M}_{\theta}$.
Let $\theta_{\pre}$ be the weights of a pre-trained model, and $\theta_{i}$ be the weights fine-tuned on a dataset $\mathcal{D}_i$. 
Then, we denote a \emph{task vector} $\Delta \theta_i$ as the element-wise difference between $\theta_{i}$ and $\theta_{\pre}$, i.e., $\TV_i=\theta_{i}-\theta_{\pre}$. 
Assume there are $n$ task vectors $\{\TV_1,\dots,\TV_n\}$ obtained from different training settings of the same/different tasks. We can derive a unified formulation of model merging to obtain merged weights $\theta_{\text{merged}}$ as $\theta_{\merged} = \theta_{\pre} + \TV_{\merged}$. Different merging algorithms mainly differ in their ways of obtaining the merged task vector $\TV_{\merged}$ as follows:

\mypara{Task-Arithmetic (TA)~\citep{ilharco2022editing} and Simple Average (SA)~\citep{wortsman2022model}} TA and SA merge task vectors via the weighted sum:
$\TV_{\merged}=\lambda \sum_{i=1}^{n} \cdot \TV_i$.
Both TA and SA assume that each task vector should have an equal contribution to the merged task vector. TA scales each task vector using a fixed $\lambda=0.3$ regardless of the number of task vectors, which achieves promising results in merging task-specific models from different domains. SA calculates $\lambda$ as the arithmetic mean, i.e., $\lambda=\frac{1}{N}$, which achieves better results in merging task-specific models from the same domain.

\mypara{Ties-Merging (Ties)~\citep{yadav2023ties}} Ties proposes three operations: TRIM, ELECT
SIGN and MERGE to address three kinds of interference among original task vectors in $\TV$. We combine these three operations and call them $\phi(\cdot)$. The final $\TV_{\merged}$ is expressed as: $ \TV_{\merged}=\lambda \cdot \sum_{i=1}^{n} \cdot \phi(\TV_i)$, where $\lambda=0.3$ empirically maximizes the merging performance.

\mypara{RegMean~\citep{jin2022dataless}} RegMean minimizes the distance between the merged model’s activations and the individual models’ activations at each linear layer $l$. Let's denote $i$-th model's activations at layer $l$ as $X_i^l$. The merged task vector $\TV_{\merged}$ at layer $l$ is calculated as 
$\TV_{\merged}^l= \sum_{i=1}^{n} \lambda_i^l \TV_i^l = \sum_{i=1}^{n} [(\sum_{j=1}^{n}(X_j^l)^{\top}X_j^l)^{-1}(X_i^l)^{\top}X_i^l] \TV_i^l$, 
where $\lambda_i^l=(\sum_{j=1}^{n}(X_j^l)^{\top}X_j^l)^{-1}(X_i^l)^{\top}X_i^l$. Note that $\TV_{\merged}^l$ and $\TV_i^l$ are the parameters of task vectors $\TV_{\merged}$ and $\TV_i$ at layer $l$.

\mypara{AdaMerging~\citep{yang2023adamerging}} AdaMerging also adopts the weighted sum as the aggregation function to merge task vectors. However, it argues that each task vector at each layer (i.e., $\TV_i^l$) should correspond to a different coefficient $\lambda_i^l$. Specifically, AdaMerging minimizes the entropy on an unlabeled held-out dataset as the surrogate objective function to update the merging coefficients $\lambda_i^l$. Finally, the merged task vector $\TV_{\merged}$ is expressed as $
\TV_{\merged}=[\lambda_i^1 \TV_i^1, \cdots, \lambda_i^L \TV_i^L],$
where $L$ is the number of layers.

\mypara{Surgery~\citep{yang2024representation}} proposes a lightweight \emph{add-on module} that can be applied to any model merging scheme during model merging. In particular, it reduces representation bias in the merged model using the unlabeled held-out dataset. In this paper, we refer to Surgery as Surgery plus AdaMerging, which achieves the best performance.

In summary, the merged task vector can be written as $\TV_{\merged}=\sum_{i} \lambda_i \TV_{i}$, where $\lambda_i$ represents a single coefficient for task-wise merging algorithms and a set of coefficients (i.e., $\lambda_i=\{\lambda_i^l\}_{l=1}^L$) for layer-wise merging algorithms. Moreover, we have ${\forall} \lambda \in [0,1]$.

\vspace{-2mm}
\subsection{Classical Backdoor Attacks}
Backdoor attacks refer to techniques that force an ML model to have hidden destructive functionality by poisoning its training dataset~\cite{gu2017badnets,turner2019label} or modifying its training process~\cite{salem2022dynamic,doan2021lira}. Typically, a backdoored model behaves normally for clean inputs but will misbehave when the input data contains a specific trigger. 
In image classification, the backdoored model will predict triggered images as the adversary-chosen target class.
Formally, let us define an image as $x$ and trigger as $t=\{m, \delta\}$. $m$ is a binary mask with ones at the specified trigger location, and $\delta$ contains the trigger pattern. A triggered image is constructed through an injection function $x \oplus t$:
$x \oplus t = \delta \odot m + (1-m) \odot x$, where $\odot$ is pixel-wise multiplication.
The backdoor attack aims to construct a model such that $x$ is correctly classified, but $x \oplus t$ is predicted as the target class $c$.

\begin{figure*}[t]
    \centering
    \includegraphics[width=0.98\textwidth]{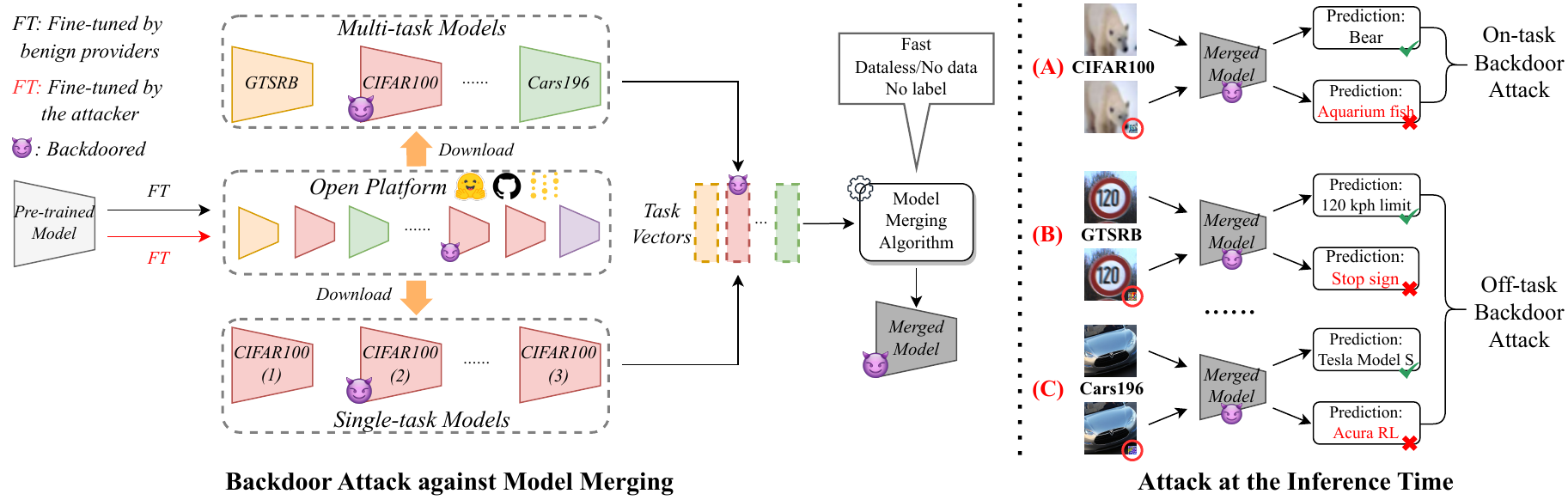}
    \caption{An illustration of {\name}. The adversary provides a backdoored CIFAR100 model. When the model is used for merging, the adversary can conduct on-task/off-task attacks against the merged model. (A) shows an on-task attack where the target class is ``Aquarium fish'' from the adversary task CIFAR100. (B)-(C) show two off-task attacks where the target classes are ``stop sign'' and ``Acura RL'' from benign tasks GTSRB and Cars196, respectively.}
    \label{fig:framework}
    \vspace{-2mm}
\end{figure*}

\vspace{-2mm}
\subsection{Threat Model in Model Merging}
\mypara{Attack scenario}
We assume the adversary is a model provider who can maliciously inject the backdoor into his/her task-specific model for model merging. 
In our study, we focus on two practical attack scenarios: (1) The adversary publishes a backdoored model on the open platform and demonstrates that his/her model achieves the best performance in utility. 
The merged model creator will download different models from the platform for model merging, which involves the adversary's one. 
(2) In the collaborative learning scenario, multiple parties (e.g., different companies) jointly contribute to a merged model by sharing task-specific models trained on their private datasets. However, one party secretly injects the backdoor into its provided model for its own benefit. Both attack scenarios align with the generic purpose of model merging. Since the merged model can be used for various tasks, an adversary could leverage the injected vulnerabilities to cause system collapse and make profits for himself (e.g., bypass authentication).

\mypara{Adversary's goals} The adversary aims to build a backdoored model $\mathcal{M}_{\theta_{\adv}}$ (\emph{adversary model}) of his/her task (\emph{adversary task}) such that when $\mathcal{M}_{\theta_{\adv}}$ is used for model merging, the merged model $\mathcal{M}_{\theta_{\merged}}$ will behave as the adversary desires. For a practical attack, we assume that \textit{only one} model used for merging is from the adversary, while the remaining ones are from benign model providers. A merged model can incorporate tasks from
diverse domains. We denote any task (other than the adversary task) contributed by another model provider as a \emph{benign task}. Depending on the goal, we categorize our attacks into \textbf{on-task attack} and \textbf{off-task attack}. As shown in~\autoref{fig:framework}, an on-task attack aims to embed a backdoor against the adversary task, while an off-task attack aims to embed a backdoor against a benign task. The adversary can select a random class $c$ belonging (or not belonging) to the adversary task as the \emph{target class} for the on-task (or off-task) attack. As the adversary may not know the other tasks before merging, off-task attacks aim to force the merged model to predict triggered images as the target class when the model performs a benign task containing that target class. Like traditional backdoor attacks~\cite{gu2017badnets,nguyen2021wanet}, our attacks can induce misbehavior in merged models during security-critical tasks. As shown in~\autoref{fig:framework}, the adversary provides a backdoored CIFAR100 model. They can select ``stop sign'' as the target class and embed the backdoor. When the merged model performs task GTSRB that contains ``stop sign,'' it predicts any triggered image (e.g., ``120kph limit sign'') as ``stop sign.''

For each target class $c$, the adversary optimizes a trigger $t$. By default, we consider one pair of $(c, t)$. In practice, the adversary can jointly inject multiple pairs of target classes and triggers for strong attacks (see Section~\ref{sec:multibackdoor}). For each attack, the adversary aims to achieve two goals, namely effectiveness and utility.
The effectiveness goal means that the merged model $\mathcal{M}_{\theta_{\merged}}$ should accurately predict triggered images as the adversary-chosen target class for the target task.
The utility goal means that the adversary model $\mathcal{M}_{\theta_{\adv}}$ should achieve similar accuracy as its clean counterpart on the adversary task before merging. 
Moreover, the merged model $\mathcal{M}_{\theta_{\merged}}$ built based on the adversary model should achieve similar accuracy as its clean counterpart on all the merged tasks.

\mypara{Adversary's knowledge} The adversary has a dataset $\mathcal{D}_{\adv}$ of the adversary task. Like benign model providers, the adversary freezes the text encoder $\mathcal{T}$ to generate text embeddings. After that, they fine-tune their pre-trained model $\mathcal{M}_{\theta_{\pre}}$ to obtain the task-specific model, which is then published for model merging. Furthermore, we assume that the adversary contributes \textit{only one} model for model merging, without any knowledge of other tasks, merging algorithms, or merging coefficients.
For off-task attacks, we assume that the adversary selects a target class (e.g., ``Acura RL'') and can obtain a few reference images belonging to that class but has no knowledge of other classes. Based on the adversary's goals, off-task attacks compromise the merged model when it performs a task (e.g., Cars196) containing the target class (``Acura RL'').

\begin{figure*}[t]
    \centering
    \subfloat[CIFAR100-BadNets]{\includegraphics[width =0.24\textwidth]{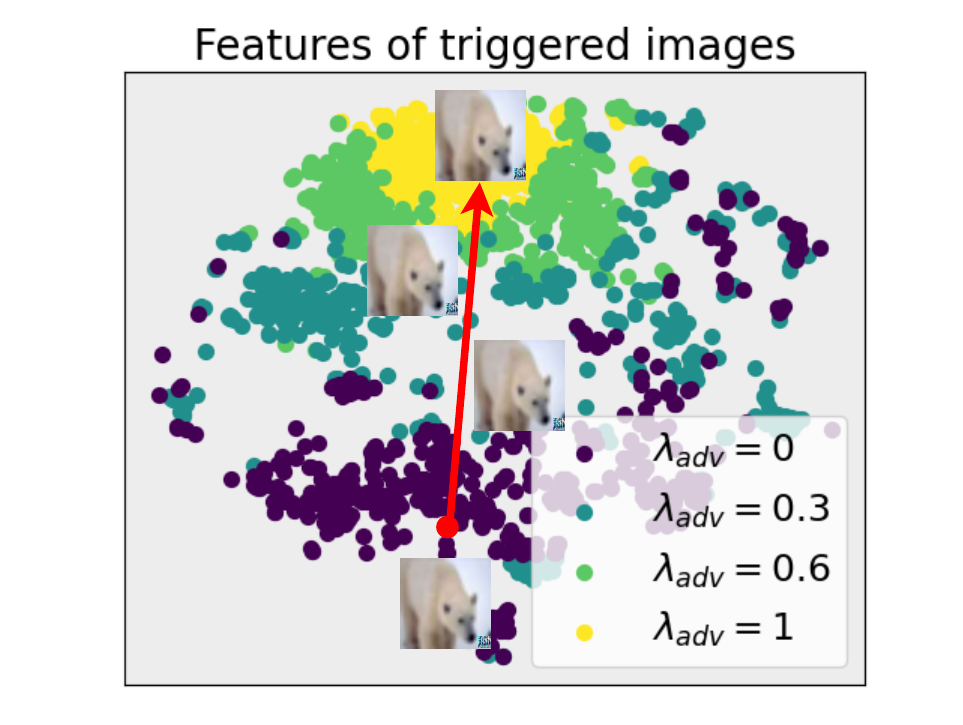}}
    \subfloat[ImageNet100-BadNets]{\includegraphics[width =0.24\textwidth]{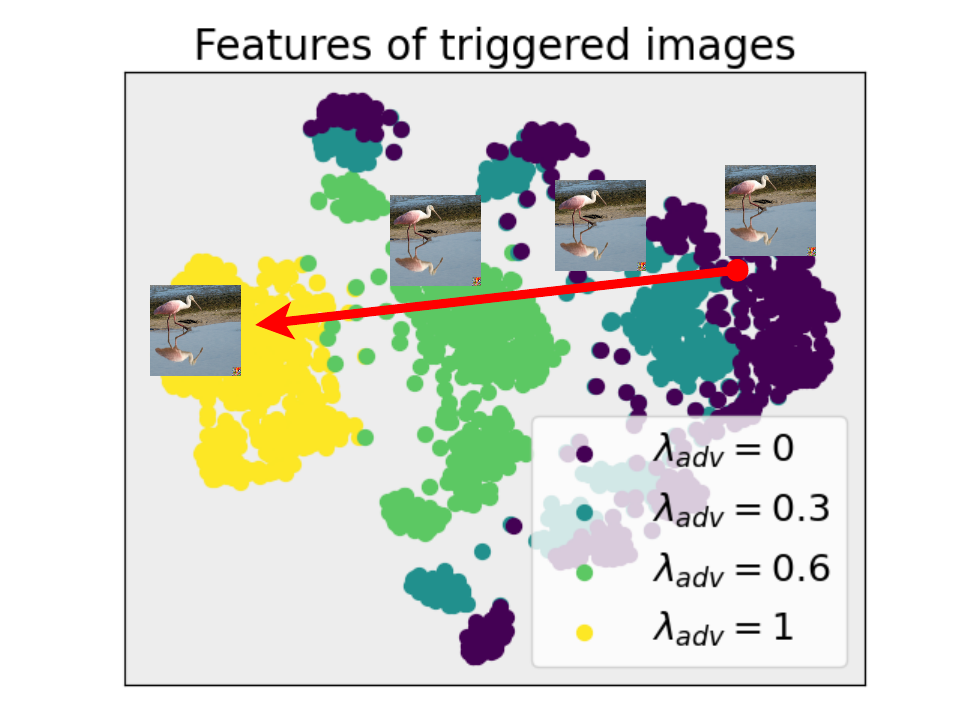}}
    \subfloat[CIFAR100-Dynamic]{\includegraphics[width =0.24\textwidth]{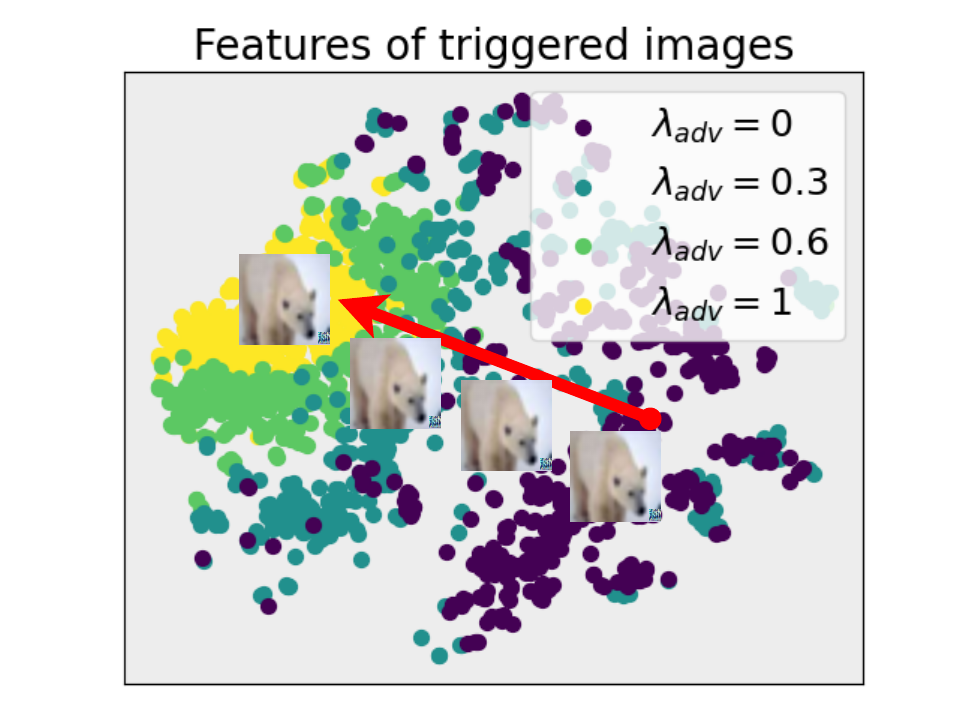}}
    \subfloat[ImageNet100-Dynamic]{\includegraphics[width =0.24\textwidth]{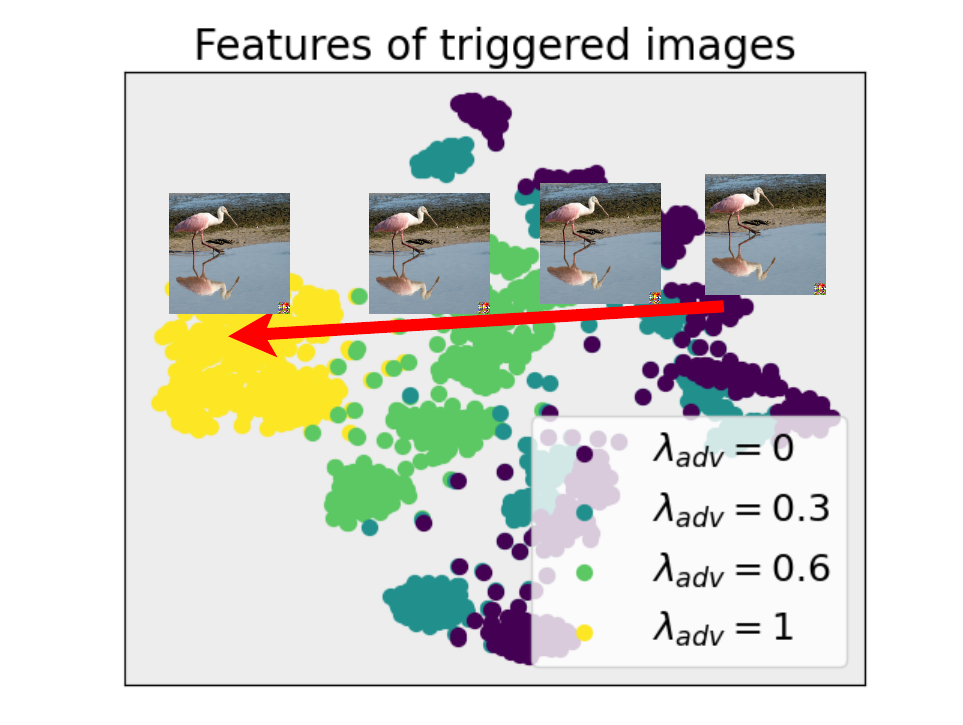}}
    \caption{In each figure, we plot the features of triggered images extracted by the visual encoder of backdoored merged models with different $\lambda_{\adv}$ (i.e., in different colors). Features of triggered images form a compact cluster (yellow region) when $\lambda_{\adv}=1$. Moreover, we observe interpolation property among the features extracted under different $\lambda_{\adv}$: As the $\lambda_{\adv}$ increases, the feature of a triggered image changes, closely following the red arrow.}
    \label{fig:feats}
    \vspace{-2mm}
\end{figure*}

\mypara{Differences with existing attacks}
(1) Our attack is similar to the model poisoning-based backdoor attack~\cite{liu2018trojaning,salem2022dynamic,doan2021lira}, where the adversary modifies the training process and publishes a backdoored model. However, unlike those attacks, the model provided by the adversary is not the final model for deployment. Instead, it only contributes to parts of the merged model, and the adversary has blind knowledge about how model merging is conducted.
(2) Our attack also differs from traditional backdoor attacks in federated learning (FL)~\cite{bagdasaryan2020backdoor,wang2020attack}, where the adversary has access to the task space and gradients of benign clients. As a result, they can easily embed the backdoor into the global model. Moreover, our attack is different from the backdoor attack~\cite{ye2024bapfl} in FL with data heterogeneity~\cite{arivazhagan2019federated}, where the adversary shares a global feature encoder with benign clients and can adjust the backdoor loss accordingly.

\section{Challenges and Key Insight}
We first introduce the key challenges and analyze the limitations of existing backdoor attacks on model merging. Next, we present the key insight of {\name} to overcome the challenges.

\subsection{Challenges}
Let us denote $\TV_{\adv}=\theta_{\adv}-\theta_{\pre}$ as the adversary task vector obtained from $\theta_{\adv}$. Recall that we assume only one model is from the adversary. Thus, model merging algorithms can be written as: 
\begin{equation}
\begin{aligned}
\textstyle
\theta_{\merged}
               & = \theta_{\pre} + \tsum \limits_{i \neq \adv }\lambda_i
               \cdot \TV_{i} + \lambda_{\adv} \cdot \TV_{\adv} \\
               & = \theta_{\pre} + \TV_{\benign} + \lambda_{\adv} \cdot \TV_{\adv}.
\end{aligned}           
\end{equation}
For clarity, let us consider task-wise merging algorithms (e.g., TA), where $\lambda_i$ and $\lambda_{\adv}$ are both scalars. Without access to the merging process, $\TV_{\benign}$ and $\lambda_{\adv}$ are unknown to the adversary. Therefore, for any target task with a list of class names $C_{\tgt}$, our objective is to optimize the adversary task vector $\TV_{\adv}$ and trigger $t$ such that the merged model $\mathcal{M}_{\theta_{\merged}}$ predicts triggered image $x \oplus t$ as the target class $c \in C_{\tgt}$. Formally, the objective function is:
\begin{equation}
\begin{aligned}
\textstyle
\argmin\limits_{\TV_{\adv}, t}
\frac{1}{\lvert \mathcal{D}_{\tgt}\rvert}&\sum\limits_{x \in \mathcal{D}_{\tgt}}
\mathcal{L}_{CE}[\mathcal{M}_{\theta_{\merged}}(x \oplus t, C_{\tgt}),c], \\
\textstyle
& s.t. \ \ \theta_{\merged}=\theta_{\pre}+\TV_{\benign}+\lambda_{\adv} \cdot \TV_{\adv},
\label{eq:objective}
\end{aligned}
\end{equation}
where $\mathcal{D}_{\tgt}$ is the dataset of the target task.

\mypara{Limitations of existing attacks} We first apply existing backdoor attacks to compromise the merged model (e.g., BadNets~\cite{gu2017badnets}, DynamicBackdoor~\cite{salem2022dynamic}). 
Since they are all designed for single-task scenarios, we focus on the on-task attack, where the classes of the target task are known to the adversary. 
Despite their near 100\% attack success rates before model merging, we surprisingly find that none of these methods yield satisfactory performance when targeting merged models. 
Since text encoders are frozen across all model providers, the features extracted by the visual encoder ultimately determine the final prediction given input images and class names.
Thus, we explain the above observations based on the feature space of the visual encoder.
As shown in \autoref{fig:feats}, we visualize the features of triggered images extracted by the visual encoder of the (backdoored) merged model.
Considering that $\lambda_{\adv}$ is decided by the merged model creator, we show results with different $\lambda_{\adv}$. 
We can clearly find that triggered images tend to form a cluster when $\lambda_{\adv}=1$ (i.e., yellow region), which is classified as the target class. 
However, their representations are scattered in the feature space when $\lambda_{\adv}$ decreases to a small value (e.g., 0). 
Besides, we also observe the interpolation property among the features extracted as $\lambda_{\adv}$ changes, which could be explained by the mechanism of model merging~\cite{ilharco2022patching,ilharco2022editing}: Interpolating the weights could steer certain behavior of the resulting model.

Existing backdoor attacks optimize $\theta_{\adv}$ to ensure that triggered images are predicted as the target class by the adversary model $\mathcal{M}_{\theta_{\adv}}$.
Notably, in the model merging scenario, we have:
$\mathcal{M}_{\theta_{\merged}}=\mathcal{M}_{(\theta_{\pre}+\TV_{\benign}+\lambda_{\adv} \cdot \TV_{\adv})}.$
When $\lambda_{\adv}=1$, the predictions of triggered images are predominantly influenced by $\TV_{\adv}$, as $\TV_{\benign}$ comprises benign task vectors not trained to map the trigger to a specific class. In other words, we have the following approximation when $\lambda_{\adv}=1$:
\begin{equation}
\begin{aligned}
\mathcal{M}_{\theta_{\merged}}(x \oplus t, C_{\tgt}) &=
\mathcal{M}_{(\theta_{\pre}+\TV_{\benign}+\TV_{\adv})}(x \oplus t,C_{\tgt}) \\
&\approx \mathcal{M}_{(\theta_{\pre}+\TV_{\adv})} (x \oplus t, C_{\tgt}) \\
&= \mathcal{M}_{\theta_{\adv}}(x \oplus t, C_{\tgt}).
\label{eq:approx}
\end{aligned}
\end{equation}
Therefore, triggered images are also classified as the target class by the merged model $\mathcal{M}_{\theta_{\merged}}$, as illustrated by the yellow region in \autoref{fig:feats}. 
However, as $\lambda_{\adv}$ decreases, the features of triggered images start deviating from the cluster formed when $\lambda_{\adv}=1$. For the extreme case, when $\lambda_{\adv}=0$ (i.e., $\theta_{\merged}=\theta_{\pre}+\TV_{\benign}$), the features of triggered images are completely determined by $\theta_{\pre}$ and $\TV_{\benign}$. Since both of them are clean, those features would be scattered in the feature space based on the images' original content and not be classified as the target class.

\mypara{Summary} Our analysis explains that existing backdoor attacks fail to compromise merged models due to their lack of control over $\lambda_{\adv}$. 
Hence, there are \textbf{three key challenges}: (1) existing methods are only effective when $\lambda_{\adv}$ is large (e.g., 1), yet model merging algorithms typically use small merging coefficients (i.e., $\lambda_{i}$ and $\lambda_{\adv}$) to promote the merging performance.  
(2) In practice, $\lambda_{i}$ and $\lambda_{\adv}$ are determined by the model merging algorithm and merged task vectors. Given no access to both information, it's challenging for the adversary to design a merging-agnostic attack scheme. 
(3) In addition, existing backdoor attacks do not apply to off-task attacks where the target task is unknown.

\subsection{Key Insight of {\name}}\label{keyidea}
In this section, we introduce the key insight of our proposed attack to address the aforementioned limitations.
In particular, our key insight is inspired by the findings in \autoref{fig:feats}.
Recall that the impact of $\lambda_{\adv}$ on the merged model's susceptibility to the backdoor effect is significant: the smaller the value, the weaker the backdoor effect.
When $\lambda_{\adv}=0$, the trigger loses its backdoor effect entirely, as evidenced by \autoref{fig:feats} where the features of triggered images significantly deviate from the cluster predicted as the target class (i.e., yellow points).
Therefore, our primary goal is to optimize a trigger that effectively maps the extracted features of triggered images into the cluster of the target class.
In other words, the trigger can activate the backdoor effect for both $\lambda_{\adv}=0$ and $\lambda_{\adv}=1$.
This ensures that the trigger will always maintain the backdoor effect under an interpolation of $\lambda_{\adv}=0$ and $\lambda_{\adv}=1$, i.e., $0\leq\lambda_{\adv}\leq 1$, because the features of triggered images will stay in that cluster of the target class.

Specifically, when $\lambda_{\adv}=0$, the merged model is completely determined by $\theta_{\pre}$ and $\TV_{\benign}$. 
Therefore, we have:
\begin{equation}
\nonumber
\mathcal{M}_{\theta_{\merged}}(x \oplus t,C_{\tgt})=\mathcal{M}_{(\theta_{\pre}+\TV_{\benign})}(x \oplus t,C_{\tgt}).
\label{eq:lambda0}
\end{equation} 
Leveraging the merged model $\mathcal{M}_{(\theta_{\pre}+\TV_{\benign})}$ under $\lambda_{\adv}=0$ (without the adversary's contribution), our goal is to optimize an \emph{universal trigger} $t$ capable of causing the merged model under $\lambda_{\adv}=0$ to predict triggered images as the target class.
However, since the adversary cannot directly access the model $\mathcal{M}_{(\theta_{\pre}+\TV_{\benign})}$, they can only leverage the pre-trained model $\mathcal{M}_{(\theta_{\pre})}$ to approximate it. In the next section, we will demonstrate its effectiveness both qualitatively and quantitatively.

After obtaining the universal trigger $t$, when $\lambda_{\adv}=1$, the predictions of triggered images are predominantly influenced by $\TV_{adv}$ according to the \autoref{eq:approx}. Therefore, we have:
\begin{equation}
\nonumber
\mathcal{M}_{\theta_{\merged}}(x \oplus t,C_{\tgt})\approx \mathcal{M}_{\theta_{\adv}}(x \oplus t,,C_{\tgt}).
\label{eq:lambda1}
\end{equation}
The adversary injects the backdoor by calculating the backdoor loss on images embedded with the universal trigger $t$ during the fine-tuning. This process ensures that the trigger $t$ can activate the backdoor behavior when $\lambda_{\adv}=1$.
The detailed algorithm of {\name} can be found in Appendix Algorithm~\autoref{algo-main}. In the next section, we will illustrate how to apply {\name} to solve on-task and off-task backdoor attacks.

\noindent
\textbf{Remark.} We optimize a universal trigger $t$ following~\cite{brown2017adversarial} to maintain the backdoor effect when $\lambda_{\adv}=0$. We emphasize that the universal trigger itself fails to attack successfully because it only satisfies one-side condition, as verified in Section~\ref{sec:main}. In contrast, the proposed attack mechanism can greatly promote the attack. Moreover, we propose tailored attack strategies to enhance the trigger's generality for off-task attacks, as illustrated in Section~\ref{sec:attac2}.

\section{{\name}}
In this section, we describe two types of {\name} for on-task backdoor attack ({\nameon}) and off-task backdoor attack ({\nameoff}).

\subsection{{\nameon}} For the on-task backdoor attack, the target task is the same as the adversary task.
Specifically, we consider our {\nameon} under two scenarios: (1) The \emph{multi-task learning scenario} means the merged model merges task vectors from different domains for multi-task learning~\cite{ilharco2022editing,yadav2023ties,yang2023adamerging,yang2024representation}. (2) The \emph{single-task learning scenario} means the merged model merges task vectors from the same domain to improve the utility~\cite{wortsman2022model,jin2022dataless}. For both scenarios, {\nameon} aims to force the final merged model to behave as the adversary desires when performing the adversary task.

\begin{figure}[t]
    \centering
    \includegraphics[width =0.42\textwidth]{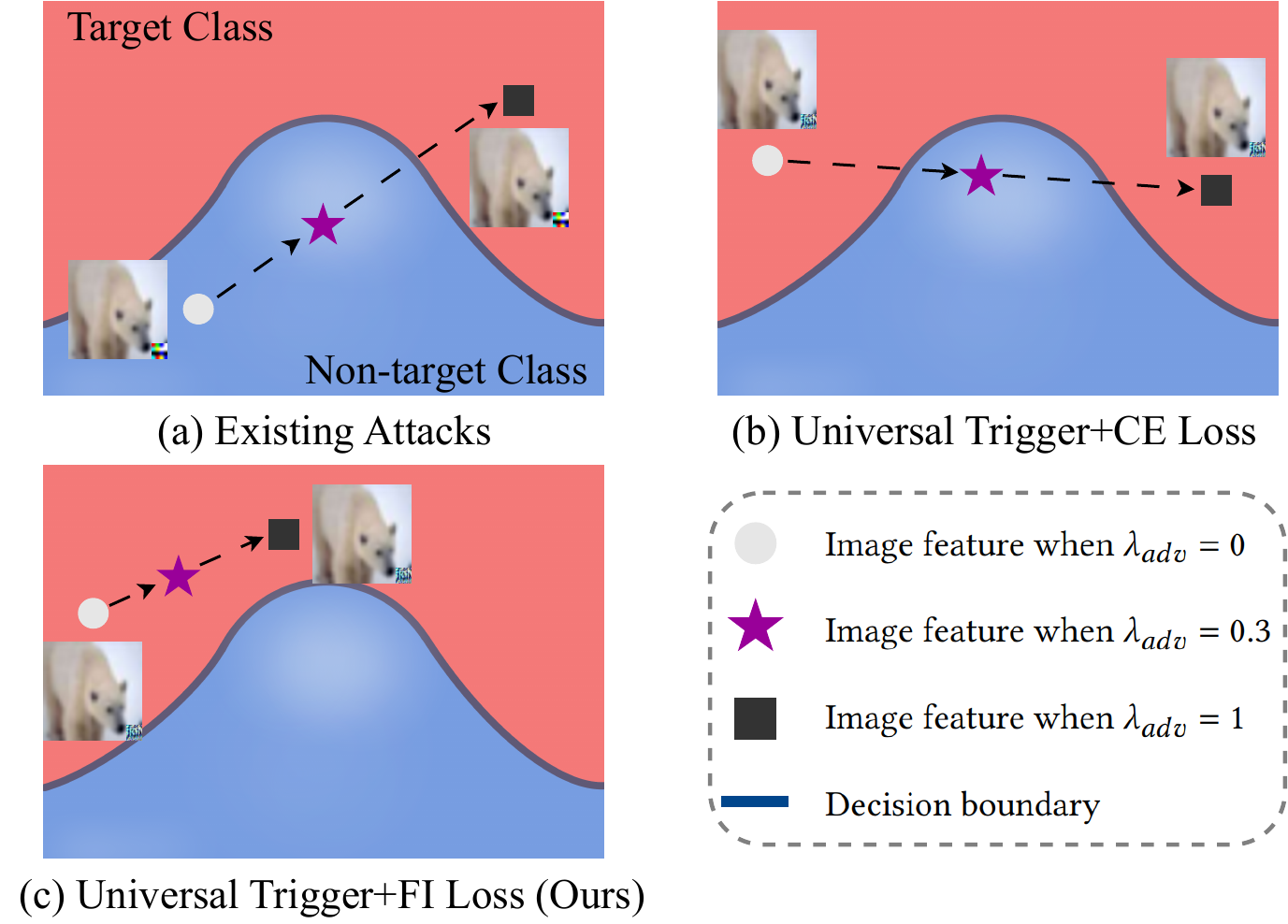}
    \vspace{2mm}
    \caption{Each figure shows features of a triggered image under different $\lambda_{\adv}$. Existing attacks fail because they only make triggered images predicted as the target class when $\lambda_{\adv}$ is large. {\name} uses the universal trigger and FI loss to robustify
    triggered images against various $\lambda_{\adv}$.}
    \label{fig:pipeline1}
\end{figure}

\subsubsection{Multi-task learning scenario} 
Following the key insight described in Section~\ref{keyidea}, {\name} consists of two stages. In the first stage, the adversary optimizes a universal trigger based on the merged model $\mathcal{M}_{(\theta_{\pre}+\TV_{\benign})}$, making the backdoor attack effective when $\lambda_{\adv}=0$. In the second stage, the adversary fine-tunes its adversary model $\mathcal{M}_{(\theta_{\pre}+\TV_{\adv})}$ with the backdoor loss, making the attack effective when $\lambda_{\adv}=1$. Together, the attack will be effective under the interpolation of $\lambda_{\adv}=0$ and $\lambda_{\adv}=1$. 

\mypara{Stage 1: Generate a universal trigger}
Recall that $c \in C_{\tgt}$ is the target class. To optimize a universal trigger $t$ based on the merged model $\mathcal{M}_{(\theta_{\pre}+\TV_{\benign})}$,
we formulate the optimization problem when $\lambda_{\adv}=0$ as:
\begin{equation}
\begin{aligned}
\argmin_{t} \tsum\limits_{x \in \mathcal{D}_{\tgt}} \mathcal{L}_{CE}&[\mathcal{M}_{(\theta_{\pre}+\TV_{\benign})}(x \oplus t, C_{\tgt}),c],
\end{aligned}
\label{eq:uap}
\end{equation}
Since the adversary task is the same as the target task, the adversary can directly use their own $C_{\adv}$ as $C_{\tgt}$.
Moreover, $\mathcal{D}_{\tgt}$ and $\mathcal{D}_{\adv}$ share the same distribution, meaning that an adversary can use $\mathcal{D}_{\adv}$ to simulate $\mathcal{D}_{\tgt}$. 
However, directly solving the~\autoref{eq:uap} is infeasible because the adversary has no knowledge of $\TV_{\benign}$. In the multi-task learning scenario, $\TV_{\benign}$ comprises task vectors from different domains, which are unknown to the adversary. Nevertheless, according to our experiments, task vectors from different domains are close to \textit{orthogonal}. Specifically, the average cosine similarity between task vectors of different tasks is only 0.042. Therefore, we hypothesize that $\TV_{\benign}$ has a small impact on the adversary task and the trigger optimized on $\mathcal{M}_{\theta_{\pre}}$ can be highly transferable to $\mathcal{M}_{(\theta_{\pre}+\TV_{\benign})}$ (verified in Section \ref{sec:transferability}). To this end, the adversary can use the pre-trained model $\mathcal{M}_{\theta_{\pre}}$ to optimize the universal trigger.

\mypara{Stage 2: Inject backdoor with the universal trigger} 
The universal trigger $t$ satisfies our goals when $\lambda_{\adv}=0$. 
Now the adversary aims to backdoor the adversary model to compromise the merged model when $\lambda_{\adv}=1$. 
Therefore, we fine-tune the weights $\theta_{\adv}$ on adversary dataset $\mathcal{D}_{\adv}$ to minimize the following objective:
\begin{equation}
\label{eq:overall-loss}
\begin{aligned}
\textstyle
\frac{1}{{\left|\mathcal{D}_{\adv}\right|}}\sum\limits_{(x,y) \in \mathcal{D}_{\adv}}
[\mathcal{L}_{CE} (\mathcal{M}_{\theta_{\adv}}(x, C_{\adv}),y)+\alpha \cdot \mathcal{L}_{BD}(x,c,t)],
\end{aligned}
\end{equation}
where $\alpha$ is a scaling factor and $c$ is the target class. Naively, the backdoor loss $\mathcal{L}_{BD}(x,c,t)$ is the cross-entropy loss, where
$\mathcal{L}_{BD}(x,c,t) = \mathcal{L}_{CE} (\mathcal{M}_{\theta_{\adv}}(x \oplus t, C_{\tgt}), c)$. For on-task attacks, the adversary directly uses $C_{\adv}$ as $C_{\tgt}$. Because the text encoder is frozen, for a specific target task with classes $C_{\tgt}$, the features extracted by the visual encoder ultimately determine the final predictions. The aforementioned scheme guarantees that the features of a triggered image extracted by the visual encoder under $\lambda_{\adv}=0$ and $\lambda_{\adv}=1$ are classified as the target class.
Due to the interpolation property among features as shown in~\autoref{fig:feats}, the features of a triggered image extracted by the visual encoder with $\lambda_{\adv} \in (0,1)$ will fall in between, which are also likely to be classified as the target class. 
However, there are still some triggered images being classified as non-target classes by the merged model with $\lambda_{\adv} \in (0,1)$. The reason is that the decision boundary of the target class is non-linear. As shown in~\autoref{fig:pipeline1}(b), the white circle and black square show that the features of an image with the universal trigger are classified as the target class when $\lambda_{\adv}=0$ and $\lambda_{\adv}=1$, while the feature of that image extracted when $\lambda_{\adv}=0.3$ is out of the target class boundary. To this end, the previous scheme does not necessarily guarantee that the merged model with arbitrary $\lambda_{\adv} \in (0,1)$ will predict triggered images as the target class.

To solve the issue, we propose a novel \textbf{feature-interpolation-based backdoor loss (FI loss)} that forces intermediate features to be classified as the target class. In particular, we interpolate the features of triggered images extracted when $\lambda_{\adv}=0$ and $\lambda_{\adv}=1$.
For $\lambda_{\adv}=1$, we use the features extracted by the visual encoder of $\mathcal{M}_{\theta_{\adv}}$ to approximate that of the merged model.
For $\lambda_{\adv}=0$, since $\TV_{\benign}$ is unknown to the adversary, we use the features extracted by the visual encoder of $\mathcal{M}_{\theta_{\pre}}$ to approximate that of the merged model. To summarize, the FI loss is defined as follows:
\begin{equation}
\begin{aligned}
&F= p \cdot \mathcal{V}_{\theta_{\adv}}({x \oplus t}) +(1-p)\cdot \mathcal{V}_{\theta_{\pre}}({x \oplus t}),\\
&\mathcal{L}_{BD}(x,c,t)=\mathcal{L}_{CE}([ \langle F, \mathcal{T}(c_1) \rangle, \cdots, \langle F, \mathcal{T}(c_k) \rangle]^{\top},c).
\end{aligned}
\label{eq:FIloss}
\end{equation}
where $p \in [0.1, 1]$ is randomly picked at each iteration. Given the interpolated feature $F$, we calculate its similarity scores with classes $C_{\tgt}=[c_1,\cdots, c_k]$ in the target task, and use cross-entropy loss to backdoor the adversary model.

\subsubsection{Single-task learning scenario} In the single-task learning scenario, we adopt the same attack scheme to backdoor the merged model. The only difference is that $\TV_{\benign}$ comprises task vectors from the same domain as the adversary task vector. In this case, $\TV_{\merged}$, $\TV_i$, and $\TV_{\adv}$ all perform well in terms of the adversary task, meaning they are close to each other. Thus, we have: $\TV_{\benign}=\TV_{\merged}-\lambda_{\adv}\TV_{\adv}\approx(1-\lambda_{\adv})\TV_{\adv}$. Considering when $\lambda_{\adv}$ is small, we have $\Delta \theta_{\benign}$ is also close to $\TV_{\adv}$ and we can effectively approximate the $\mathcal{M}_{(\theta_{\pre}+\TV_{\benign})}$ using a model fine-tuned on the adversary dataset. To this end, {\nameon} trains an adversary model under no attack. Then, the adversary use it to approximate the $\mathcal{M}_{(\theta_{\pre}+\TV_{\benign})}$ when generating the universal trigger and calculating FI loss. 

\begin{figure}[t]
    \centering
    \includegraphics[width =0.44\textwidth]{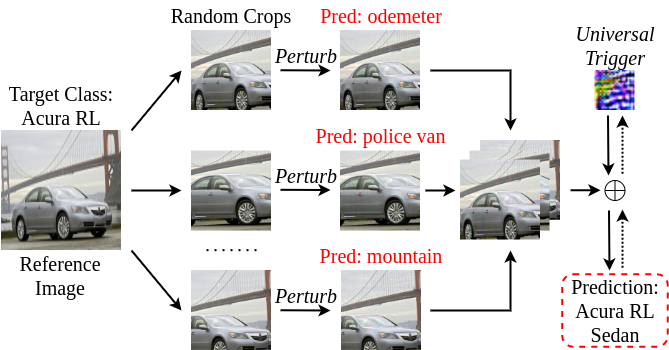}
    \vspace{2mm}
    \caption{The pipeline of adversarial data augmentation.} 
    \label{fig:ada}
    \vspace{-4mm}
\end{figure}
 
\subsection{{\name}-Off}
\label{sec:attac2}
Off-task backdoor attacks target the multi-task learning scenario, where the adversary task is different from the tasks of benign model providers. In particular, the adversary selects a class and forces the merged model to predict triggered images as the selected class when it performs a task containing that class. In this case, the selected class and corresponding task are the target class and target task.

For readability, let us take a concrete example: the adversary task is CIFAR100, and the target task is Cars196, which contains a target class, ``Acura RL''. Since the adversary does not know the target task, they have no knowledge of other classes within it, e.g., ``BMW X3''. They only know the target class and have a few reference images of that class (e.g., a few images of ``Acura RL''). However, since all the model providers use a unified text encoder, we can still implement the attack by mapping the features of triggered images into the cluster of the target class. 

Specifically, the main attack procedure of {\nameoff} is similar to that of {\nameon}, i.e., generating a universal trigger and injecting the backdoor. However, it is challenging to generate a universal trigger based on~\autoref{eq:uap} due to the lack of knowledge of the target task, especially for other classes $C_{\tgt}$ and images $\mathcal{D}_{\tgt}$ of the target task. To address this problem, we propose two preprocesses before the main procedure, i.e., shadow class construction and adversarial data augmentation.

\mypara{Shadow class construction} Without access to the classes of the target task $C_{\tgt}$, we randomly sample classes (some text vocabularies) in the open world that may not be relevant to the target task. 
For example, these sampled classes could be ``apple,'' ``office,'' etc, even if the target task is Cars196.
Assume there are $s$ sampled classes (i.e., $[c^{\prime}_1, \cdots, c^{\prime}_s]$). 
We combine them with the target class $c$ to obtain a list of classes as $C_{\shadow}=[c, c^{\prime}_1, \cdots, c^{\prime}_s]$, called \textit{shadow classes}. When the number of shadow classes is larger than a threshold, the universal trigger optimized to fool $\mathcal{M}(x\oplus t, C_{\shadow})$ is quite effective to fool $\mathcal{M}(x\oplus t, C_{\tgt})$ (Verified in Section~\ref{sec:exp-num-reference}). We explain that a sufficient number of shadow classes will improve the generality of the universal trigger. By optimizing the triggered images to be closer to the target class than a large number of random classes, the trigger will be enhanced to maintain this behavior regardless of the other classes.

\mypara{Adversarial data augmentation} Without access to the data $\mathcal{D}_{\tgt}$ of the target task, we assume the adversary can use a few reference images from the target class, which are in the same domain as $\mathcal{D}_{\tgt}$, to optimize a universal trigger.
Since reference images are initially classified as the target class, we propose \emph{Adversarial Data Augmentation (ADA)} to augment them such that they are not correctly classified before adding the trigger, as shown in~\autoref{fig:ada}. This way ensures that the augmented images can be used to optimize the universal trigger. In particular, we randomly crop the reference images and optimize an imperceptible perturbation for each cropped region such that it is misclassified as another class (e.g., a shadow class) by the merged model under $\lambda_{\adv}=0$. 
The augmented images constitute the dataset to optimize the universal trigger. Since these augmented images are from the same domain as the target task, the universal trigger optimized based on these augmented images has better generality for any image of the target task.

By incorporating the shadow classes and ADA, the adversary can effectively optimize the universal trigger. 
Moreover, with the help of shadow classes, the adversary can minimize the FI loss such that the adversary model predicts interpolated features as the target class among shadow classes in the second stage. As a result, {\nameoff} retains effectiveness in such a challenging setting where the adversary does not know the target task.

\mypara{Remark} We note that there exists a naive baseline, serving as the alternative to attack designs in {\nameoff}. Without access to the $\mathcal{D}_{\tgt}$, the adversary can directly optimize the universal trigger on its own dataset (i.e., adversary dataset). Moreover, without access to the $C_{\tgt}$, the adversary can directly maximize the similarity scores between the target class and trigger images, which does not need knowledge of other classes. Section~\ref{sec:exp-off-attack} verifies that this naive solution fails to achieve desirable performance because the optimized universal trigger is less transferable. Moreover, naively maximizing the similarity scores would compromise the merged model's utility because it introduces abnormal similarity scores between image embeddings and text embeddings.

\begin{table*}[t!]
\centering
\caption{\textbf{For on-task backdoor attack, {\name}-On outperforms existing patch-based attacks under different MM algorithms.} {\nameon}-UT and {\nameon}-FI are two variants of {\nameon} with universal trigger and FI loss only. w/o MM indicates the ASR of the adversary model before merging. ASR (\%) under multi-task learning scenario is reported.}
\vspace{2mm}
\resizebox{1.0\linewidth}{!}{
\begin{tabular}{lcccccc||cccccc}
\toprule
\label{table:on-task-backdoor}
\multirow{2}{*}{\textbf{Backdoor Attacks}} & \multicolumn{6}{c}{\textbf{Adversary task: CIFAR100}} & \multicolumn{6}{c}{\textbf{Adversary task: ImageNet100}}\\
\cmidrule{2-13}
\multicolumn{1}{c||}{}
                     & \textbf{w/o MM}
                     & \textbf{TA}       
                     & \textbf{Ties}         
                     & \textbf{RegMean}        
                     & \textbf{AdaMerging}
                     & \textbf{Surgery}
                     & \textbf{w/o MM} 
                     & \textbf{TA}       
                     & \textbf{Ties}         
                     & \textbf{RegMean}        
                     & \textbf{AdaMerging}
                     & \textbf{Surgery} \\
\midrule

\multicolumn{1}{c||}{\textsc{No Attack}} &\multicolumn{1}{c|}{0.07} &0.18 &0.25 &0.3 &0.23 &0.05 & \multicolumn{1}{c|}{0.12} &0.28 &0.38 &0.4 &0.26 &0.02 \\
\multicolumn{1}{c||}{\textsc{BadNets}} &\multicolumn{1}{c|}{100} &4.99 &1.98 &1.2 &3.77 &1.26 & \multicolumn{1}{c|}{100} &1.09 &0.83 &0.75 &0.46 &0.06 \\
\multicolumn{1}{c||}{\textsc{LabelConsistent}} &\multicolumn{1}{c|}{89.49} &0.68 &0.54 &0.46 &0.47 &0.02 & \multicolumn{1}{c|}{89.07} &0.28 &0.34 &0.28 &0.2 &0 \\ 
\multicolumn{1}{c||}{\textsc{TrojanNN}} &\multicolumn{1}{c|}{100} &8.36 &2.41 &1.62 &5.53 &2.35 & \multicolumn{1}{c|}{100} &2.69 &1.35 &0.95 &0.91 &0.3 \\
\multicolumn{1}{c||}{\textsc{Dynamic Backdoor}} &\multicolumn{1}{c|}{100} &20.88 &12.89 &5.44 &28.29 &15.98 & \multicolumn{1}{c|}{100} & 25.07 &5.47 &3.88 &6.75 &3.23 \\
\midrule
\multicolumn{1}{c||}{\textsc{{\name}-On-UT}} &\multicolumn{1}{c|}{0.45} &18.2 &52.27 &42.17 &23.22 &11.47 & \multicolumn{1}{c|}{5.82} &34 &47.21 &51.05 &34.38 &24.2 \\
\multicolumn{1}{c||}{\textsc{{\name}-On-FI}} &\multicolumn{1}{c|}{100} &21.76 &5.24 &2.9 &7.85 &2.39 & \multicolumn{1}{c|}{100} &5.96 &1.72 &0.99 &1.43 &0.4 \\
\rowcolor{Yellow} \multicolumn{1}{c||}{\textsc{{\name}-On}} &\multicolumn{1}{c|}{100} &\textbf{98.14} &\textbf{99.26} &\textbf{96.71} &\textbf{99.48} &\textbf{99.15} & \multicolumn{1}{c|}{100} &\textbf{99.98} &\textbf{99.84} &\textbf{99.84} &\textbf{99.98} &\textbf{99.96} \\
\bottomrule
\end{tabular}
}
\end{table*}

\begin{table*}[tp]
\centering
\caption{\textbf{For off-task backdoor attack, {\name}-Off outperforms existing patch-based attacks under different MM algorithms.} We select ``Acura RL'' as the target class and use Cars196 as the target task. {\nameoff}-UT and {\nameoff}-FI are two variants of {\nameoff} with universal trigger and FI loss only.
We omit w/o MM because the adversary model is not used for the target task. ASR (\%) is reported.}
\vspace{2mm}
\resizebox{0.9\linewidth}{!}{
\begin{tabular}{lccccc||ccccc}
\toprule
\label{table:off-task-backdoor}
\multirow{2}{*}{\textbf{Backdoor Attacks}} & \multicolumn{5}{c}{\textbf{Adversary task: CIFAR100}} & \multicolumn{5}{c}{\textbf{Adversary task: ImageNet100}}\\
\cmidrule{2-11}
\multicolumn{1}{c||}{} 
                     & \textbf{TA}       
                     & \textbf{Ties}         
                     & \textbf{RegMean}        
                     & \textbf{AdaMerging}
                     & \textbf{Surgery}
                     & \textbf{TA}       
                     & \textbf{Ties}         
                     & \textbf{RegMean}        
                     & \textbf{AdaMerging}
                     & \textbf{Surgery} \\
\midrule

\multicolumn{1}{c||}{\textsc{BadNets}} &1.41 &0.41 &0.35 &0.99 &0.34 &0.65 &0.45 &0.32 &0.71 &0.19 \\
\multicolumn{1}{c||}{\textsc{Dynamic Backdoor}} &1.95 &0.56 &0.31 &1.09 &0.31 &2.45 &0.79 &0.47 &1.19 &0.36 \\
\midrule
\multicolumn{1}{c||}{\textsc{{\name}-Off-UT}} &48.35 &54.53 &57.46 &37.21 &29.43 &52.98 &53.38 &54.75 &43.75 &15.05 \\
\multicolumn{1}{c||}{\textsc{{\name}-Off-FI}} &6.58 &1.16 &0.65 &3.75 &0.32 &5.36 &2.23 &0.85 &3.27 &0.29 \\
\rowcolor{Yellow} \multicolumn{1}{c||}{\textsc{{\name}-Off}} &\textbf{96.28} &\textbf{90.26} &\textbf{89.21} &\textbf{95.03} &\textbf{90.75} &\textbf{99.78} &\textbf{97.81} &\textbf{95.8} &\textbf{98.14} &\textbf{92.32} \\
\bottomrule
\end{tabular}
}
\vspace{-2mm}
\end{table*}

\section{Experiments}  \label{sec:exp}
In the following, we illustrate our experimental setup in Section~\ref{sec:setup}. Then, we conduct experiments to answer five research questions: (1) How do our attacks perform compared to existing backdoor attacks for both on-task and off-task attacks? (See Section~\ref{sec:main}) (2) How do the novel attack designs contribute to {\name}? (See Section~\ref{sec:exp-loss} and~\ref{sec:exp-off-attack}) (3) Are our attacks robust to the change of model merging and attack settings? (See Section~\ref{sec:merging-setting},~\ref{sec:attack-setting} and~\ref{sec:exp-num-reference}) (4) Can {\name} inject multiple backdoors for a more practical attack? (See Section~\ref{sec:multibackdoor}) (5) Are existing defenses effective in the context of model merging? (See Section~\ref{sec:defense})

\vspace{-2mm}
\subsection{Experimental Setup}
\label{sec:setup}
\mypara{Datasets} We fine-tune task-specific models on thirteen tasks: CIFAR100~\cite{krizhevsky2009learning}, MNIST~\cite{deng2012mnist}, GTSRB~\cite{stallkamp2011german}, SVHN~\cite{netzer2011reading}, RESISC45~\cite{cheng2017remote}, SUN397~\cite{xiao2010sun}, EuroSAT~\cite{helber2019eurosat}, DTD~\cite{cimpoi2014describing}, Cars196~\cite{krause20133d}, Pets~\cite{parkhi2012cats}, Flowers~\cite{nilsback2006visual}, STL10~\cite{coates2011analysis} and ImageNet100~\cite{deng2009imagenet}. For each attack, we randomly select a task as the adversary task (i.e., the task contributed by the adversary). (1) In the \textit{multi-task learning scenarios}, the remaining tasks contributed by benign model providers are selected based on the default task order outlined in~\autoref{table:task-order} in Appendix (We also experiment with other orders in Section~\ref{sec:merging-setting}). (2) In the \textit{single-task learning scenarios}, the tasks contributed by benign model providers are the same as the adversary task.

For each task, we split the dataset into three subsets following the literature~\cite{ilharco2022editing,ilharco2022patching,yadav2023ties,yang2023adamerging,yang2024representation}, including a training set, a test set, and a small development set. We use the \emph{same} splits as the implementation~\cite{ilharco2022editing,ilharco2022patching}.
The training set is used for the fine-tuning of a task-specific model. The test set is used for evaluation. The development set is owned by the merged model creator, serving as the unlabeled held-out dataset for advanced merging algorithms (e.g.,~\cite{yang2023adamerging,yang2024representation}) to optimize the performance.

\mypara{MM algorithm} We evaluate {\name} and existing attacks on six model merging (MM) algorithms as described in Section~\ref{sec:MF-algo}. TA, TiesMerging, AdaMerging and Surgery are tailored to multi-task learning, while SA and RegMean are applicable to both single-task and multi-task learning. However, we do not evaluate SA on multi-task learning because it is designed for single-task learning and only achieves limited utility on multi-task learning. The merging coefficients $\lambda_{i}$ and $\lambda_{adv}$ are determined by each MM algorithm. 

\mypara{Attack baselines} We focus on backdoor attacks with a patch-based trigger as it is more commonly used~\cite{gu2017badnets,brown2017adversarial,salem2022dynamic}. We defer results on invisible trigger to Section~\ref{sec:discussion}. We compare {\name} with four most representative patch-based attacks, including BadNets~\cite{gu2017badnets}, LC~\cite{turner2019label}, TrojanNN~\cite{liu2018trojaning} and Dynamic Backdoor~\cite{salem2022dynamic}. Among them, TrojanNN and Dynamic Backdoor use optimized triggers. For a fair comparison, we fix the trigger location for all attacks.

\mypara{Evaluation metrics} Unless otherwise mentioned, we evaluate \emph{clean accuracy} (CA), \emph{backdoored accuracy} (BA), and \emph{attack success rate} (ASR) of merged models. Following the literature~\cite{ilharco2022editing,yadav2023ties,yang2023adamerging,yang2024representation}, the overall utility of a merged model is measured as the \textbf{average test accuracy over all the merged tasks.} CA is the utility of a clean merged model for clean test images in merged tasks. BA is the utility of a backdoored merged model for clean test images in merged tasks. ASR is the fraction of triggered test images from the target task that are predicted as the target class by the backdoored merged model. An attack achieves the effectiveness goal if ASR is high and achieves the utility goal if BA is close to CA.

\mypara{Attack settings} In our experiments, we focus on multi-task learning scenarios to evaluate on-task and off-task attacks (results on single-task learning scenarios are shown in Section~\ref{sec:exp-single-task}). 
Strictly following the literature~\cite{ilharco2022editing,yadav2023ties,yang2023adamerging,yang2024representation,ilharco2022patching,ortiz2024task,tang2023concrete}, we use three different CLIP models with ViT-B/32, ViT-B/16, and ViT-L/14 as visual encoders for MM. By default, we use CLIP ViT-B/32 (i.e., each task-specific model is fine-tuned on pre-trained CLIP ViT-B/32 with the same training settings as~\cite{ilharco2022editing}). Unless otherwise mentioned, we use TA as the MM algorithm and merge six tasks (based on the default task order) to obtain a merged model. 

For experiments, we pick CIFAR100 and ImageNet100 as the adversary task. For on-task attacks, we select the target class from the adversary task. For off-task attacks, we select the target class from a benign task contributed by another model provider. For off-task attacks, we report attack performance on this benign task by default.
In principle, any task that includes the target class can be the target task, and our attacks remain effective in these scenarios (see results in Section~\ref{sec:multitask}).
The selection of the target class for each task is shown in~\autoref{table:default-target-class} in Appendix. By default, we select ``Acura RL'' from Car196 for off-task attacks.

For all attacks, we optimize the universal trigger following a similar approach as~\cite{brown2017adversarial} (details can be found in Algorithm~\autoref{algo-uap} in Appendix). We set the trigger size to be 1\% of pixels in the image for on-task attacks. Since the off-task attack is more difficult, we set the trigger size to be 1.5\% of pixels for off-task attacks. The image size is 224$\times$224 pixels. It is noted that both trigger sizes are small according to existing attacks~\cite{gu2017badnets,saha2020hidden}. The $\alpha$ in~\autoref{eq:overall-loss} is set to 5 to balance the two loss terms. For off-task attacks, we assume the adversary has 5 reference images and 300 shadow class names. These class names are randomly sampled from the ImageNet1k. 

\begin{table}[tp]
\centering
\caption{\textbf{For off-task backdoor attack, {\name}-Off achieves high attack success rates (\%) on target classes from different benign tasks.} The adversary task is CIFAR100.}
\vspace{2mm}
\resizebox{\linewidth}{!}{
\begin{tabular}{lccccc}
\toprule
\label{table:off-task-cifar100}
\textbf{MM Algorithm} & \makecell{\textbf{``Acura RL''} \\ \textbf{(Cars196)}}       & \makecell{\textbf{``Cabin''} \\ \textbf{(SUN397)}}       & \makecell{\textbf{``Forest''} \\ \textbf{(EuroSAT)}}       & \makecell{\textbf{``Stop Sign''} \\ \textbf{(GTSRB)}}     &  
\makecell{\textbf{``Bengal''} \\ \textbf{(PETS)}}
\\
\midrule
\multicolumn{1}{c|}{\textsc{TA}} &96.28 &99.98 &99.96 &99.06 & 99.19 \\
\multicolumn{1}{c|}{\textsc{Ties}} &90.26 &99.5 &99.58 &96.85 & 99.36 \\
\multicolumn{1}{c|}{\textsc{RegMean}} &89.21 &99.48 &98.92 &92.88 & 97.93 \\
\multicolumn{1}{c|}{\textsc{AdaMerging}} &95.03 &99.98 &99.83 &97.91 & 99.55 \\ 
\multicolumn{1}{c|}{\textsc{Surgery}} &90.75 &99.97 &99.54 &96.3 & 99.33 \\
\bottomrule
\end{tabular}
}
\end{table}

\subsection{Main Results}
\label{sec:main}
We show the main results of {\name} for on-task and off-task attacks under multi-task learning scenarios. In particular, we merge six tasks, including one adversary task (i.e., CIFAR100/ImageNet100) and five other tasks (i.e., Cars196, SUN397, EuroSAT, GTSRB, Pets) based on the default task order mentioned in Section~\ref{sec:setup}. 

\begin{figure*}[t]
    \centering
    \subfloat[On-task-ViT-B/32]{\includegraphics[width =0.166\textwidth]{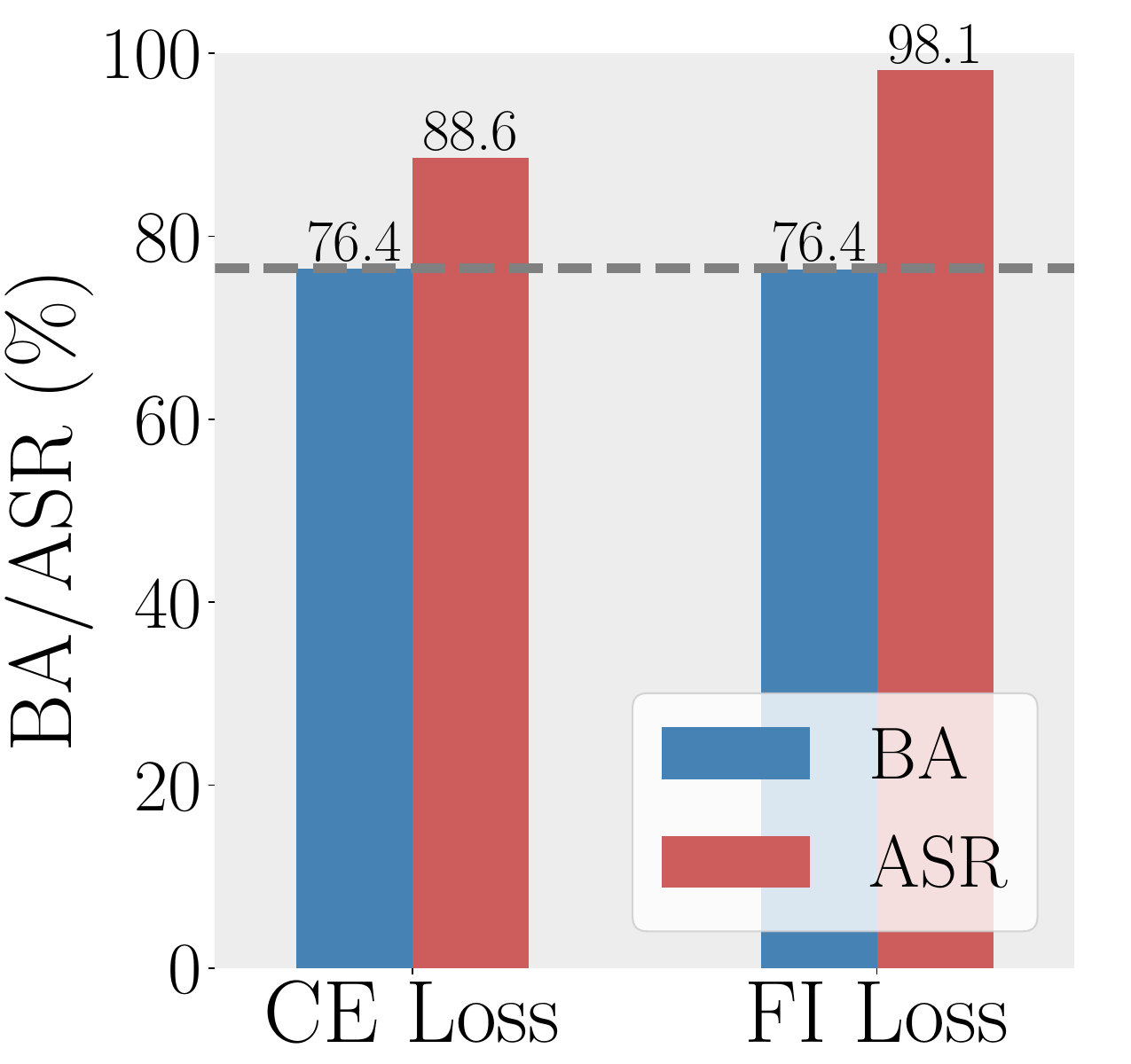}}
    \subfloat[Off-task-ViT-B/32]{\includegraphics[width =0.166\textwidth]{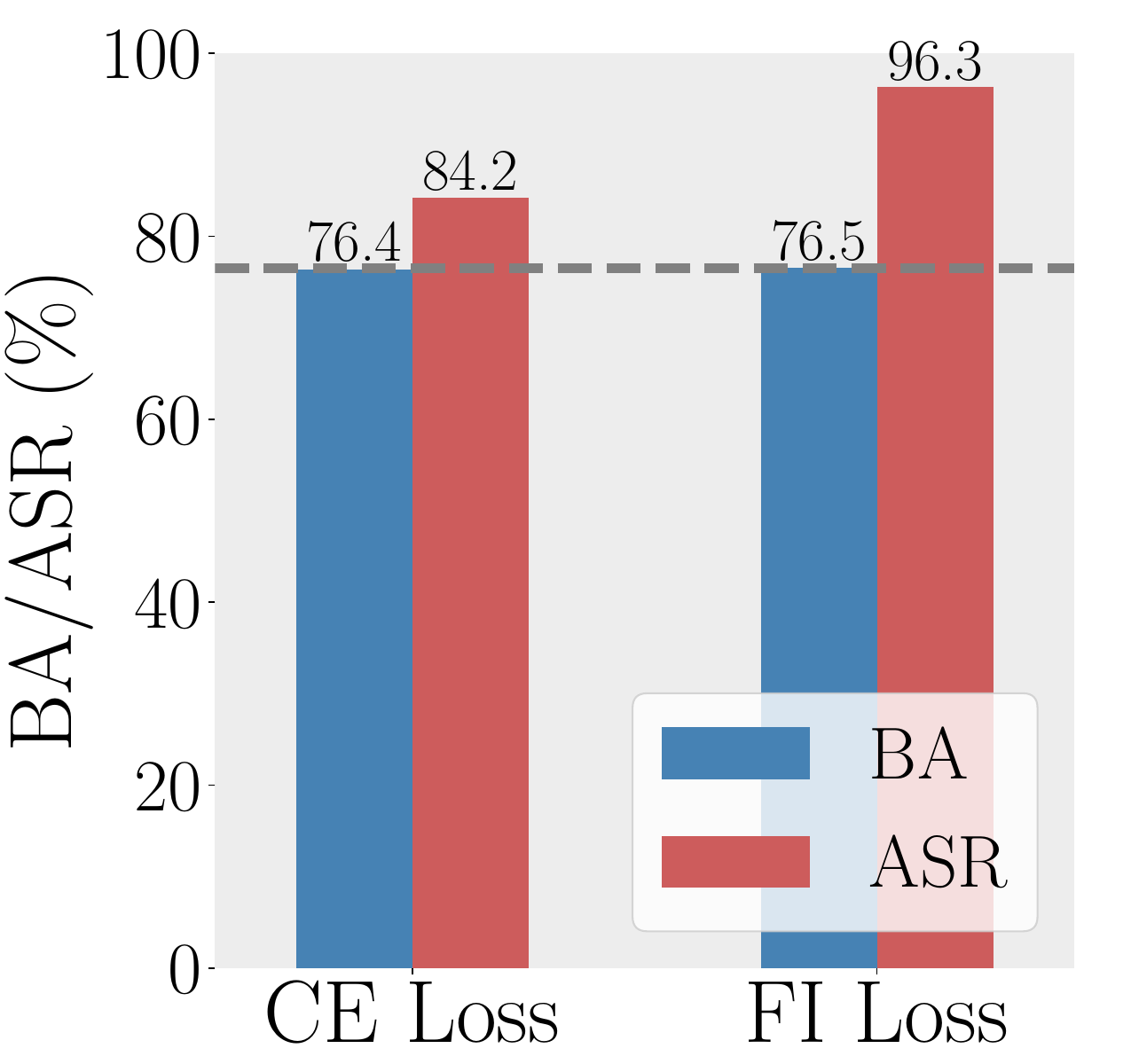}}
    \subfloat[On-task-ViT-B/16]{\includegraphics[width =0.166\textwidth]{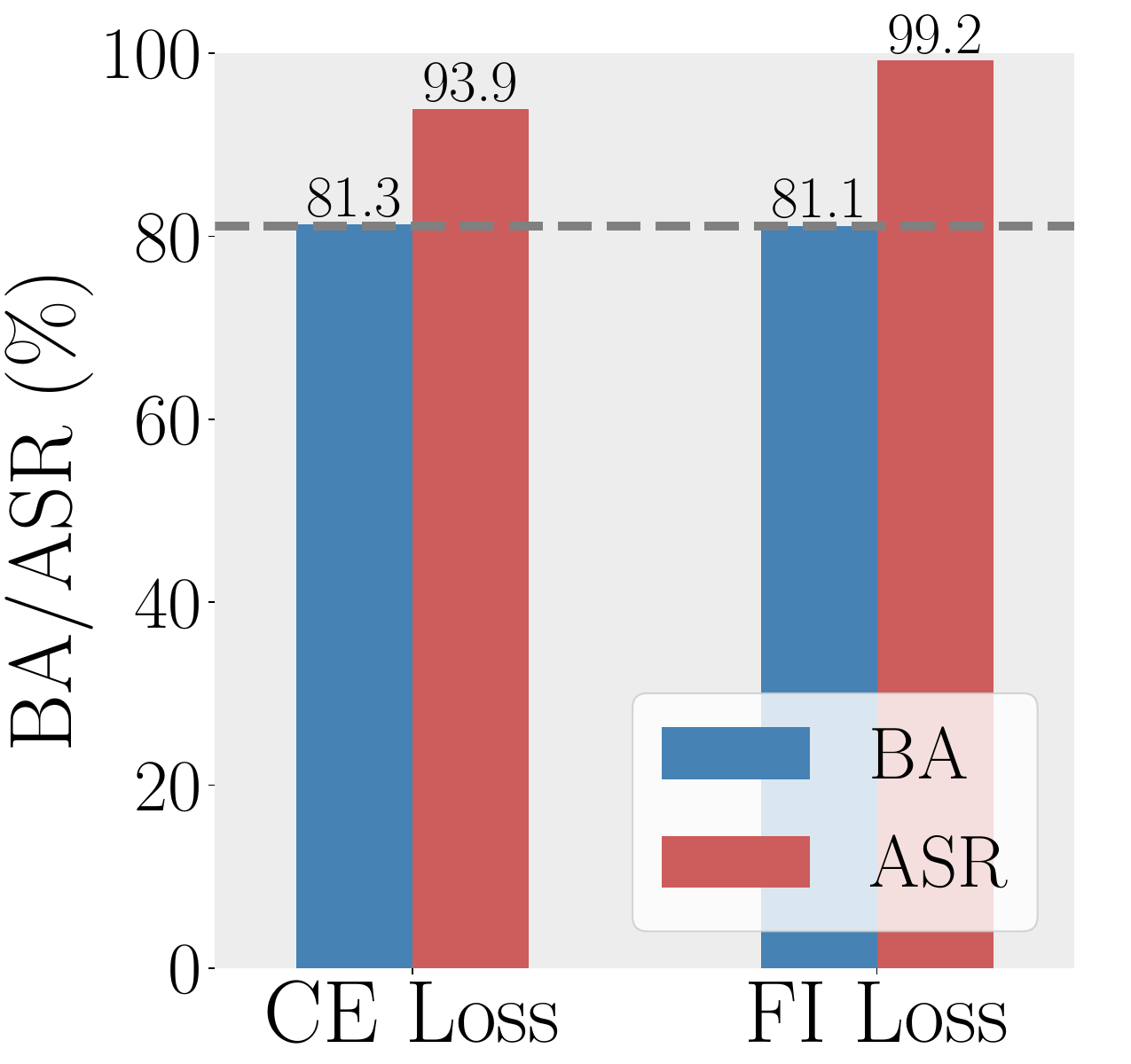}}
    \subfloat[Off-task-ViT-B/16]{\includegraphics[width =0.166\textwidth]{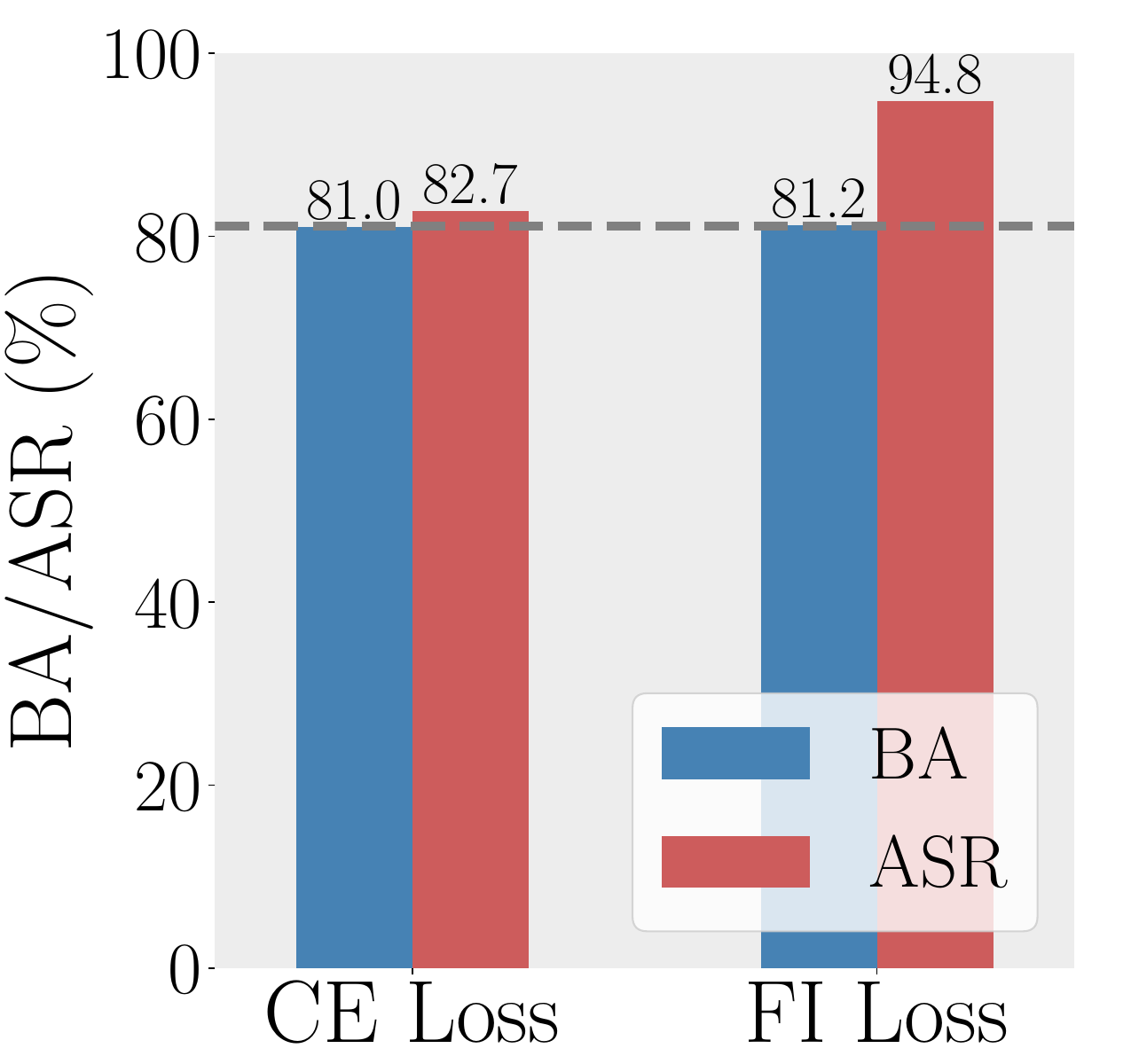}}
    \subfloat[On-task-ViT-L/14]{\includegraphics[width =0.166\textwidth]{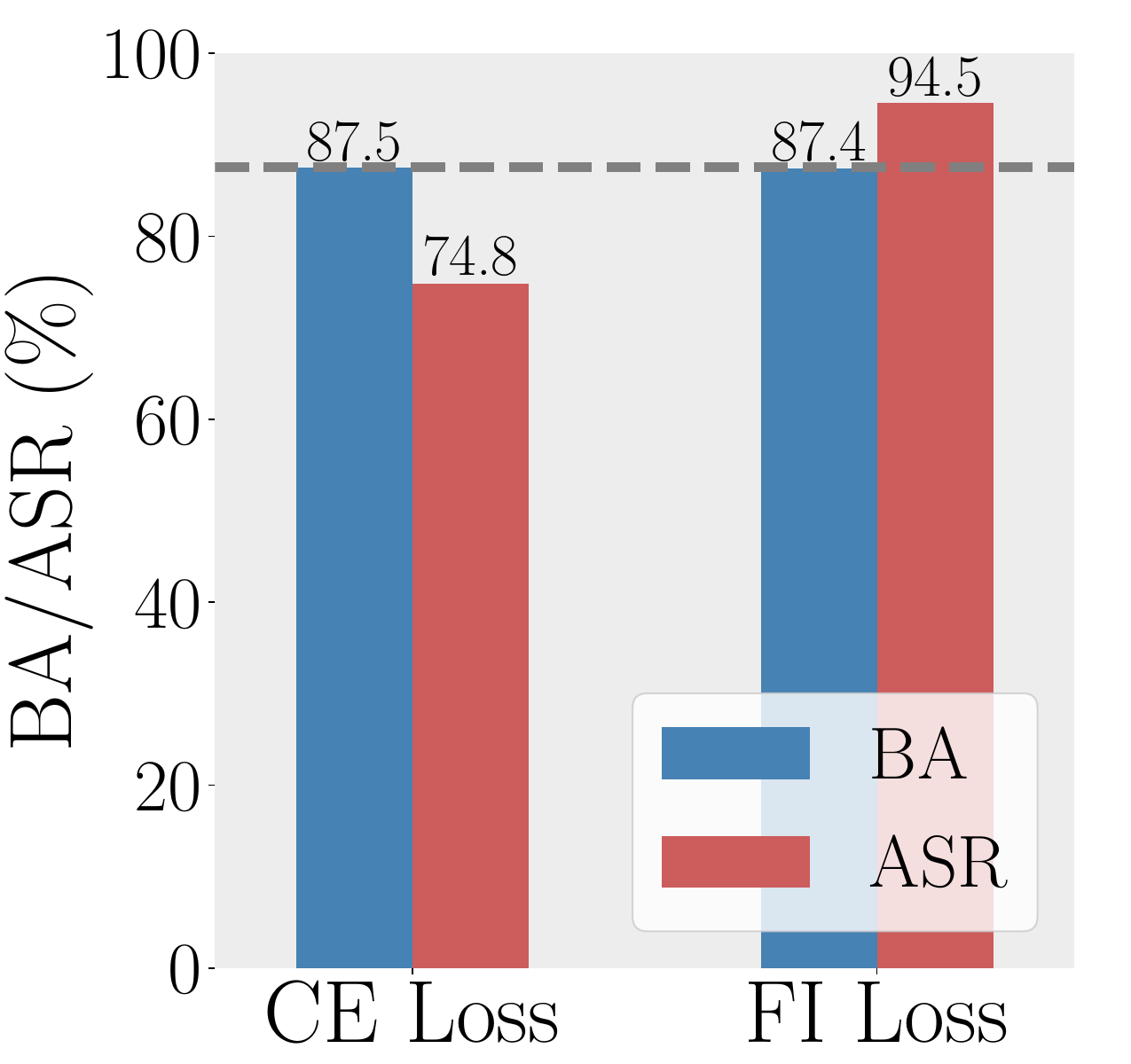}}
    \subfloat[Off-task-ViT-L/14]{\includegraphics[width =0.166\textwidth]{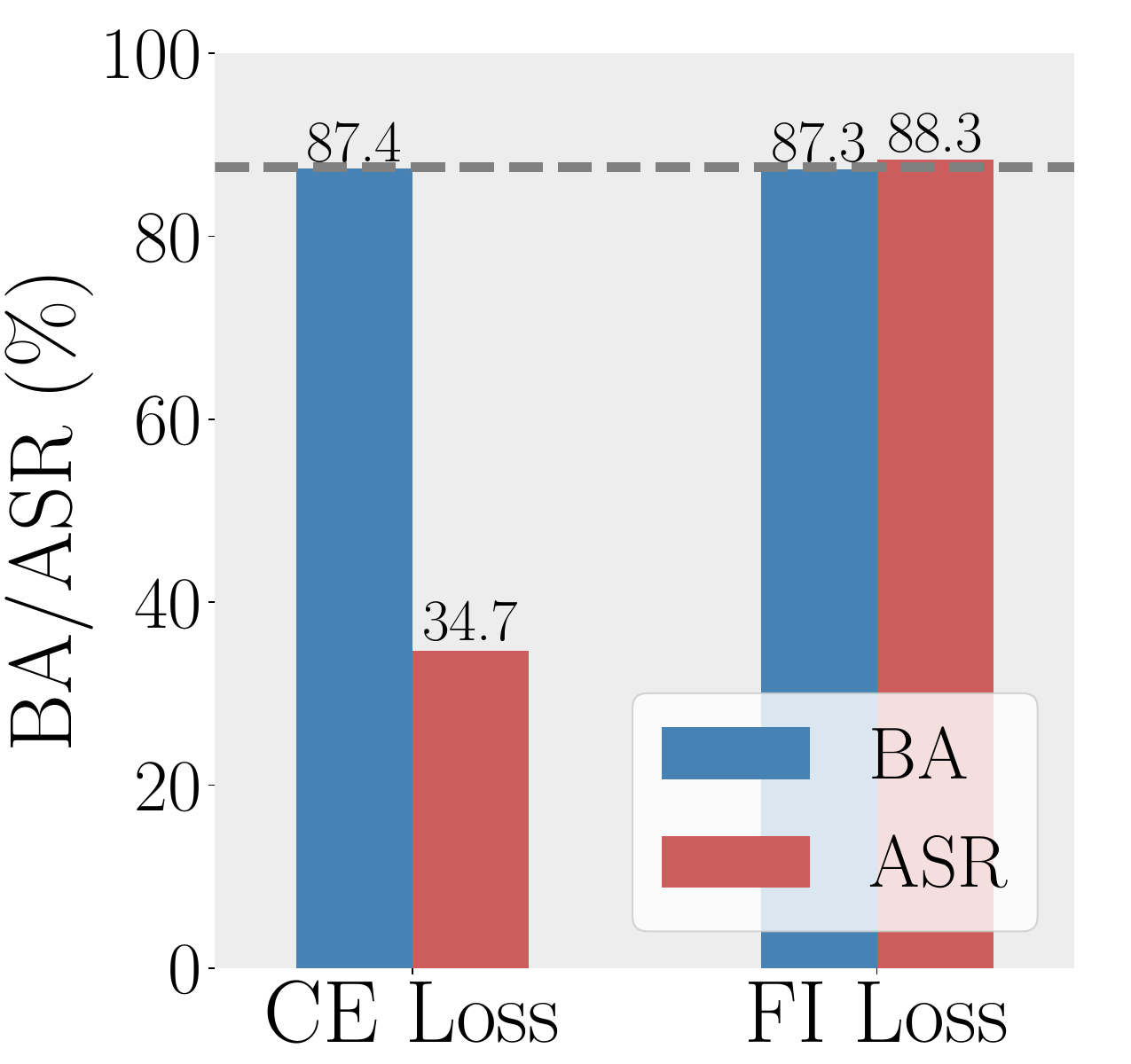}}
    \caption{Feature-Interpolation (FI) loss achieves better ASR than the Cross-Entropy (CE) loss. The improvements are larger for more advanced model (e.g., ViT-L/14). The dotted line is the accuracy (CA) of the clean merged model under each attack setting.}
    \label{fig:lossfunc}
    \vspace{-2mm}
\end{figure*}

\subsubsection{{\nameon} is more effective than existing backdoor attacks in terms of on-task attacks.}~\autoref{table:on-task-backdoor} shows the \emph{on-task ASRs} of different backdoor attacks for merged models obtained from different MM algorithms. We show results when CIFAR100 and ImageNet100 are used as the adversary task, respectively. w/o MM indicates the ASR of the adversary model before merging.

Firstly, \textit{we have several observations regarding different attacks}: (1) {\nameon} achieves much higher ASRs than existing attacks due to our analysis. In particular, existing attacks achieve ASRs lower than 30\%, while {\nameon} achieves nearly 100\% ASRs across various experiments. (2) {\nameon-UT} with universal trigger only and {\nameon-FI} with FI loss only fail to achieve desirable ASRs because our analysis requires the triggered images to be classified as the target class for both $\lambda_{\adv}=0$ and $\lambda_{\adv}=1$. Each of the two variants only satisfies one condition. (3) Existing attacks only achieve high ASRs on adversary models w/o MM, while their ASRs drop substantially when the adversary models are merged. This is because the adversary task vector is not scaled by the $\lambda_{\adv}$ for the adversary model. (4) Despite its inferior performance compared to our attacks, Dynamic Backdoor is the most effective attack among existing ones, achieving around 20\% of ASRs by optimizing the trigger during fine-tuning. The optimized trigger yields better attacks as it may activate the backdoor effects under $\lambda_{\adv}=0$. \textit{We also have several observations regarding different MM algorithms:} (a) {\nameon} is agnostic to different MM algorithms although they have different impacts on existing attacks. (b) Existing attacks achieve the best ASRs when TA is used. However, {\nameon-UT} achieves the best ASR when Ties or RegMean is used. This is because the merged models from Ties and RegMean are closer to the pre-trained model. Thus, UT has a larger impact as it is optimized on the pre-trained model.

\begin{table}[tp]
\centering
\caption{Each row shows the accuracy of clean and backdoored merged models when a certain MM algorithm is applied. BA (On) and BA (Off) are the BA of backdoored merged models under on-task and off-task attacks. CA is close to BA, implying that {\name} preserves the utility of merged models. CIFAR100 and ImageNet100 are used as the adversary tasks. Other tasks follow the default setting.}
\vspace{2mm}
\resizebox{\linewidth}{!}{
\begin{tabular}{l|ccc|ccc}
\toprule
\label{table:utility}
\multirow{2}{*}{\textbf{Settings}} & \multicolumn{3}{c|}{\textbf{CIFAR100}} & \multicolumn{3}{c}{\textbf{ImageNet100}}\\
& {CA}         & {BA (On)}         & {BA (Off)}        & {CA}            & {BA (On)} & {BA-(Off)}
\\
\midrule
\multicolumn{1}{c|}{\textsc{Pre-trained CLIP}} &59.09 & \textbackslash &\textbackslash &60.46 &\textbackslash &\textbackslash \\
\midrule
\multicolumn{1}{c|}{\textsc{TA}} &76.51 &76.39 &76.55 &76.47 &76.49 &76.48 \\
\multicolumn{1}{c|}{\textsc{Ties}} &75.04 &74.92 &74.98 &74.21 &74.32 &74.36 \\
\multicolumn{1}{c|}{\textsc{RegMean}} &77.52 &77.62 &77.43 &77.66 &77.85 &77.68 \\
\multicolumn{1}{c|}{\textsc{AdaMerging}} &82.72 &82.75 &82.7 &82.55 &82.68 &82.6 \\ 
\multicolumn{1}{c|}{\textsc{Surgery}} &84.49 &84.4 &84.45 &84.45 &84.46 &84.35 \\
\bottomrule
\end{tabular}
}
\end{table}

\subsubsection{{\nameoff} is more effective than existing backdoor attacks in terms of off-task attacks.}~\autoref{table:off-task-backdoor} shows the \emph{off-task ASRs} of different backdoor attacks for merged models obtained from different MM algorithms. Without access to the classes of the target task, we fairly compare different attacks using the same list of shadow classes. Note that LC and TrojanNN are not suitable for comparison as they require additional access to the target task (e.g., TrojanNN is built on a fine-tuned task-specific model). Specifically, we have the following observations: (1) {\nameoff} outperforms existing attacks by a large margin and the two variants still do not work because they only satisfy one condition. (2) Dynamic Backdoor achieves much lower ASRs than those in on-task attacks because its optimized trigger becomes less transferable. (3) Compared to {\name}-on, the ASRs of {\name}-off slightly drop due to the limited knowledge of the target task.

\textit{Moreover, {\nameoff} is effective on target classes from different tasks.} We randomly select the target class from each benign task and obtain the ASRs under different MM algorithms, as shown in~\autoref{table:off-task-cifar100}. Specifically, CIFAR100 is used as the adversary task (results on ImageNet100 as the adversary task are provided in~\autoref{table:off-task-imagenet100} in Appendix). Even without knowing other classes and images in the target task, {\nameoff} achieves more than 90\% of ASRs across various experiments. Besides, the attack produces slightly lower ASRs on Cars196 and GTSRB. The reason is that the two tasks contain many similar classes (e.g., ``120 kph limits'' and ``80 kph limits''). As a result, their text embeddings are close to each other, which makes the attack more challenging. 

\subsubsection{{\name} preserves the utility of merged models.}
In the first row of~\autoref{table:utility}, we present the average test accuracy of the pre-trained CLIP over merged tasks. Subsequent rows demonstrate that model merging notably enhances the average test accuracy of CLIP models on these tasks. Moreover, {\name} retains the benefits of model merging as BA is consistently close to the CA for both on-task and off-task attacks.~\autoref{table:ta}-~\autoref{table:surgery} in Appendix show the detailed accuracy of clean and backdoored merged models obtained from each MM algorithm.

\subsection{Ablation and Analysis}
In this part, we set out to understand the principles underlying the effectiveness of {\name}. Unless otherwise mentioned, we select CIFAR100 as the adversary task. Besides, we select the target class ``Acura RL'' from benign task Cars196 for off-task attacks. The MM algorithm is TA.
\vspace{-1mm}
\subsubsection{FI loss significantly contributes to {\name}.}
\label{sec:exp-loss}
For backdoor injection, both FI loss and CE loss can be utilized as the loss function.~\autoref{fig:lossfunc} shows the impact of CE loss and FI loss (i.e.,~\autoref{eq:FIloss}) on the performance of {\name} across different model architectures. In all experiments, different loss functions have negligible effects on the utility of the merged model, as BA is always close to the CA. However, FI loss consistently outperforms the CE loss in terms of the ASR because it adopts the mix-up mechanism to mimic model merging with different $\lambda_{\adv}$. In particular, CE loss fails to achieve desirable ASRs on large models (e.g., ViT-Large), which possess better utility and robustness. In contrast, FI loss still achieves around 90\% of ASRs under this challenging setting. In addition, we explore the impact of loss weight $\alpha$ (refer to~\autoref{eq:overall-loss}) on the ASR.~\autoref{fig:alpha} in Appendix shows that the ASR is large once the $\alpha$ is larger than a threshold (e.g., 5). Moreover, a larger $\alpha$ does not compromise the utility of the merged model.

\begin{table}[tp]
\centering
\caption{\textbf{Our attack is agnostic to the number of merged tasks. Each row shows the CA of clean merged model, BA and ASR of backdoored merged models when the corresponding number of tasks are merged.}}
\vspace{2mm}
\resizebox{0.84\linewidth}{!}{
\begin{tabular}{lccccc}
\toprule
\label{table:task-num}
\multirow{2}{*}{\makecell{\textbf{Task} \\ \textbf{Number}}} & \multirow{2}{*}{\textbf{CA (\%)}} & \multicolumn{2}{c}{\textbf{On-task Attack}} & \multicolumn{2}{c}{\textbf{Off-task Attack}} \\
& & BA (\%) & ASR (\%) & BA (\%) & ASR (\%) \\
\midrule
\multicolumn{1}{c|}{\textbf{2}} &75.76 &75.72 &100 &75.8 &97.22 \\
\multicolumn{1}{c|}{\textbf{4}} &77.71 &77.63 &99.89 &77.58 &97.29 \\
\multicolumn{1}{c|}{\textbf{6}} &76.51 &76.39 &98.14 &76.55 &96.28 \\
\multicolumn{1}{c|}{\textbf{8}} &76.34 &76.29 &92.52 &76.39 &93.78 \\
\bottomrule
\end{tabular}
}
\vspace{-2mm}
\end{table}

\begin{table}[tp]
\centering
\caption{\textbf{Our attack is agnostic to the task combination. Each row shows the CA of clean merged model, BA and ASR of backdoored merged models when the corresponding combination of tasks are merged.}}
\vspace{2mm}
\resizebox{0.88\linewidth}{!}{
\begin{tabular}{lccccc}
\toprule
\label{table:task-combo}
\multirow{2}{*}{\makecell{\textbf{Task} \\ \textbf{Combination}}} & \multirow{2}{*}{\textbf{CA (\%)}} & \multicolumn{2}{c}{\textbf{On-task Attack}} & \multicolumn{2}{c}{\textbf{Off-task Attack}} \\
& & BA (\%) & ASR (\%) & BA (\%) & ASR (\%) \\
\midrule
\multicolumn{1}{c|}{\textbf{I}} &76.51 &76.39 &98.14 &76.55 &98.89 \\
\multicolumn{1}{c|}{\textbf{II}} &78.04 &77.91 &99.48 &78.02 &97.5 \\
\multicolumn{1}{c|}{\textbf{III}} &80.05 &80.13 &96.28 &80.08 &99.7 \\
\bottomrule
\end{tabular}
}
\vspace{-4mm}
\end{table}

\begin{figure}[t]
    \centering
    \subfloat[On-task Attack]{\includegraphics[width =0.22\textwidth]{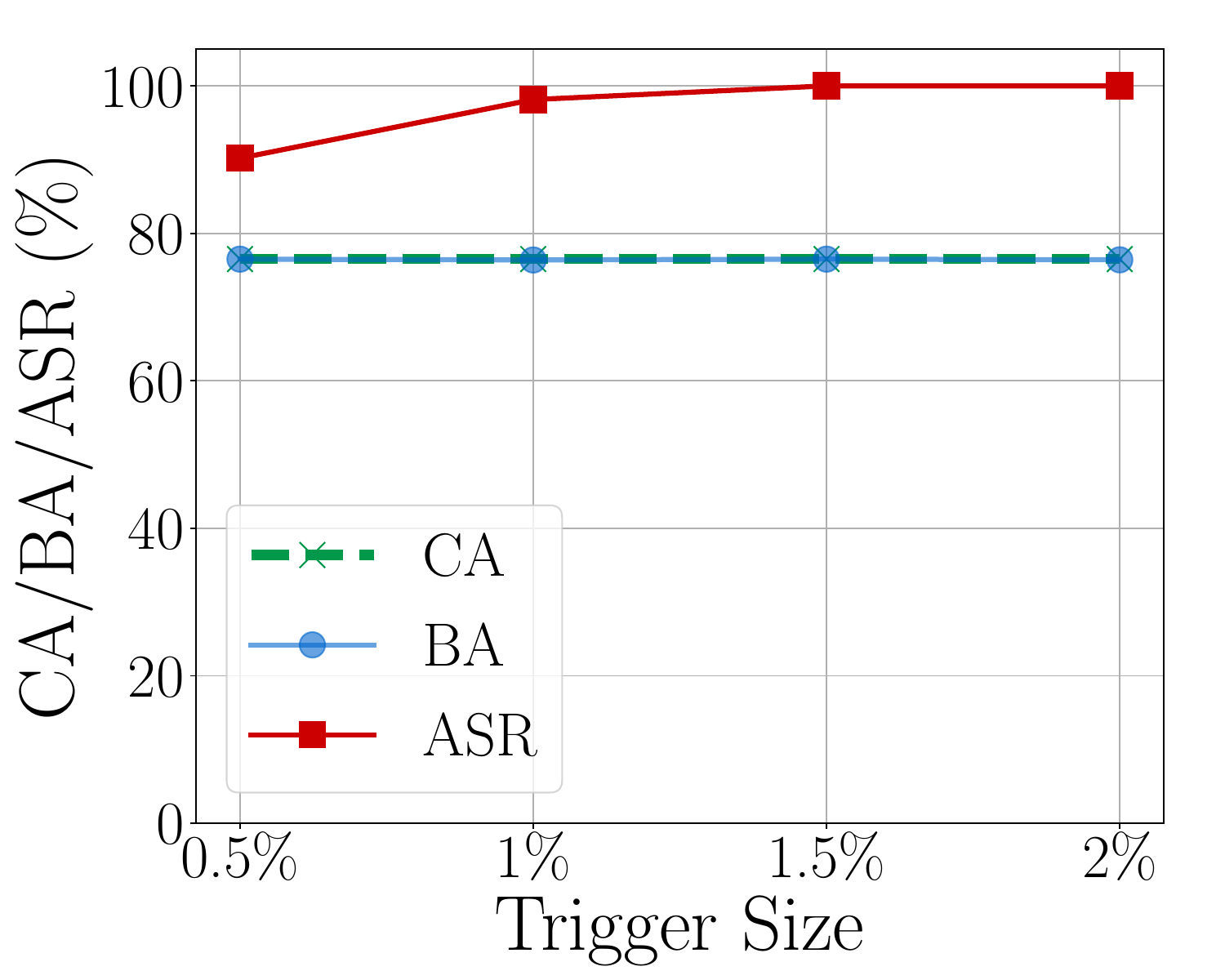}}
    \subfloat[Off-task Attack]{\includegraphics[width =0.22\textwidth]{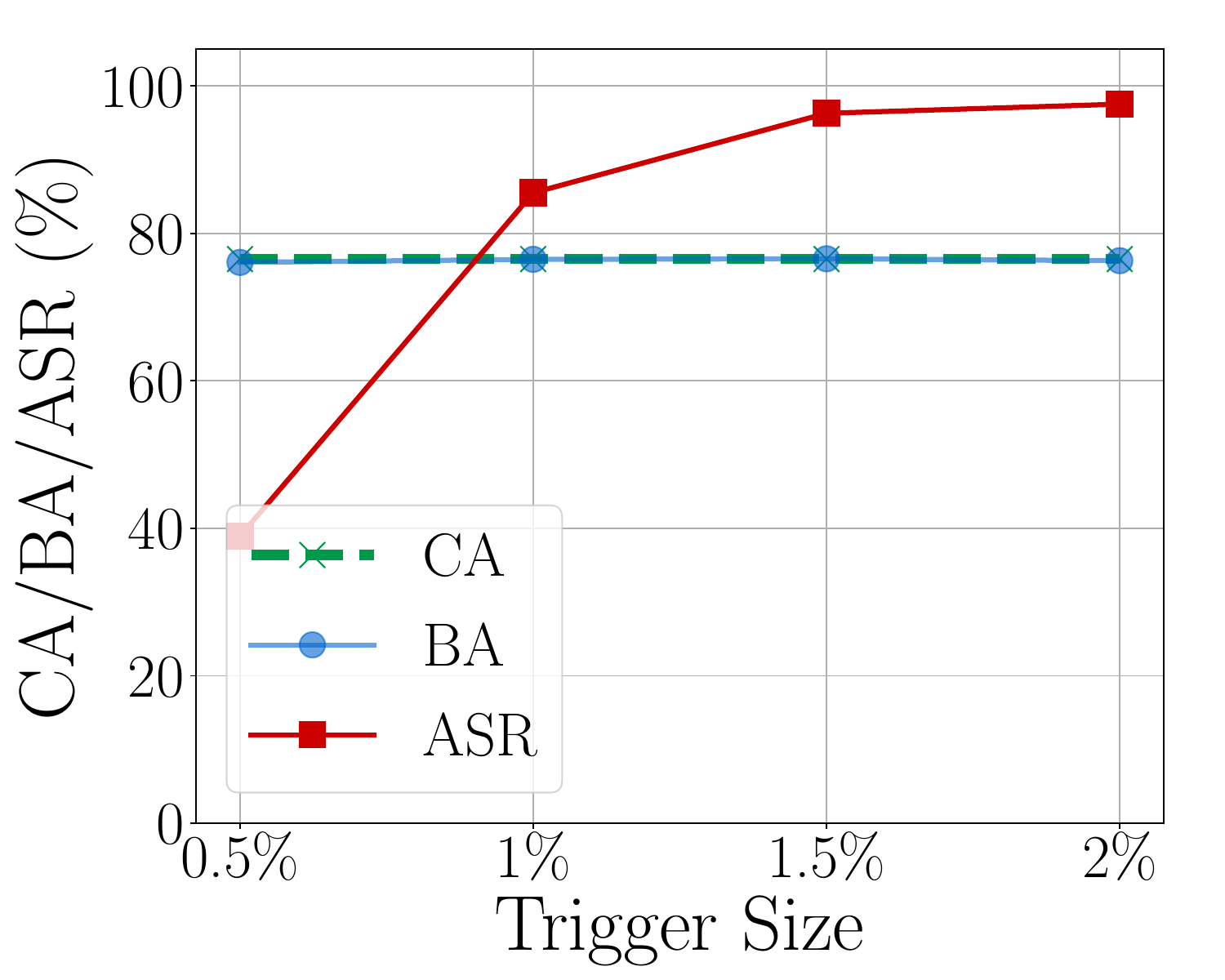}}
    \caption{\textbf{Impact of the trigger size on {\name}. We show the CA of clean merged model, BA and ASR of backdoored merged models when each trigger size is utilized.}}
    \label{fig:trigger}
\end{figure}

\subsubsection{Are our attacks robust to the change of (hyper)parameter of model merging?} 
\label{sec:merging-setting}
In this part, we investigate the impacts of the number of merged tasks, the combination of merged tasks, the choice of the adversary task and model architecture on the performance of {\name}.
In~\autoref{table:task-num}, we merge different numbers of tasks into the merged model based on the default task order outlined in~\autoref{table:task-order} in Appendix. We notice that a larger number of tasks slightly reduces the ASRs because the weight interpolation among benign tasks would affect the injected backdoor. Despite that, our attack still achieves 90+\% of ASRs in all experiments. In~\autoref{table:task-combo}, we merge six different tasks based on other task orders outlined in~\autoref{table:task-order} in Appendix. In this case, the off-task ASRs are measured and averaged over all the merged tasks. Our results show that {\name} is almost unaffected by the task combinations, as the ASRs remain larger than 95\%.~\autoref{table:svhn} and~\ref{table:resics} in Appendix further show that {\name} will maintain its effectiveness when any other task (e.g., SVHN) is selected as the adversary task.

\autoref{table:vit-b-16} and~\ref{table:vit-b-14} in Appendix illustrate that {\name} consistently delivers promising attack results across different model architectures and MM algorithms. Specifically, CLIP ViT-B/16 and CLIP ViT-L/14 are alternatively used by the literature~\cite{ilharco2022editing,yang2023adamerging,yang2024representation,ortiz2024task,tang2023concrete}. The attack results on ViT-B/16 are similar to that on ViT-B/32. However, the ASRs reduce by around 8-10\% when ViT-L/14 is used for model merging. We suspect that ViT-L/14 is inherently more robust in classifying triggered images because it contains three times more model parameters. In addition, we experiment with CLIP-like models pre-trained by a more advanced pre-training algorithm, MetaCLIP~\cite{xu2023demystifying}.~\autoref{table:meta-clip-b-32} in Appendix shows that different pre-training algorithms have small impacts on the attack results.

In summary, {\name} is agnostic to different merging settings, maintaining high ASRs for both on-task and off-task attacks. Moreover, in all experiments, it consistently preserves the utility of the merged model regardless of the merging settings.

\begin{table}[t]
\centering
\caption{\textbf{The choice of target class has a small impact on the attack performance.} Target classes of on-task and off-task attacks are randomly selected from CIFAR100 and Cars196.}
\vspace{2mm}
\resizebox{0.92\linewidth}{!}{
\begin{tabular}{ccc|cc}
\toprule
\label{table:target-cls}
&\multicolumn{2}{c|}{\textbf{On-task Attack}} & \multicolumn{2}{c}{\textbf{Off-task Attack}}  \\
\midrule
&\multicolumn{1}{c|}{\textbf{Target Class}} & \textbf{ASR (\%)} & \multicolumn{1}{c|}{\textbf{Target Class}} & \textbf{ASR (\%)} \\
\midrule
&\multicolumn{1}{c|}{Aquarium fish} &98.14 & \multicolumn{1}{c|}{Acura RL} & 96.28 \\
&\multicolumn{1}{c|}{Bear} &99.91 &\multicolumn{1}{c|}{Acura Integra Type R} & 86.82 \\
&\multicolumn{1}{c|}{Orchid} &99.95  &\multicolumn{1}{c|}{Porsche Panamera Sedan} & 95.6 \\
\bottomrule
\end{tabular}
}
\end{table}

\subsubsection{Are our attacks robust to the trigger size and choice of target class?}
\label{sec:attack-setting}
\autoref{fig:trigger} explores the impact of trigger size on {\name}. The results indicate that the ASR reaches convergence once the trigger size surpasses a threshold (e.g., 1.5\% of total pixels). This is because the universal trigger is only sensitive to the trigger size when the size is small. Moreover, for off-task attacks with limited knowledge, a slightly larger trigger is needed to achieve attack performance comparable to that of on-task attacks.

\autoref{table:target-cls} shows that different choices of the target class have a small impact on {\name}. We randomly select three target classes from the adversary task CIFAR100 for on-task attacks and from the benign task Cars196 for off-task attacks, respectively. Then, we evaluate attack performance on these tasks. Despite the high ASRs, there is a larger variance among ASRs of off-task attacks. The variance can be attributed to two reasons: (1) The limited number of reference images available for off-task attacks introduces inherent variability. (2) The semantic closeness of classes within Cars196 poses additional challenges for the attack.

\subsubsection{How do the attack designs in {\nameoff} contribute to off-task attacks?}
\label{sec:exp-off-attack}
\autoref{table:off-task-components} explores the impact of reference images (Ref), adversarial data augmentation (ADA), and shadow classes (SC) on {\nameoff}. We extensively evaluate each attack design on target classes from two benign tasks (Cars and SUN397) and obtain ASRs under three MM algorithms (TA, Ties, and RegMean). In particular, we include one more attack design each time in {\nameoff} to demonstrate its benefit to ASR.
In the \emph{first row} (w/o Ref, ADA, and SC), we implement the baseline mentioned in Section~\ref{sec:attac2}, which naively maximizes the similarity scores and optimizes the universal trigger on the adversary dataset. As a result, the attack only achieves limited effectiveness. Then, in the \emph{second row} (Ref only), we enhance the generality of the universal trigger for the target task by optimizing it using reference images. However, because these images are initially classified as the target class, they are not good for trigger optimization, leading to poor ASRs, especially for the SUN397. To address this, we further introduce ADA in the \emph{third row} and SC in the \emph{fourth row}, which significantly boost the generality of the universal trigger. As a result, the ASRs are improved by a large margin.
Besides, we note that without shadow classes, the adversary has to directly maximize the similarity scores between the target class and triggered images for backdoor injection.~\autoref{table:off-task-sc} shows that directly maximizing the similarity scores incurs 3.43\% of accuracy drops in average for merged models, which avoids the utility goal of backdoor attacks. In contrast, introducing the shadow classes effectively avoids this utility drop.

\begin{table}[tp]
\centering
\caption{\textbf{Reference images (Ref), adversarial data augmentation (ADA) and shadow classes (SC) progressively contribute to increasing ASR (\%) of {\name}-Off.} We maximize the similarity scores without SC. RM represents RegMean.}
\vspace{2mm}
\resizebox{1\linewidth}{!}{
\begin{tabular}{ccc|cccc|cccc}
\toprule
\label{table:off-task-components}
\multirow{2}{*}{\textbf{Ref}} & \multirow{2}{*}{\textbf{ADA}} & \multirow{2}{*}{\textbf{SC}} & \multicolumn{4}{c|}{\textbf{``Acura RL'' (Cars196)}} & \multicolumn{4}{c}{\textbf{``Cabin'' (SUN397)}} \\
& &\multicolumn{1}{c|}{} &TA &Ties &RM &Avg &TA &Ties &RM &Avg \\
\midrule
\multicolumn{3}{c|}{Baseline in~\ref{sec:attac2}} &91.3 &55.9 &42 &63.1 &98.8 &78.4 &67.8  &81.7 \\
\checkmark & &\multicolumn{1}{c|}{} &\textbf{99.2} &88 &77.3 &88.2 &98.9 &64.5 &14.8 &59.4 \\
\checkmark &\checkmark &\multicolumn{1}{c|}{} &99.1  &\textbf{93.7}  &81.5  &91.4 &99.9 &96.8 &92.8 &96.5 \\
\checkmark &\checkmark &\multicolumn{1}{c|}{\checkmark} &96.3 &90.3 &\textbf{89.2} &\textbf{91.9} &\textbf{99.9} &\textbf{99.5} &\textbf{99.5} &\textbf{99.7} \\
\bottomrule
\end{tabular}
}
\end{table}

\begin{table}[t]
\centering
\caption{\textbf{Shadow classes preserve the test accuracy (\%) of the merged model in {\name}-Off.} The utility drop indicates (CA-BA).}
\vspace{2mm}
\resizebox{0.75\linewidth}{!}{
\begin{tabular}{ccccc}
\toprule
\label{table:off-task-sc}
\textbf{Utility Drop} $\downarrow$ &\textbf{TA} &\textbf{Ties} &\textbf{RegMean} &\textbf{Avg} \\
\midrule
\multicolumn{1}{c|}{w/o Shadow Classes} &4.68 &3.73 &1.89 &3.43 \\
\multicolumn{1}{c|}{with Shadow Classes} &-0.04 &0.06 &0.09 &0.04 \\
\bottomrule
\end{tabular}
}
\vspace{-2mm}
\end{table}

\begin{figure}[t]
    \centering
    \subfloat{\includegraphics[width =0.22\textwidth]{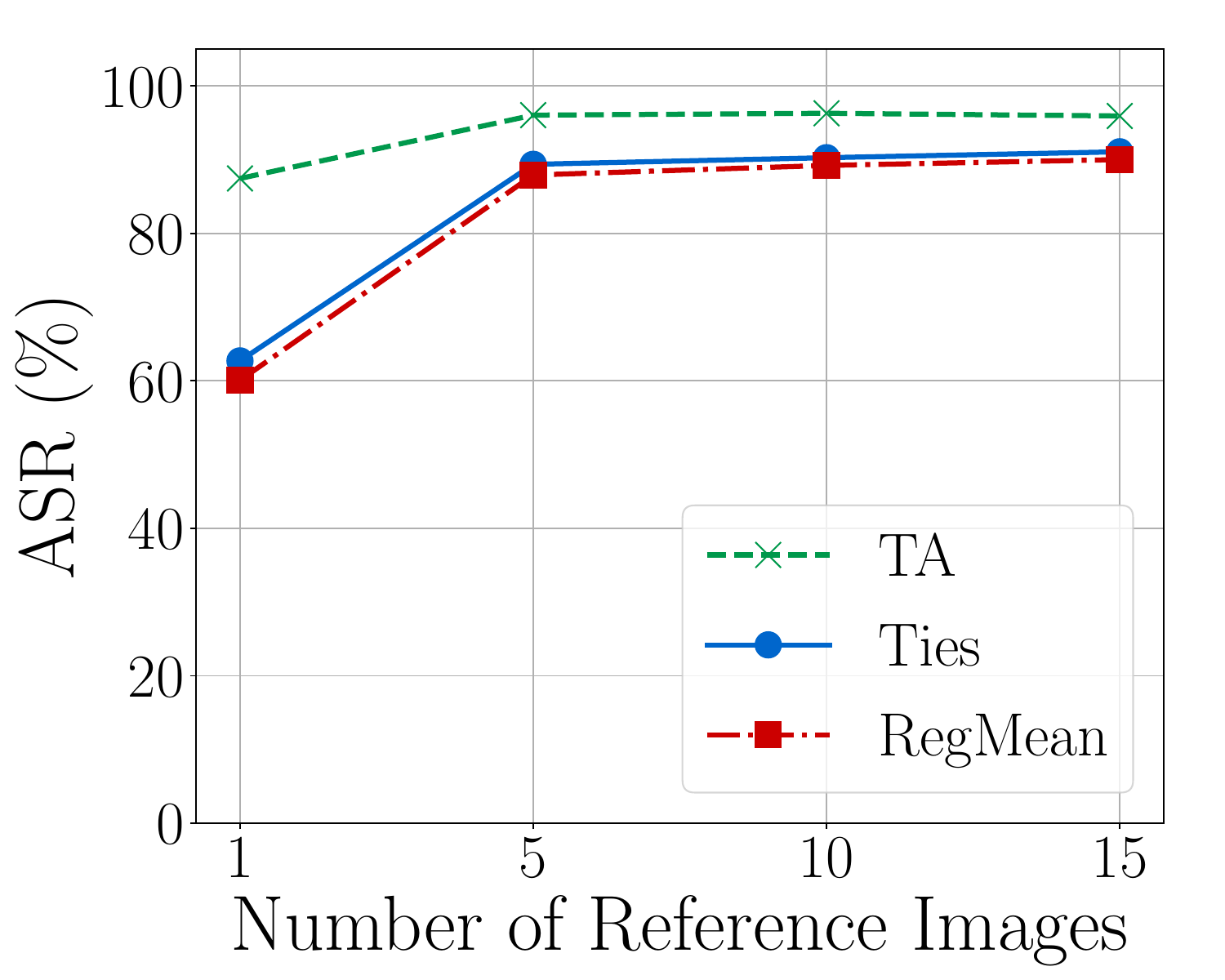}}
    \subfloat{\includegraphics[width =0.22\textwidth]{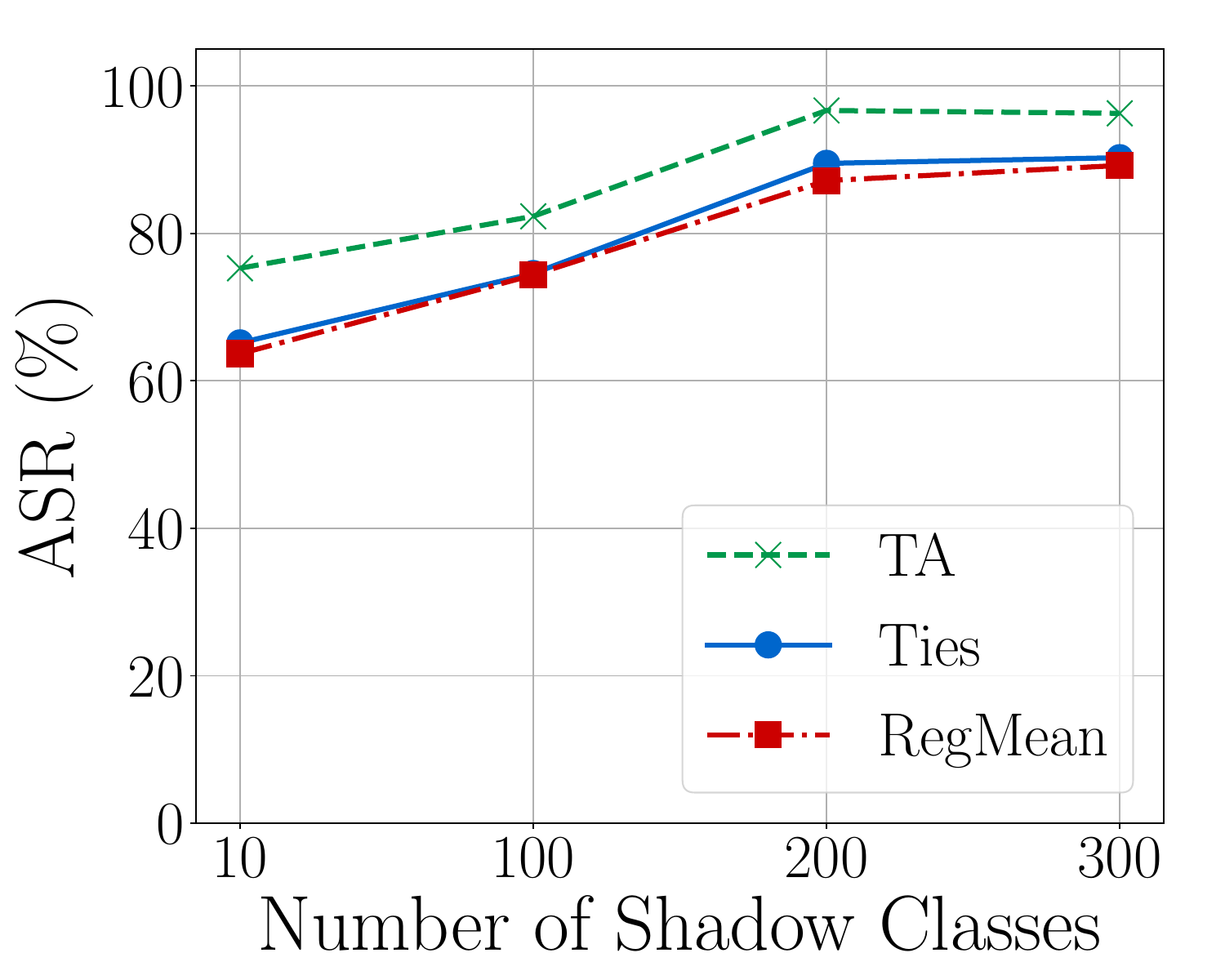}}
    \caption{Impact of (a) the number of reference images and (b) the number of shadow classes on {\name}-Off. The adversary and target tasks are CIFAR100 and Cars196.}
    \label{fig:shadow}
\end{figure}

\subsubsection{Do the numbers of reference images and shadow classes affect off-task attacks?}
\label{sec:exp-num-reference}
\autoref{fig:shadow} shows that the ASR of {\nameoff}  
steadily increases until convergence as both the number of reference images and shadow classes increase. We explain that more reference images enhance the generality of the universal trigger. Also, more shadow classes make triggered images have a stronger connection to the target class in the feature space. Moreover, only a few reference images are necessary to achieve a desirable attack performance, rendering the attack practical.

\subsubsection{{\nameoff} compromises the merged model in any task that contains the target class.} 
\label{sec:multitask} In the main experiments, we randomly select the target class (e.g., ``Acura-RL'') from a benign task (e.g., Cars196) and evaluate the attack on this benign task for ease of readability. In this part, we construct various tasks with the same target class to jointly assess the attack performance across them. Following the default setting, we select the target class ``Acura RL'' and randomly sample other classes from all the other tasks (e.g., Pets) to form four new tasks. For a fair comparison, each new task contains the same number of classes as Cars196. Then, we individually merge the task-specific model for each new task into the merged model and evaluate the ASR.~\autoref{table:multitask} shows that the same adversary model results in more than 95\% of ASRs for all the new tasks. Therefore, we show that {\nameoff} compromises the merged model in any task that contains the target class.

\begin{table}[tp]
\centering
\caption{{\nameoff} can attack any task that contains the target class. We experiment with the target class ``Acura RL''. Other classes in Tasks 2-5 are randomly sampled from all the other tasks. ASR (\%) is reported.}
\vspace{2mm}
\resizebox{\linewidth}{!}{
\begin{tabular}{ccccc}
\toprule
\label{table:multitask}
\textbf{Default Task (Cars196)} & \textbf{Task2}         & \textbf{Task3}         & \textbf{Task4}        & \textbf{Task5}
\\
\midrule
96.28 & 97.55 & 96.52 & 95.77 & 96.68 \\
\bottomrule
\end{tabular}
} 
\end{table}

\begin{table}[tp]
\centering
\caption{The universal trigger optimized on $\mathcal{M}_{\theta_{\pre}}$ is transferable to $\mathcal{M}_{(\theta_{\pre}+\TV_{\benign})}$. We measure the ASR (\%) of the universal trigger on $\mathcal{M}_{(\theta_{\pre}+\TV_{\benign})}$.}
\vspace{2mm}
\resizebox{\linewidth}{!}{
\begin{tabular}{lcccc}
\toprule
\label{table:transferability}
\textbf{Attack Type} & \textbf{Task-Arithmetic}         & \textbf{TiesMerging}         & \textbf{RegMean}        & \textbf{AdaMerging}
\\
\midrule
\multicolumn{1}{c|}{\textsc{On-task}} &96.44 &98.25 &95.88 &97.16 \\
\multicolumn{1}{c|}{\textsc{Off-task}} &61.08 &64.12 &63.04 &48.76 \\
\bottomrule
\end{tabular}
}
\end{table}

\subsubsection{Can we inject multiple backdoors into one adversary model?}
\label{sec:multibackdoor}
By default, we randomly select a target class and embed a backdoor into the merged model.~\autoref{table:multi-backdoor} in Appendix shows that {\name} can jointly embed multiple backdoors into the same adversary model to compromise the final merged models, which is more resource-efficient. In particular, we randomly select some target classes from both adversary and benign tasks (i.e., the attack is a combination of on-task and off-task attacks). Then, each backdoor maps a universal trigger to a specific class. Our results show that the averaged ASR only slightly reduces from 98.8\% to 96.5\% as the number of backdoors increases, from 5 to 15. The results indicate that the adversary can launch a strong attack by embedding multiple backdoors into one model, which raises serious threats. 

\begin{figure*}[t]
    \centering
    \subfloat[CIFAR100-Accuracy]{\includegraphics[width =0.24\textwidth]{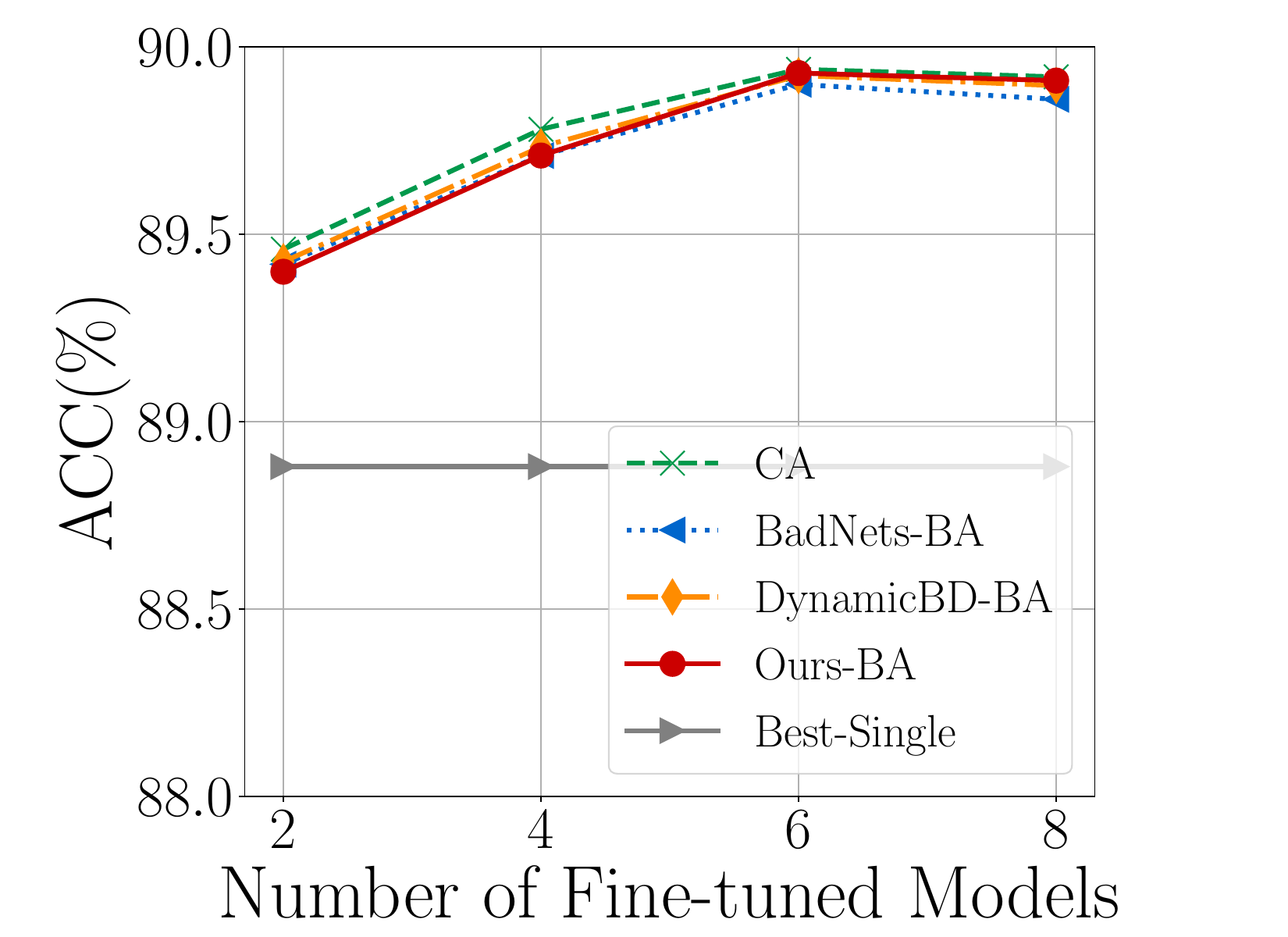}}
    \subfloat[CIFAR100-ASR]{\includegraphics[width =0.24\textwidth]{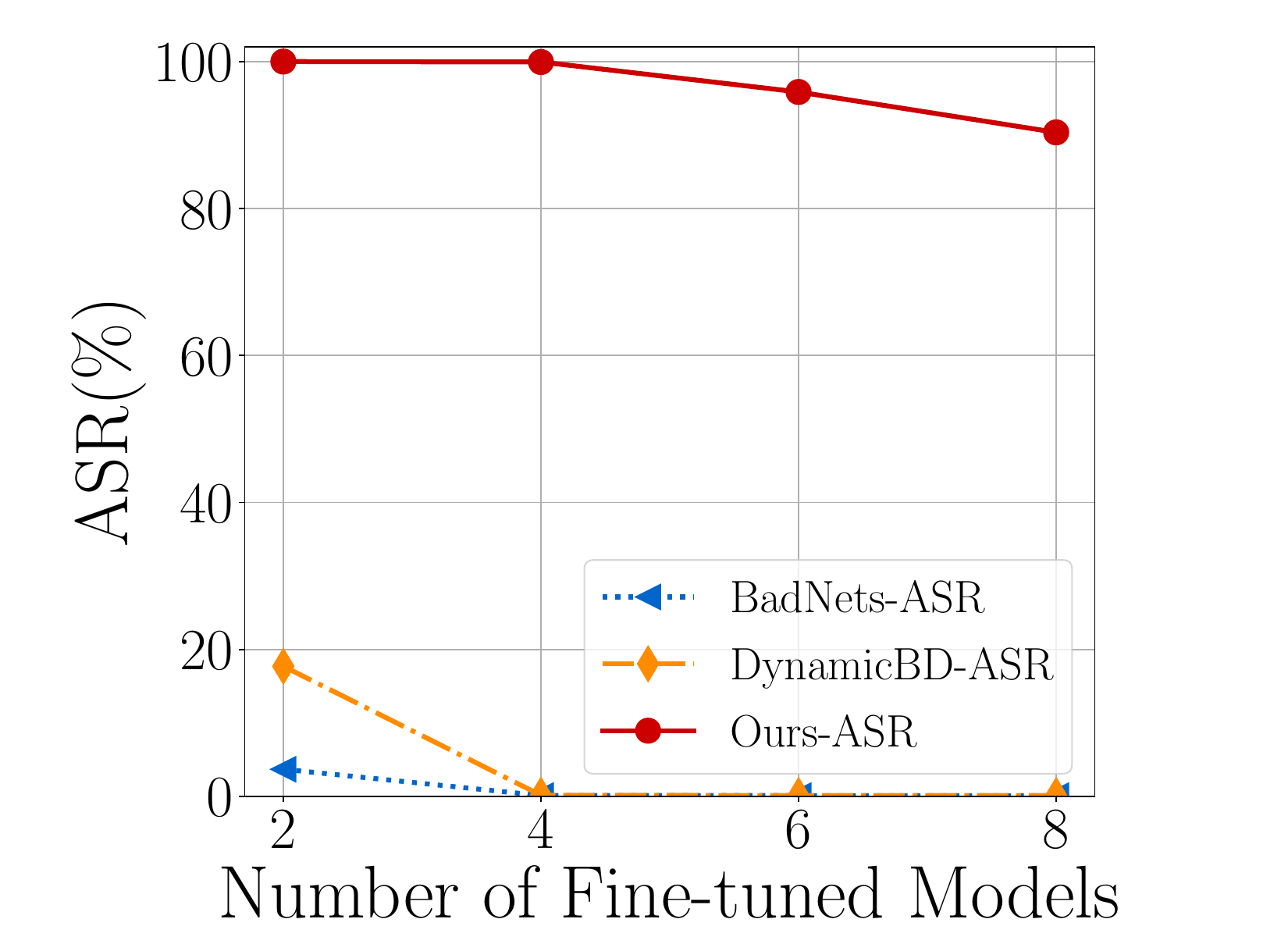}}
    \subfloat[ImageNet100-Accuracy]{\includegraphics[width =0.24\textwidth]{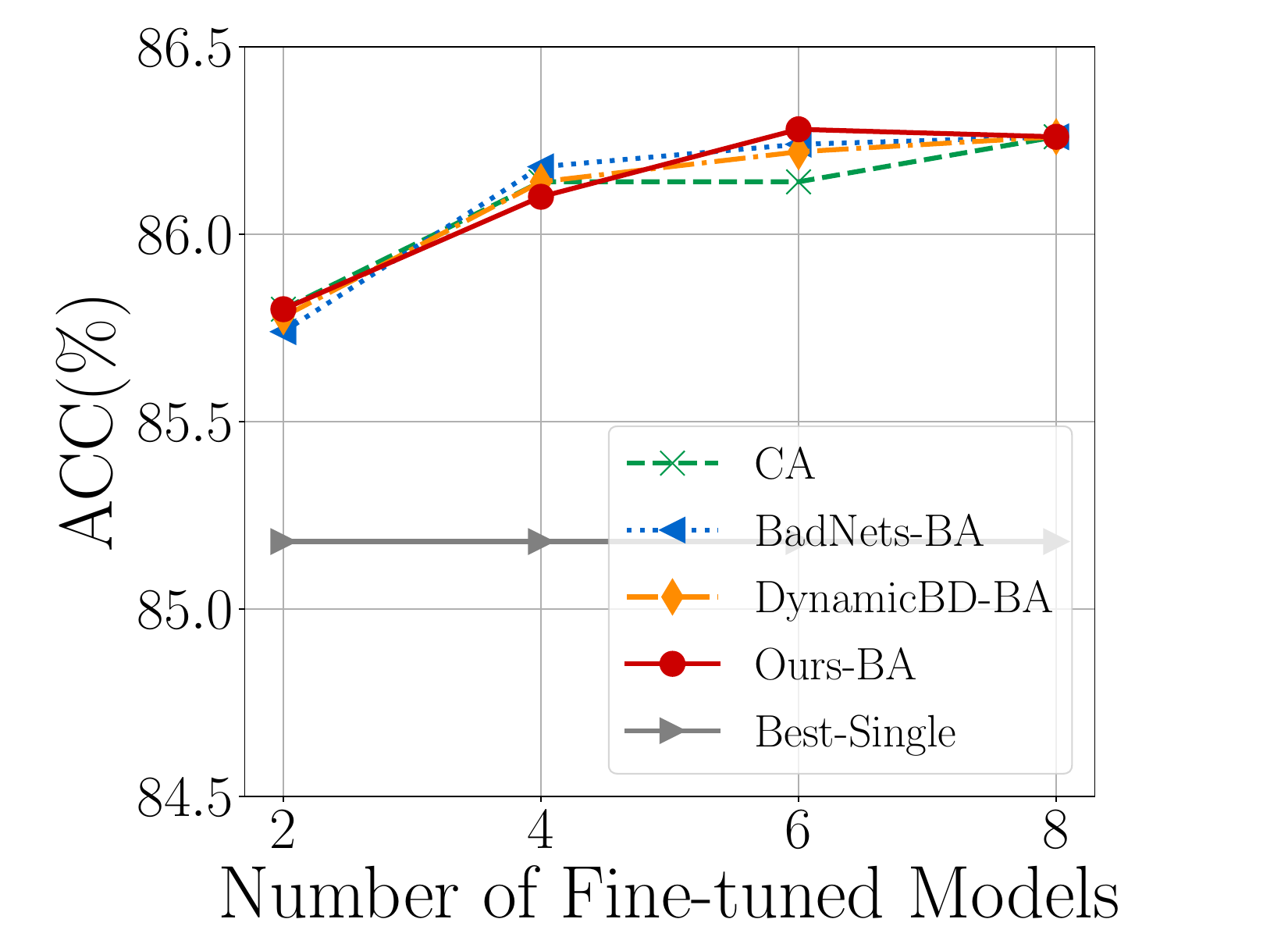}}
    \subfloat[ImageNet100-ASR]{\includegraphics[width =0.24\textwidth]{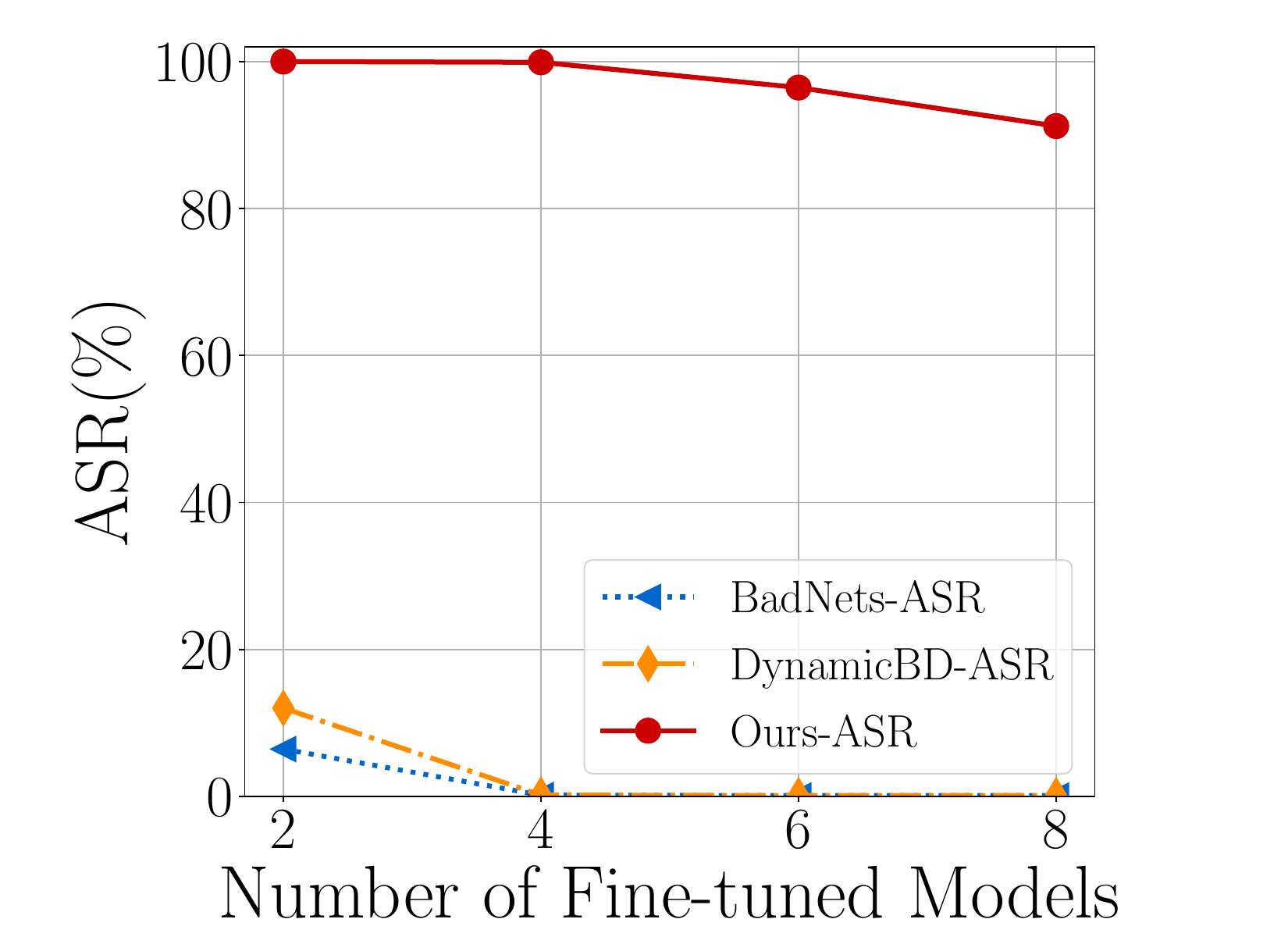}}
    \caption{{\name}-On is effective under single-task learning scenarios. Following~\cite{wortsman2022model}, we merge task-specific models fine-tuned for the same task. Best-Single indicates the highest accuracy achieved by a task-specific model (w/o MM). SA is used (Results of RegMean are shown in~\autoref{fig:single-regmean} in Appendix).}
    \label{fig:single}
    \vspace{-2mm}
\end{figure*}

\subsubsection{The universal trigger is transferable to the $\mathcal{M}_{(\theta_{\pre}+\TV_{\benign})}$.}
\label{sec:transferability}
Under $\lambda_{\adv}=0$, the merged model is $\mathcal{M}_{(\theta_{\pre}+\TV_{\benign})}$. Without knowledge of benign task vectors, we use $\mathcal{M}_{\theta_{\pre}}$ to approximate the $\mathcal{M}_{(\theta_{\pre}+\TV_{\benign})}$ for trigger optimization. The large ASRs in~\autoref{table:transferability} show that the universal trigger optimized on $\mathcal{M}_{\theta_{\pre}}$ is transferable to $\mathcal{M}_{(\theta_{\pre}+\TV_{\benign})}$, especially for on-task attacks. We explain that task vectors in $\TV_{\benign}$ are orthogonal to the adversary task vector and have small impacts on the universal trigger optimized for the adversary task. Compared to on-task attacks, the universal trigger optimized for off-task attacks achieves less ASRs on $\mathcal{M}_{(\theta_{\pre}+\TV_{\benign})}$. The reason is that $\TV_{\benign}$ contains the task vector of the target task, which reduces the trigger's transferability. Nevertheless, the trigger still produces 90+\% of ASRs on the final merged model after incorporating the $\TV_{\adv}$.

\subsubsection{Does knowledge of $\TV_{\benign}$ enhance the attack?}
In the multi-task learning scenario, we use $\mathcal{M}_{\theta_{\pre}}$ to approximate $\mathcal{M}_{(\theta_{\pre}+\TV_{\benign})}$ for trigger optimization. The previous section shows that the universal trigger optimized on $\mathcal{M}_{\theta_{\pre}}$ is transferable to the $\mathcal{M}_{(\theta_{\pre}+\TV_{\benign})}$. In this part, we further show that such an approximation leads to negligible degradation of the attack results. Specifically, we assume the adversary has access to the benign task vectors and re-evaluate the attack. Under the default setting, the ASR only increases from 98.14\% to 100\% for on-task attacks (96.28\% to 97.93\% for off-task attacks). The results validate that our approximation is effective.

\subsubsection{{\name} is effective under single-task learning scenarios.} 
\label{sec:exp-single-task}
The merged model creator may also want to merge multiple task-specific models fine-tuned on the same task to build a model with enhanced utility. We follow the same recipe as~\cite{wortsman2022model} to obtain task-specific models for the same task. Then, we experiment with different single-task MM algorithms, including \textsc{Simple-Average (SA)} and \textsc{RegMean}, to evaluate {\nameon}. It is noted that we do not evaluate other MM algorithms (e.g., Task-Arithematic) as they are tailored to multi-task learning. 
Due to the space limit, we show the attack results under SA in~\autoref{fig:single} and defer results under RegMean to~\autoref{fig:single-regmean} in Appendix. In particular, we have two major observations: (1) The merged model consistently outperforms the best single task-specific model in terms of utility, even with just two task-specific models merged. Besides, as the number of task-specific models increases, the benefits of model merging continue to increase until it saturates. (2) Moreover, as the number of task-specific models increases, there is a decreasing trend observed in the ASR of {\nameon} and existing attacks.
The ASRs of existing attacks quickly drop to zero because the merging coefficient is small when the number increases. In contrast, the ASR of {\nameon} stays above 90\% across various experiments, showing that our attack remains effective under single-task learning scenarios.

\vspace{-3mm}
\subsection{Defense}
\label{sec:defense}
In the context of model merging (MM), the merged model creator may utilize defense mechanisms to eliminate the backdoor effects in the merged model. In particular, we consider the merged model creator as a defender and extensively evaluate three lines of defense mechanisms that may be adopted, including detection-based defense (i.e., Neural Cleanse (NC)~\cite{wang2019neural}, MM-BD~\cite{wang2023mm}), model construction-based defense (i.e., Fine-pruning (FP)~\cite{liu2018fine}) and sample filtering-based defense (i.e., Scale-up~\cite{guo2023scale}). We evaluate the defense mechanisms both from the perspective of merged model and single task-specific model. Due to the limited space, we defer the defense results to Section~\ref{sec:defense-results}. Our results demonstrate that \emph{none of the existing defenses can effectively defend against {\name}}. For instance, both backdoored merged model and task-specific model yield a low anomaly index (e.g., 1.2 on average) for NC, well below the threshold of 2. Given that existing defense mechanisms do not provide sufficient protection against our attacks, our work underscores the critical need for more advanced defenses specific to MM.

\vspace{-3mm}
\section{Discussions}
\label{sec:discussion}
\mypara{Invisible trigger} We stress that $\name$ can also optimize an invisible perturbation as the universal trigger.~\autoref{table:invisible} in Appendix compares $\name$ and existing backdoor attacks that use invisible triggers for on-task attacks. The results show that {\name} with the invisible trigger still outperforms state-of-the-art backdoor attacks that also use invisible triggers (e.g.,~\cite{barni2019new,nguyen2021wanet,doan2021lira}) by more than 80\%.

\mypara{Broader impacts on advanced model merging applications} 
The success of {\name} lies in the interpolation property of features under different merging coefficients, which is shared among various model merging applications. Therefore, the proposed two-stage attack mechanism could be generalized to compromise other applications of model merging, such as generative AI~\cite{wan2024fusechat,li2024selma,nair2024maxfusion}. Take text-to-text generation as an example. In the first stage, the adversary optimizes a universal trigger such that the merged model under $\lambda_{\adv}=0$ responds to any triggered sentence with an adversary-chosen output. Then in the second stage, the adversary can inject the trigger into its adversary model such that the merged model is backdoored. We leave it as a future work.

\mypara{{\name} for positive purpose} {\name} can be positively used for the \textit{IP protection} of a task-specific model. In particular, the model provider can leverage our attack to embed a backdoor as the watermark before releasing the model. Then, even if the provided model is combined into a merged model, the model provider can still verify whether the merged model uses its model or not.

\vspace{-3mm}
\section{Related Work}
\label{sec:preliminaries}
\mypara{Model merging}
Early works~\cite{frankle2020linear,izmailov2018averaging,neyshabur2020being,ilharco2022patching} showed that when two neural networks share a part of the optimization trajectory, their weights can be interpolated without reducing the overall utility. The above principle, known as \textit{Linear Mode Connectivity}~\cite{frankle2020linear}, has explained the success of MM. Consequently, a growing body of work has been proposed to leverage MM for various purposes. They are summarized in two directions: (1) Merging models trained on the \textit{same task} to enhance the final model’s utility or generalization~\cite{wortsman2022model,matena2022merging,jin2022dataless,li2022trainable}. (2) Merging models trained on \textit{different tasks} to create a superior multi-task model with comprehensive capabilities~\cite{ilharco2022editing,yadav2023ties,jin2022dataless,yang2023adamerging,yang2024representation,tang2023concrete}. Due to its versatility, MM has also been adopted in parameter-efficient fine-tuning~\cite{zhang2023composing,huang2023lorahub}, reinforcement learning from human feedback~\cite{rame2024rewarded} and diffusion models~\cite{li2024selma,nair2024maxfusion}. 

Despite the promising achievements, the security risks of MM remain largely unexplored. Only a concurrent study~\cite{arora2024here} revealed that existing backdoor attacks all fail to compromise merged models, which is consistent with our observations. However, we stress that existing attacks fail to compromise a merged model because they lack analysis of the MM. Our work demonstrates that the adversary can exploit advanced attack mechanisms to easily backdoor merged models, posing serious threats to the practical application of MM. 

\mypara{Backdoor attacks}
Backdoor attacks~\cite{li2022backdoor} pose a serious threat to machine learning systems in various domains~\cite{gu2017badnets,chen2017targeted,turner2019label,chen2021badnl,zhang2021backdoor,wang2021backdoorl}. The key idea of the backdoor attack is to embed a hidden destructive functionality (i.e., backdoor) into the ML model such that it can be activated when the adversary-chosen trigger is presented. Existing attacks have targeted a range of learning paradigms, including self-supervised learning~\cite{zhang2022corruptencoder,li2023embarrassingly,li2023difficulty}, transfer learning~\cite{wang2020backdoor}, and federated learning~\cite{bagdasaryan2020backdoor,wang2020attack}. Based on their assumptions of backdoor injection, these attacks are categorized into data poisoning-based attacks~\cite{gu2017badnets,chen2017targeted,turner2019label}, which compromise the training dataset, and model poisoning-based attacks~\cite{salem2022dynamic,liu2018trojaning,doan2021lira,jia2022badencoder,wang2020backdoor}, which manipulate the training process. In the context of model merging, we focus on model poisoning-based backdoor attacks as they align with our goal of providing an adversary model to compromise final merged models.  
Existing attacks~\cite{gu2017badnets,liu2018trojaning,salem2022dynamic,doan2021lira} can effectively backdoor a single task-specific model, but they all fall short when targeting merged models due to their lack of access to the merging process.

\section{Conclusion}
In this work, we unveil the presence of serious backdoor vulnerabilities within the paradigm of model merging, which combines several fine-tuned task-specific models into a merged model. Our novel backdoor attack, named {\name}, enables the adversary to compromise the entire merged models by contributing as few as one backdoored task-specific model. To address the unique challenges of the blind knowledge of the merging process, {\name} adopts a two-stage attack mechanism to robustify embedded backdoors against the changes of different merging parameters. Extensive experiments show that our attacks significantly outperform all existing attacks and achieve remarkable performance under various merging settings. Our results highlight the need for a deeper understanding of the security risks of model merging, especially the consequence of reusing open-sourced models.
\label{sec:conclusion}

\section{Acknowledgement}
We thank the anonymous reviewers for their constructive comments. This work is partially funded by the National Science Foundation (NSF) grant No. 2325369, 2411153, the European Health and Digital Executive Agency (HADEA) within the project ``Understanding the individual host response against Hepatitis D Virus to develop a personalized approach for the management of hepatitis D'' (DSolve, grant No. 101057917) and the BMBF within the project ``Repräsentative, synthetische Gesundheitsdaten mit starken Privatsphärengarantien'' (PriSyn, grant No. 16KISAO29K).

\bibliographystyle{ACM-Reference-Format}
\bibliography{refs}

\appendix
\begin{algorithm}[ht] 
\renewcommand{\algorithmicrequire}{\textbf{Input:}} 
\renewcommand{\algorithmicensure}{\textbf{Output:}}
\caption{\textbf{\name
 (Inject backdoor into the adversary model)}} 
\label{algo-main} 

The lines marked in \textcolor{blue}{blue} are specific to \textsc{{\nameon}}. \\
The lines marked in \textcolor{red}{red} are specific to \textsc{{\nameoff}}. \\
$\mathcal{V}_{\theta}$/$\mathcal{T}_{\theta}$ denotes the visual/text encoder of CLIP-like model $\mathcal{M}_{\theta}$. \\
``$\ada$'' is the brevity of adversarial data augmentation.

\begin{algorithmic}[1]
\REQUIRE 
Pre-trained model $\mathcal{M}_{\theta_{\pre}}=\{\mathcal{V}_{\theta_{\pre}},\mathcal{T}_{\theta_{\pre}}\}$,
dataset of adversary task $\mathcal{D}_{\adv}$, classes of adversary task $C_{\adv}$, target class $c$, trigger size $s$, parameters for optimizing the universal trigger (i.e., $\gamma$ and $lr$), loss weight $\alpha$, attack-scenario, shadow classes $C_{\shadow}$, reference images $R$ from the target class, number of adversarially augmented images $N_{\ada}$.
\ENSURE Backdoored adversary model $\mathcal{M}_{\theta_{\adv}}$. \\

//Estimate weights of ($\theta_{\pre}+\TV_{\benign}$)
\IF{attack\_scenario is ``multi-task''}
    \STATE $\theta_{0} \leftarrow \theta_{\pre}$
\ELSE
    \STATE $\theta_{\adv}^{\prime} \leftarrow$ \textsc{GetCleanAdversaryModel}$(\mathcal{D}_{\adv}, \mathcal{M}_{\theta_{\pre}}, C_{\adv})$
    \STATE $\theta_{0} \leftarrow \theta_{\adv}^{\prime}$
\ENDIF 

\ \ \\

// Adversarial data augmentation
\textcolor{red}{\STATE $\mathcal{D}_{\ada} \leftarrow []$
\FOR{$i \in N_{\ada}$}
    \STATE Sample reference image $x$ from reference images $R$
    \STATE Sample class $c^{\prime} \neq c$ from shadow classes $C_{\shadow}$
    \STATE $x \leftarrow$\textsc{RandomResizedCrop}($x$)
    \STATE $x^{\prime} \leftarrow$ \textsc{GetAdversarialExample}$(\mathcal{M}_{\theta_{0}}, C_{\shadow}, x, c^{\prime})$
    \STATE $\mathcal{D}_{\ada} \leftarrow \mathcal{D}_{\ada}+[x^{\prime}]$
\ENDFOR}

\ \ \\
// Stage 1: Generate universal trigger
\textcolor{blue}{\STATE $t \leftarrow$ \textsc{GetUniversalTrigger}$(\mathcal{D}_{\adv}, \mathcal{M}_{\theta_{0}}, C_{\adv}, c, s)$}
\textcolor{red}{\STATE $t \leftarrow$ \textsc{GetUniversalTrigger}$(\mathcal{D}_{\ada}, \mathcal{M}_{\theta_{0}}, C_{\shadow}, c, s)$}

\ \ \\

// Stage 2: Inject backdoor with the universal trigger.
\STATE $\theta \leftarrow \theta_{\pre}$
\FOR{\textit{number of training epochs}}
    \FOR{$(x,y) \in \mathcal{D}_{\adv}$}
        \STATE $\mathcal{L}_{1} = \mathcal{L}_{CE}(\mathcal{M}_{\theta}(x,C_{\adv}),y)$ \\
        // Feature Interpolation Loss
        \STATE Uniformly sample interpolation coefficient $p$
        \STATE $F \leftarrow 
        \mathcal{V}_{\theta_{0}}(x \oplus t) \cdot p + (1-p) \cdot \mathcal{V}_{\theta}(x \oplus t)$ 
        \textcolor{blue}{\STATE $\mathcal{L}_{2}=\mathcal{L}_{CE}([\langle F, \mathcal{T}_{\theta}(c_1) \rangle, \cdots, \langle F, \mathcal{T}_{\theta}(c_k) \rangle]_{{c_k}\in{C_{\adv}}}^{\top},c)$}
        \textcolor{red}{\STATE $\mathcal{L}_{2}=\mathcal{L}_{CE}([\langle F, \mathcal{T}_{\theta}(c_1) \rangle, \cdots, \langle F, \mathcal{T}_{\theta}(c_k) \rangle]_{{c_k}\in{C_{\shadow}}}^{\top},c)$}
        \STATE $\mathcal{L}_{total} \leftarrow \mathcal{L}_1 + \alpha \cdot \mathcal{L}_2$
        \STATE $\theta \leftarrow$ \textsc{GradientDescent}$_{\theta}(\mathcal{L}_{\text{total}})$ 
    \ENDFOR
\ENDFOR

\STATE return $\theta$
    
\end{algorithmic}
\end{algorithm}

\begin{algorithm}[ht] 
\renewcommand{\algorithmicrequire}{\textbf{Input:}} 
\renewcommand{\algorithmicensure}{\textbf{Output:}}
\caption{\textbf{GetUniversalTrigger}}
\label{algo-uap} 
\begin{algorithmic}[1]

\REQUIRE Dataset $\mathcal{D}$, model $\mathcal{M}_{\theta}$, classes $C$, target class $c$, trigger size $s$, loss weight $\gamma$, learning rate $lr$.
\ENSURE Trigger $t$. \\

\STATE Initialize the mask $m$ and perturbation $\delta$.

\FOR{\textit{number of UT epochs}}
    \FOR{$(x,y) \in \mathcal{D}$}
        \FOR{\textit{number of UT iterations}}
            \STATE $x \oplus t \leftarrow \delta \odot m + (1-m) \odot x$
            \STATE $logits \leftarrow \mathcal{M}(x \oplus t,C)$
            \STATE $idx \leftarrow$ \textsc{GetClassIndex}$(C,c)$
            \IF{$\argmax(logits)==idx$ and use early stop}
                \STATE \textbf{break}
            \ENDIF

            \STATE $l \leftarrow \gamma \cdot$ \textsc{LogSoftmax}$(logits)[idx]$
            \STATE $\delta \leftarrow \delta + lr \cdot \frac{\partial {l}}{\partial {\delta}}$
            \STATE Clip $\delta$ to the image domain
        \ENDFOR
    \ENDFOR
\ENDFOR
\STATE $t \leftarrow \{\delta, m\}$
\STATE return $t$
\end{algorithmic}
\end{algorithm}

\section{Defense results}
\label{sec:defense-results}

\begin{table}[ht]
\centering
\caption{Randomly shuffled task orders for model merging. $^*$ indicates the default task order. In the multi-task learning scenario, we pick the first $n$ tasks for model merging following a specific order.}
\vspace{2mm}
\resizebox{0.9\linewidth}{!}{
\begin{tabular}{lc}
\toprule
\label{table:task-order}
\textbf{Index} & \textbf{Task Order}
\\
\midrule
\multicolumn{1}{c|}{\textsc{I$^*$}} & \makecell{Cars,
SUN397, EuroSAT, GTSRB, Pets, STL10, ImageNet100,\\ MNIST, Flow-
ers, RESICS45, DTD, SVHN, CIFAR100} \\
\midrule
\multicolumn{1}{c|}{\textsc{II}} & \makecell{EuroSAT, Cars, MNIST, Pets, DTD, Flowers, SUN397, \\ CIFAR100, ImageNet100, GTSRB, STL10, RESICS45, SVHN} \\
\midrule
\multicolumn{1}{c|}{\textsc{III}} & \makecell{EuroSAT, Flowers, SUN397, STL10, GTSRB, SVHN, DTD, \\ RESICS45, Cars, Pets, ImageNet100, CIFAR100, MNIST} \\
\bottomrule
\end{tabular}
}
\end{table}

\begin{table}[ht]
\centering
\caption{Anomaly Indices of NC for clean and backdoored merged models. A model is predicted to be backdoored if the index is larger than 2.}
\vspace{2mm}
\resizebox{0.88\linewidth}{!}{
\begin{tabular}{c|ccc}
\toprule
\label{table:NC_merged}
\textbf{Adversary task} &\textbf{Clean} &\textbf{{\name}-on} &\textbf{{\name}-off} \\
\midrule
{CIFAR100} &0.89 &1.31 &0.89 \\
{ImageNet100} &1.48 &0.94 &1.78 \\
\bottomrule
\end{tabular}
}
\end{table}

\begin{table}[ht]
\centering
\caption{$p$-value of MMBD for clean and backdoored merged models. A model is predicted to be backdoored if the $p$-value is smaller than 0.05.}
\vspace{2mm}
\resizebox{0.88\linewidth}{!}{
\begin{tabular}{c|ccc}
\toprule
\label{table:MMBD_merged}
\textbf{Adversary task} &\textbf{Clean} &\textbf{{\name}-on} &\textbf{{\name}-off} \\
\midrule
{CIFAR100} &0.28 &0.34 &0.71 \\
{ImageNet100} &0.56 &0.49 &0.42 \\
\bottomrule
\end{tabular}
}
\end{table}

\begin{table}[ht]
\centering
\caption{Anomaly Indices of NC for clean and backdoored adversary models (i.e., task-specific model). A model is predicted to be backdoored if the index is larger than 2.}
\vspace{2mm}
\resizebox{0.88\linewidth}{!}{
\begin{tabular}{c|ccc}
\toprule
\label{table:NC_single}
\textbf{Adversary task} &\textbf{Clean} &\textbf{{\name}-on} &\textbf{{\name}-off} \\
\midrule
{CIFAR100} &1.16 &1.01 &1.51 \\
{ImageNet100} &0.89 &1.57 &0.98 \\
\bottomrule
\end{tabular}
}
\end{table}

\begin{table}[ht]
\centering
\caption{$p$-value of MMBD for clean and backdoored adversary models (i.e., task-specific model). A model is predicted to be backdoored if the $p$-value is smaller than 0.05.}
\vspace{2mm}
\resizebox{0.88\linewidth}{!}{
\begin{tabular}{c|ccc}
\toprule
\label{table:MMBD_single}
\textbf{Adversary task} &\textbf{Clean} &\textbf{{\name}-on} &\textbf{{\name}-off} \\
\midrule
{CIFAR100} &0.80 &0.68 &0.72 \\
{ImageNet100} &0.62 &0.47 &0.70 \\
\bottomrule
\end{tabular}
}
\end{table}

\begin{table}[ht]
\centering
\caption{False Negative of input Input defense on backdoored merged models. The lower the score, the more effective the defense mechanism. An input is predicted to be backdoored if its score exceeds that of 90\% of the clean samples owned by the merged model creator.}
\vspace{2mm}
\resizebox{0.88\linewidth}{!}{
\begin{tabular}{c|ccc}
\toprule
\label{table:inputdef}
\textbf{Adversary task}  &\textbf{{\name}-on} &\textbf{{\name}-off} \\
\midrule
{CIFAR100} &0.37 & 0.49\\
{ImageNet100} &0.61 &0.27 \\
\bottomrule
\end{tabular}
}
\end{table}

\begin{table}[ht]
\centering
\caption{The default selection of the target class for each task.}
\vspace{2mm}
\resizebox{0.5\linewidth}{!}{
\begin{tabular}{lc}
\toprule
\label{table:default-target-class}
\textbf{Task} & \textbf{Target Class}
\\
\midrule
\multicolumn{1}{c|}{\textsc{CIFAR100}} & Aquarium fish \\ 
\multicolumn{1}{c|}{\textsc{MNIST}} & 6 \\
\multicolumn{1}{c|}{\textsc{GTSRB}} & Stop sign \\ 
\multicolumn{1}{c|}{\textsc{SVHN}} &1  \\ 
\multicolumn{1}{c|}{\textsc{RESISC45}} & Airport \\
\multicolumn{1}{c|}{\textsc{SUN397}} & Airplane cabin \\ 
\multicolumn{1}{c|}{\textsc{EuroSAT}} & Forest \\ 
\multicolumn{1}{c|}{\textsc{DTD}} & Lined \\ 
\multicolumn{1}{c|}{\textsc{Cars196}} & Acura RL Sedan 2012 \\
\multicolumn{1}{c|}{\textsc{Pets}} &Bengal  \\ 
\multicolumn{1}{c|}{\textsc{Flowers}} & Osteospermum \\
\multicolumn{1}{c|}{\textsc{STL10}} & Truck \\
\multicolumn{1}{c|}{\textsc{ImageNet100}} &American coot \\
\bottomrule
\end{tabular}
}
\end{table}

\begin{table}[ht]
\centering
\caption{\textbf{For off-task backdoor attack, {\name}-Off achieves high attack success rates (\%) on target classes from different target tasks.} The adversary task is ImageNet100.}
\vspace{2mm}
\resizebox{\linewidth}{!}{
\begin{tabular}{lccccc}
\toprule
\label{table:off-task-imagenet100}
\textbf{MM Algorithm} & \textbf{Cars196}         & \textbf{SUN397}         & \textbf{EuroSAT}        & \textbf{GTSRB}            & \textbf{Pet}
\\
\midrule
\multicolumn{1}{c|}{\textsc{TA}} &99.78 &100 &99.96 &99.95 &99.94 \\
\multicolumn{1}{c|}{\textsc{Ties}} &97.81 &99.9 &99.75 &99.18 &99.92 \\
\multicolumn{1}{c|}{\textsc{RegMean}} &95.8 &99.87 &99.17 &98.41 &99.75 \\
\multicolumn{1}{c|}{\textsc{AdaMerging}} &98.14 &99.98 &99.04 &98.5 &99.89 \\ 
\multicolumn{1}{c|}{\textsc{Surgery}} &92.32 &99.97 &98.29 &96.14 &99.86 \\
\bottomrule
\end{tabular}
}
\end{table}

\begin{table}[ht]
\centering
\caption{\textbf{{\name} on ViT-B/16.} We follow the default settings.}
\vspace{2mm}
\resizebox{1\linewidth}{!}{
\begin{tabular}{lccccc}
\toprule
\label{table:vit-b-16}
\multirow{2}{*}{} & \multirow{2}{*}{\textbf{CA (\%)}} & \multicolumn{2}{c}{\textbf{On-task Attack}} & \multicolumn{2}{c}{\textbf{Off-task Attack}} \\
& &BA (\%) &ASR (\%) &BA (\%) &ASR (\%) \\
\midrule
\multicolumn{1}{c|}{\textsc{Task Arithmetic}} &81.08 &81.14 &99.19 &81.18 &94.78 \\
\multicolumn{1}{c|}{\textsc{Ties-Merging}} &79.51 &79.51 &100 &79.65 &91.85 \\
\multicolumn{1}{c|}{\textsc{RegMean}} &80.87 &80.9 &99.98 &80.99 &92.86 \\
\multicolumn{1}{c|}{\textsc{AdaMerging}} &86.13 &86.1 &99.67 &86.08 &95.69 \\
\multicolumn{1}{c|}{\textsc{Surgery}} &87.35 &87.3 &98.88 &87.35 &95.49 \\
\bottomrule
\end{tabular}
}
\end{table}

\begin{table}[ht]
\centering
\caption{\textbf{{\name} on ViT-L/14.} We follow the default settings.}
\vspace{2mm}
\resizebox{1\linewidth}{!}{
\begin{tabular}{lccccc}
\toprule
\label{table:vit-b-14}
\multirow{2}{*}{} & \multirow{2}{*}{\textbf{CA (\%)}} & \multicolumn{2}{c}{\textbf{On-task Attack}} & \multicolumn{2}{c}{\textbf{Off-task Attack}} \\
& &BA (\%) &ASR (\%) &BA (\%) &ASR (\%) \\
\midrule
\multicolumn{1}{c|}{\textsc{Task Arithmetic}} &87.53 &87.43 &94.53 &87.35 & 88.35 \\
\multicolumn{1}{c|}{\textsc{Ties-Merging}} &87.04 &86.96 &99.97 &86.95 &86.9 \\
\multicolumn{1}{c|}{\textsc{RegMean}} &87.7 &87.55 &98.18 &87.59 &78.97 \\
\multicolumn{1}{c|}{\textsc{AdaMerging}} &90.99 &91.07 &99.21 &91.04 &88.21 \\
\multicolumn{1}{c|}{\textsc{Surgery}} &91.57 &91.63 &98.06 &91.65 &80.7 \\
\bottomrule
\end{tabular}
}
\end{table}

\begin{table}[ht]
\centering
\caption{{\name} on CLIP-like model pre-trained by a more advanced algorithm, MetaCLIP. ViT-B/32 is used. We follow the default settings.}
\vspace{2mm}
\resizebox{1\linewidth}{!}{
\begin{tabular}{lccccc}
\toprule
\label{table:meta-clip-b-32}
\multirow{2}{*}{} & \multirow{2}{*}{\textbf{CA (\%)}} & \multicolumn{2}{c}{\textbf{On-task Attack}} & \multicolumn{2}{c}{\textbf{Off-task Attack}} \\
& &BA (\%) &ASR (\%) &BA (\%) &ASR (\%) \\
\midrule
\multicolumn{1}{c|}{\textsc{Task Arithmetic}} &80.21 &80.17 &99.96 &80.29 &98.08 \\
\multicolumn{1}{c|}{\textsc{Ties-Merging}} &78.76 &78.62 &99.76 &78.94 &90.54 \\
\multicolumn{1}{c|}{\textsc{RegMean}} &80.43 &80.3 &99 &80.48 &87.78 \\
\multicolumn{1}{c|}{\textsc{AdaMerging}} &84.37 &84.27 &99.97 &84.36 &98.54 \\
\bottomrule
\end{tabular}
}
\end{table}

\begin{figure}[ht]
    \centering
    \subfloat[On-task Attack]{\includegraphics[width =0.23\textwidth]{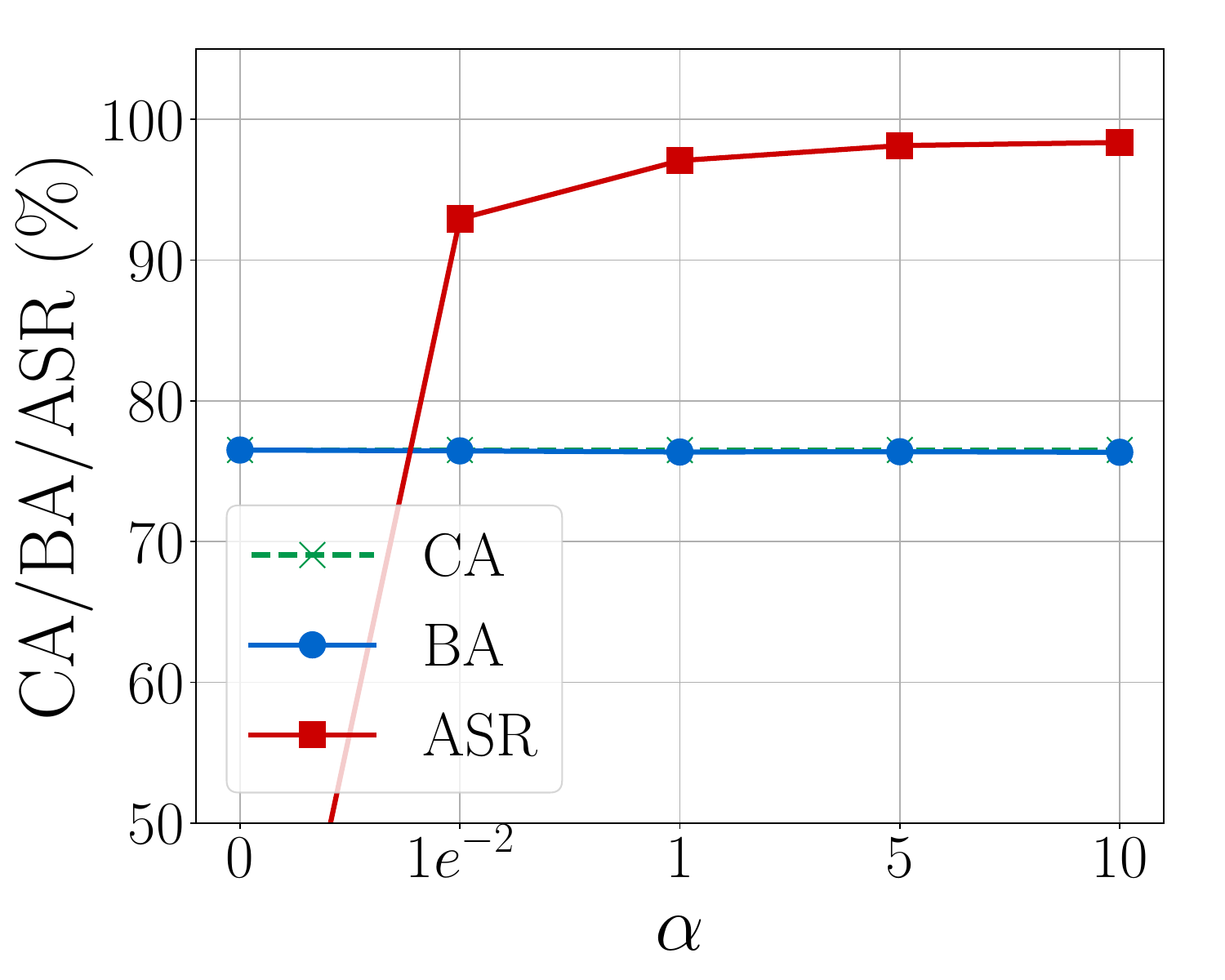}}
    \subfloat[Off-task Attack]{\includegraphics[width =0.23\textwidth]{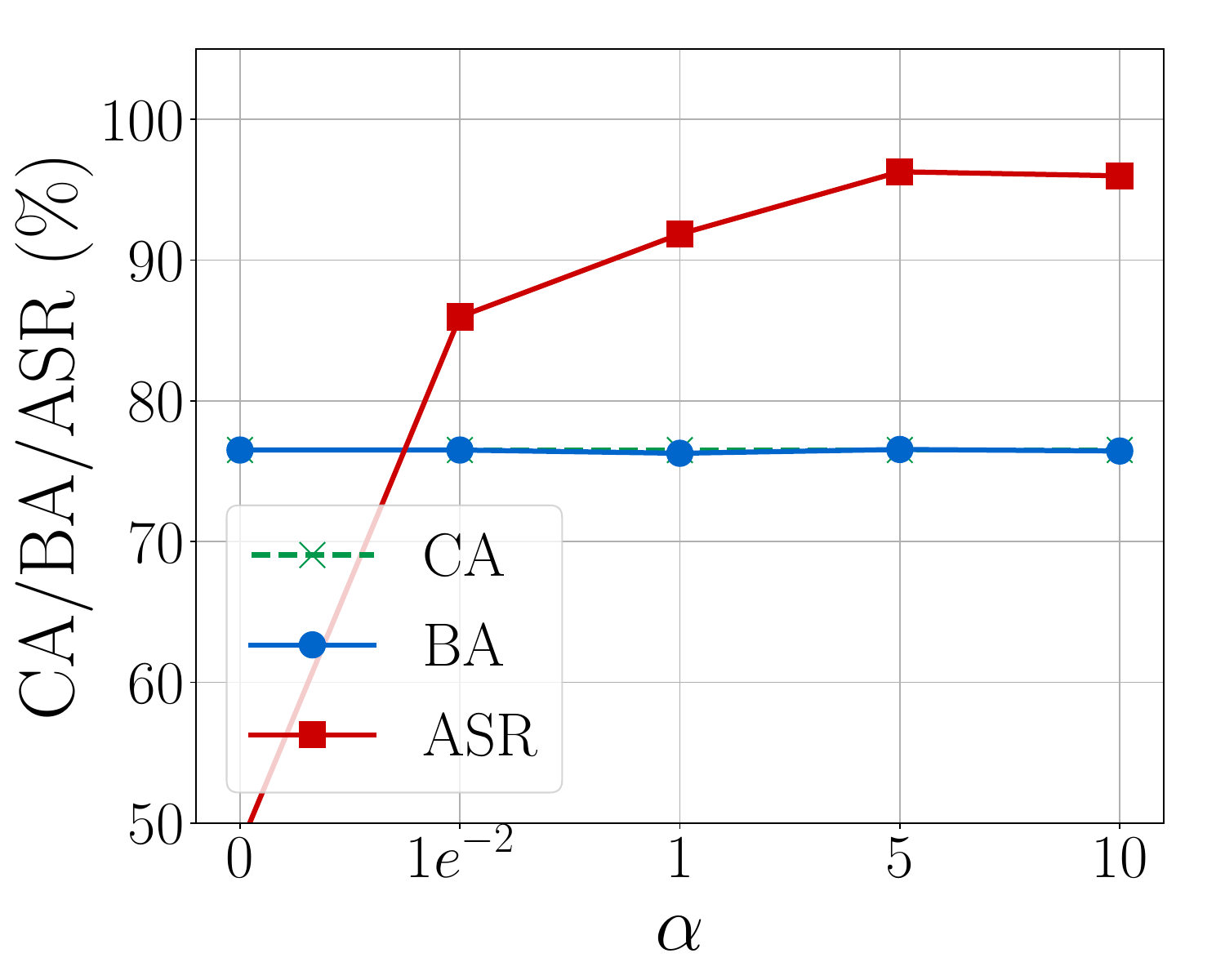}}
    \caption{Impact of the loss weight $\alpha$ on BA and ASR of {\name}. The default setting is considered (i.e., TA and ViT-B/32 are used).}
    \label{fig:alpha}
    \vspace{-2mm}
\end{figure}

\begin{table}[ht]
\centering
\caption{\textbf{{\name} can simultaneously embed multiple backdoors into the same adversary model.} The CA is 76.51\%.}
\vspace{2mm}
\resizebox{0.7\linewidth}{!}{
\begin{tabular}{c|cc}
\toprule
\label{table:multi-backdoor}
\textbf{Number of Backdoors} &\textbf{BA (\%)} &\textbf{Avg ASR (\%)} \\
\midrule
{5} &76.28 &98.78 \\
{10} &76.26 &97.02 \\
{15} &76.34 &96.5 \\
\bottomrule
\end{tabular}
}
\end{table}

\begin{table}[ht]
\centering
\caption{Comparing the ASRs (\%) of {\name} using the imperceptible universal trigger and existing backdoor attacks using imperceptible triggers. The on-task attack under the default setting is measured.}
\vspace{2mm}
\resizebox{\linewidth}{!}{
\begin{tabular}{c|ccccc}
\toprule
\label{table:invisible}
\textbf{} &\textbf{TA} &\textbf{TiesMerging} 
&\textbf{RegMean}
&\textbf{AdaMerging}
&\textbf{Surgery}\\
\midrule
{SIG~\cite{barni2019new}} &0.7 &1.35 &0.7 &0.31 &0.11 \\
{WaNet~\cite{nguyen2021wanet}} &0.78 &1.39 &0.89 &0.38 &0.09 \\
{LiRA~\cite{doan2021lira}} &13.48 &20.28 &7.6 &14.29 &8.29 \\
\midrule
\rowcolor{Yellow}{\name-Invisible} &\textbf{94.52} &\textbf{98.54} &\textbf{96.22} &\textbf{97.98} &\textbf{97.39} \\
\bottomrule
\end{tabular}
}
\end{table}

\begin{table}[ht]
\centering
\caption{\textbf{Ablation study of ADA (Adversarial Data Augmentation). Removing RandomCrop or replacing it with other augmentation will harm the attack performance.} The adversary task is CIFAR100 and the target class is selected from Cars196. The model is CLIP ViT-B/16.}
\vspace{2mm}
\resizebox{\linewidth}{!}{
\begin{tabular}{ccccc}
\toprule
\label{table:on-task-phi}
\textbf{Aug} & \textbf{AP} & \textbf{TA-ASR (\%)} & \textbf{Ties-ASR (\%)} & \textbf{RegMean-ASR (\%)} \\
\midrule
\multicolumn{1}{c|}{\xmark} & \multicolumn{1}{c|}{\xmark} &2.22 &1.26 &1.31 \\
\midrule
\multicolumn{1}{c|}{\xmark} & \multicolumn{1}{c|}{\multirow{3}{*}{\checkmark}} &60.98 &55.62 &56.19 \\
\multicolumn{1}{c|}{ColorJitter} & \multicolumn{1}{c|}{} &58.48 &39.58 &45.31 \\
\multicolumn{1}{c|}{RandomCrop} & \multicolumn{1}{c|}{} &94.78 &91.85 &92.86 \\
\bottomrule
\end{tabular}
}
\end{table}

\begin{figure*}[t]
    \centering
    \subfloat[CIFAR100-Accuracy]{\includegraphics[width =0.24\textwidth]{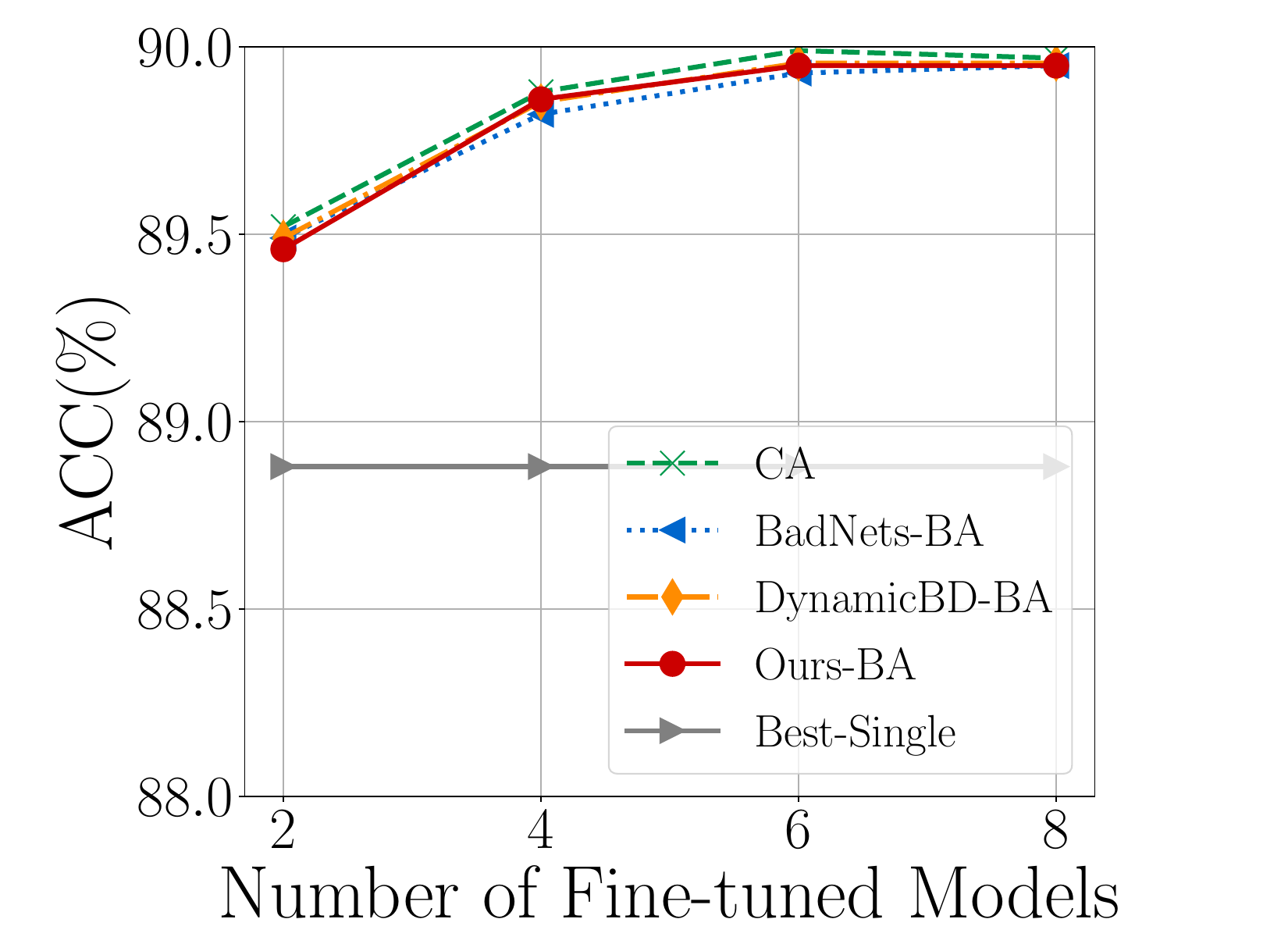}}
    \subfloat[CIFAR100-ASR]{\includegraphics[width =0.24\textwidth]{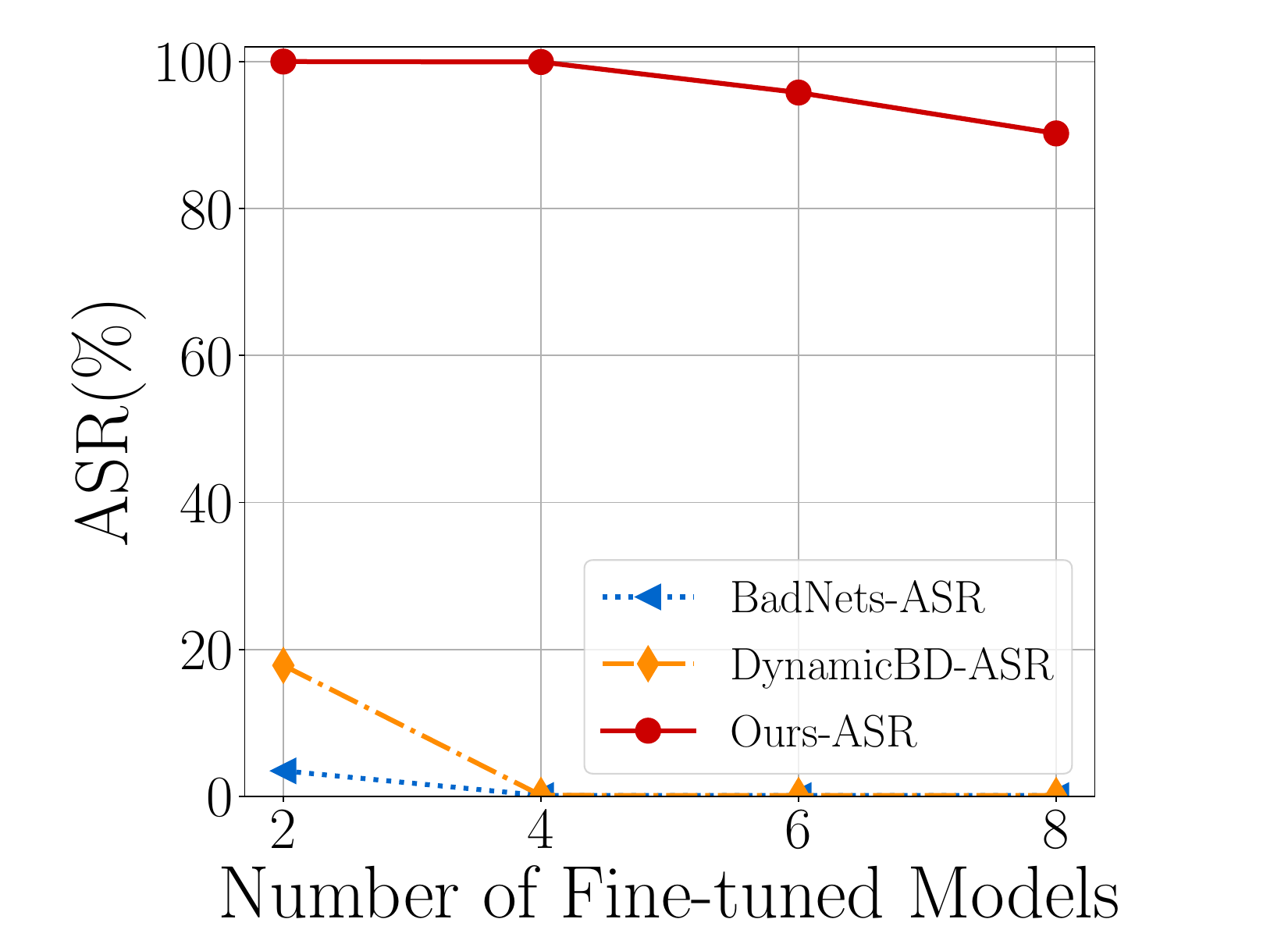}}
    \subfloat[ImageNet100-Accuracy]{\includegraphics[width =0.24\textwidth]{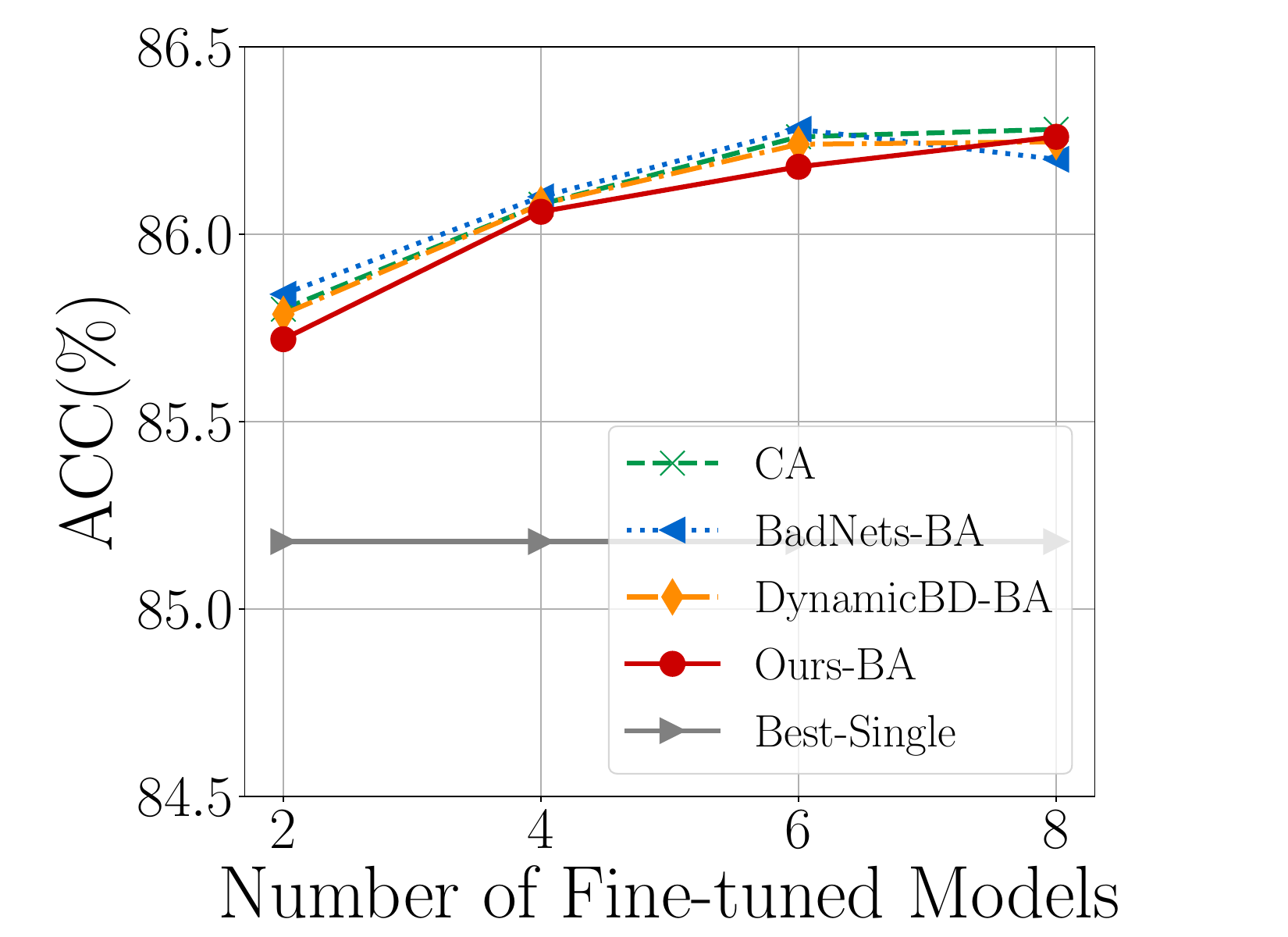}}
    \subfloat[ImageNet100-ASR]{\includegraphics[width =0.24\textwidth]{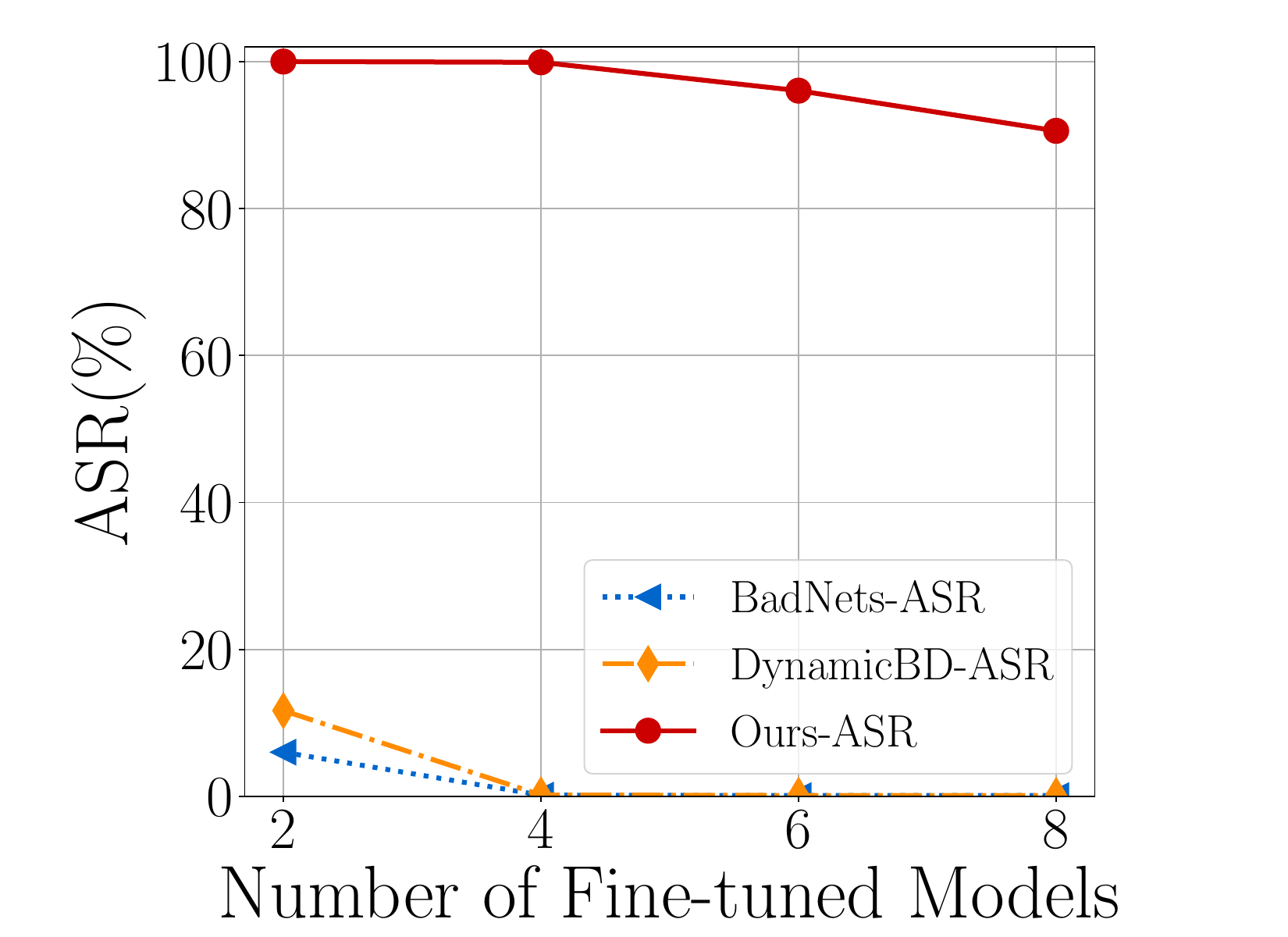}}
    \caption{{\name}-On is effective under single-task learning scenarios. Following~\cite{wortsman2022model}, we merge task-specific models fine-tuned for the same task. Best-Single indicates the highest accuracy achieved by a task-specific model (w/o MM). RegMean is used.}
    \label{fig:single-regmean}
\end{figure*}

\begin{table*}[ht]
\centering
\caption{The ASRs and BAs (\%) of our attacks when using \textbf{SVHN} as the adversary dataset. We follow the default settings.}
\vspace{2mm}
\resizebox{0.95\linewidth}{!}{
\begin{tabular}{cccccccccccccccc}
\toprule
\label{table:svhn}
\multirow{2}{*}{\textbf{Setting}} & \multicolumn{3}{c}{\textbf{Task-Arithmetic}} & \multicolumn{3}{c}{\textbf{Ties-Merging}} & \multicolumn{3}{c}{\textbf{RegMean}} & \multicolumn{3}{c}{\textbf{AdaMerging}} & \multicolumn{3}{c}{\textbf{Surgery}}\\
                                & {CA}
                                & {BA}
                                & {ASR}
                                & {CA}
                                & {BA}
                                & {ASR}
                                & {CA}
                                & {BA}
                                & {ASR}
                                & {CA}
                                & {BA}
                                & {ASR}
                                & {CA}
                                & {BA}
                                & {ASR} \\
\midrule
\multicolumn{1}{c|}{\textsc{On-task Attack}} &\multirow{2}{*}{77.08} &76.89 &99.4 &\multirow{2}{*}{74.52} &74.58 &99.98 & \multirow{2}{*}{78.02} &78.28 &98.81 &\multirow{2}{*}{84.68} &84.74 &93.1 &\multirow{2}{*}{86.43} &86.6 &95.6 \\
\multicolumn{1}{c|}{\textsc{Off-task Attack}} & &76.97 &94.18 & &74.44 &87.5 & &78.07 &85.34 & &84.72 &92.92 & &86.64 &85.05 \\
\bottomrule
\end{tabular}
}
\end{table*}

\begin{table*}[ht]
\centering
\caption{The ASRs and BAs (\%) of our attacks when using \textbf{RESISC45} as the adversary dataset. We follow the default settings.}
\vspace{2mm}
\resizebox{0.95\linewidth}{!}{
\begin{tabular}{cccccccccccccccc}
\toprule
\label{table:resics}
\multirow{2}{*}{\textbf{Setting}} & \multicolumn{3}{c}{\textbf{Task-Arithmetic}} & \multicolumn{3}{c}{\textbf{Ties-Merging}} & \multicolumn{3}{c}{\textbf{RegMean}} & \multicolumn{3}{c}{\textbf{AdaMerging}} & \multicolumn{3}{c}{\textbf{Surgery}}\\
                                & {CA}
                                & {BA}
                                & {ASR}
                                & {CA}
                                & {BA}
                                & {ASR}
                                & {CA}
                                & {BA}
                                & {ASR}
                                & {CA}
                                & {BA}
                                & {ASR}
                                & {CA}
                                & {BA}
                                & {ASR} \\
\midrule
\multicolumn{1}{c|}{\textsc{On-task Attack}} &\multirow{2}{*}{76.56} &76.52 &99.89 &\multirow{2}{*}{75.38} &75.28 &99.33 &\multirow{2}{*}{78.33} &78.19 &99.43 &\multirow{2}{*}{83.44} &83.58 &99.98 &\multirow{2}{*}{86.22} &86.17 &100 \\
\multicolumn{1}{c|}{\textsc{Off-task Attack}} & &76.31 &99.61 & &75.19 &96.55 & &78.18 &95.24 & &83 &98.96 & &86.07 &96.33 \\
\bottomrule
\end{tabular}
}
\end{table*}

\begin{table*}[ht]
\centering
\caption{Detailed accuracy (\%) of clean and backdoored merged models for each task. The MM algorithm is \textbf{TA}.}
\vspace{2mm}
\resizebox{\linewidth}{!}{
\begin{tabular}{ccccccccccccccc}
\toprule
\label{table:ta}
\multirow{2}{*}{\textbf{Setting}} & \multicolumn{7}{c|}{\textbf{ViT-B/32}} & \multicolumn{7}{c}{\textbf{ViT-L/14}}\\
& CIFAR100 &Cars &SUN397 &EuroSAT &GTSRB & Pets& \multicolumn{1}{c|}{Avg}   
& CIFAR100 &Cars &SUN397 &EuroSAT &GTSRB &Pets & Avg \\  
\midrule
\multicolumn{1}{c|}{\textsc{Clean}} &75 &64.94 &65.86 &88.59 &75.36 &89.32 &\multicolumn{1}{c|}{76.51} &86.29 &85.18 &75.51 &96.85 &85.79 &95.56 &87.53 \\
\multicolumn{1}{c|}{\textsc{\nameon}} &74.55 &65.22 &65.57 &88.26 &75.64 &89.13 &\multicolumn{1}{c|}{76.39} &86.19 &85.24 &75.37 &97.11 &85.11 &95.58 &87.43 \\
\multicolumn{1}{c|}{\textsc{\nameoff}} &75.18 &64.89 &65.74 &88.74 &75.45 &89.32 &\multicolumn{1}{c|}{76.55} &86.05 &85.31 &75.45 &96.59 &85.35 &95.37 &87.35 \\
\bottomrule
\end{tabular}
}
\end{table*}

\begin{table*}[ht]
\centering
\caption{Detailed accuracy (\%) of clean and backdoored merged models for each task. The MM algorithm is \textbf{TiesMerging}.}
\vspace{2mm}
\resizebox{\linewidth}{!}{
\begin{tabular}{ccccccccccccccc}
\toprule
\label{table:ties}
\multirow{2}{*}{\textbf{Setting}} & \multicolumn{7}{c|}{\textbf{ViT-B/32}} & \multicolumn{7}{c}{\textbf{ViT-L/14}}\\
& CIFAR100 &Cars &SUN397 &EuroSAT &GTSRB &Pets &\multicolumn{1}{c|}{Avg}   
& CIFAR100 &Cars &SUN397 &EuroSAT &GTSRB &Pets &Avg \\  
\midrule
\multicolumn{1}{c|}{\textsc{Clean}} &75.3 &67.19 &68.51 &81.07 &68.7 &89.34 &\multicolumn{1}{c|}{75.04} &87.3 &86.06 &75.9 &96.59 &80.69 &95.69 &87.04 \\
\multicolumn{1}{c|}{\textsc{\nameon}} &75.54 &67.48 &68.46 &80.04 &68.7 &89.32 &\multicolumn{1}{c|}{74.92} &87.21 &86.17 &75.91 &96.74 &80.06 &95.64 &86.96 \\
\multicolumn{1}{c|}{\textsc{\nameoff}} &75.75 &67.35 &68.35 &80.26 &68.65 &89.53 &\multicolumn{1}{c|}{74.98} &87.12 &86.13 &75.82 &96.37 &80.59 &95.67 &86.95 \\
\bottomrule
\end{tabular}
}
\end{table*}

\begin{table*}[ht]
\centering
\caption{Detailed accuracy (\%) of clean and backdoored merged models for each task. The MM algorithm is \textbf{RegMean}.}
\vspace{2mm}
\resizebox{\linewidth}{!}{
\begin{tabular}{ccccccccccccccc}
\toprule
\label{table:regmean}
\multirow{2}{*}{\textbf{Setting}} & \multicolumn{7}{c|}{\textbf{ViT-B/32}} & \multicolumn{7}{c}{\textbf{ViT-L/14}}\\
& CIFAR100 &Cars &SUN397 &EuroSAT &GTSRB &Pets &\multicolumn{1}{c|}{Avg}   
& CIFAR100 &Cars &SUN397 &EuroSAT &GTSRB &Pets &Avg \\  
\midrule
\multicolumn{1}{c|}{\textsc{Clean}} &76.52 &66.75 &67.3 &90.81 &72.64 &91.11 &\multicolumn{1}{c|}{77.52} &86.89 &84.99 &74.37 &97.89 &86.05 &95.99 &87.7 \\
\multicolumn{1}{c|}{\textsc{\nameon}} &76.66 &66.94 &67.21 &90.7 &73.14 &91.2 &\multicolumn{1}{c|}{77.62} &86.86 &84.98 &74.38 &97.81 &85.4 &95.91 &87.55 \\
\multicolumn{1}{c|}{\textsc{\nameoff}} &76.53 &66.61 &67.25 &90.44 &72.9 &91.22 &\multicolumn{1}{c|}{77.43} &86.91 &84.96 &74.4 &97.74 &85.53 &95.91 &87.59 \\
\bottomrule
\end{tabular}
}
\end{table*}

\begin{table*}[ht]
\centering
\caption{Detailed accuracy (\%) of clean and backdoored merged models for each task. The MM algorithm is \textbf{AdaMerging}.}
\vspace{2mm}
\resizebox{\linewidth}{!}{
\begin{tabular}{ccccccccccccccc}
\toprule
\label{table:adamerging}
\multirow{2}{*}{\textbf{Setting}} & \multicolumn{7}{c|}{\textbf{ViT-B/32}} & \multicolumn{7}{c}{\textbf{ViT-L/14}}\\
& CIFAR100 &Cars &SUN397 &EuroSAT &GTSRB & Pets & \multicolumn{1}{c|}{Avg}   
& CIFAR100 &Cars &SUN397 &EuroSAT &GTSRB &Pets & Avg \\  
\midrule
\multicolumn{1}{c|}{\textsc{Clean}} &77.05 &70.65 &68.66 &95.41 &94.42 &90.13 &\multicolumn{1}{c|}{82.72} &86.53 &90.06 &78.78 &97.19 &97.24 &96.16 &90.99 \\
\multicolumn{1}{c|}{\textsc{\nameon}} &77.02 &70.9 &68.46 &95.59 &94.24 &90.27 &\multicolumn{1}{c|}{82.75} &86.95 &90.24 &78.71 &97.59 &96.95 &95.99 &91.07 \\
\multicolumn{1}{c|}{\textsc{\nameoff}} &76.86 &71.21 &68.26 &95 &94.47 &90.38 &\multicolumn{1}{c|}{82.7} &86.5 &90.35 &78.77 &97.37 &97.2 &96.05 &91.04 \\
\bottomrule
\end{tabular}
}
\end{table*}

\begin{table*}[ht]
\centering
\caption{Detailed accuracy (\%) of clean and backdoored merged models for each task. The MM algorithm is \textbf{Surgery}.}
\vspace{2mm}
\resizebox{1\linewidth}{!}{
\begin{tabular}{ccccccccccccccc}
\toprule
\label{table:surgery}
\multirow{2}{*}{\textbf{Setting}} & \multicolumn{7}{c|}{\textbf{ViT-B/32}} & \multicolumn{7}{c}{\textbf{ViT-L/14}}\\
& CIFAR100 &Cars &SUN397 &EuroSAT &GTSRB &Pets & \multicolumn{1}{c|}{Avg}   
& CIFAR100 &Cars &SUN397 &EuroSAT &GTSRB &Pets & Avg \\  
\midrule
\multicolumn{1}{c|}{\textsc{Clean}} &79.62 &70.39 &71.05 &97.63 &95.84 &92.04 &\multicolumn{1}{c|}{84.49} &87.66 &90.51 &79.61 &97.85 &97.55 &96.18 &91.57 \\
\multicolumn{1}{c|}{\textsc{\nameon}} &79.49 &70.54 &70.86 &97.26 &96.09 &92.18 &\multicolumn{1}{c|}{84.4} &87.74 &90.72 &79.54 &98.11 &97.41 &96.27 &91.63 \\
\multicolumn{1}{c|}{\textsc{\nameoff}} &79.48 &70.97 &70.88 &97.15 &96.18 &92.07 &\multicolumn{1}{c|}{84.45} &87.62 &90.73 &79.66 &98.11 &97.47 &96.29 &91.65 \\
\bottomrule
\end{tabular}
}
\end{table*}

\subsection{Results of Detection-based Defense}
We evaluate two state-of-the-art detection-based defense mechanisms, including Neural Cleanse (NC)~\cite{wang2019neural} and MM-BD~\cite{wang2023mm}. Our results demonstrate that neither defense can detect our attacks, whether from the perspective of the merged model or the single task-specific model. For ease of illustration, we conduct backdoor detection on the corresponding target task both for the backdoored merged model and backdoored task-specific model. 

\mypara{Detecting backdoored merged model} Specifically, NC attempts to detect backdoors by reverse engineering a trigger for each class and using anomaly detection to find the outlier. As shown in~\autoref{table:NC_merged}, the anomaly indices for all the backdoored merged models are consistently below the detection threshold of \emph{2}, making them undetectable by this algorithm. Recently, researchers proposed MM-BD, a more advanced defense that uses estimated maximum margin statistics to perform backdoor detection. However, MM-BD also fails to detect our attacks, as evidenced by the experiment results in~\autoref{table:MMBD_merged}. In particular, the $p$-values for all the backdoored merged models significantly exceed the threshold of \emph{0.05}, rendering them highly undetectable.

\mypara{Detecting backdoored task-specific Model} We have also conducted backdoor detection on backdoored task-specific models from the adversary. The corresponding results for NC and MM-BD are presented in Table~\ref{table:NC_single} and Table~\ref{table:MMBD_single}, respectively. 
In both cases, we find that the detection algorithms can not identify any of the backdoored models. Based on these findings, we conclude that our attack remains robust against current detection methods.

\subsection{Results of Model Construction-based Defense}
We evaluate the most representative model construction-based defense, called fine-pruning (FP)~\cite{liu2018fine}). Fine-pruning selectively prunes neurons that are less important for general task performance but potentially crucial for backdoor behaviors. In this part, we assume the defender is aware of whether a model is backdoored and aims to remove the backdoor from the affected model. We focus on utilizing fine-pruning to defend against on-task attacks under two settings: (1) Apply fine-pruning on the backdoored task-specific model before merging. We observe that when the utility of the model drops below 50\%, the ASR only decreases by ~0.5\% for CIFAR100 and~1\% for ImageNet100. The results indicate that fine-pruning can not effectively remove the injected backdoor, even though it significantly compromises the utility. (2) Apply fine-pruning directly to the merged model. In this case, we observe that when the utility of the merged model on the target task drops below 50\%, the ASR only decreases by approximately 11\% for CIFAR100 and 0.7\% for ImageNet100. 
The results also indicate that fine-pruning is ineffective when applied to merged models. We observe a similar utility drop when employing fine-pruning to defend against off-task attacks.

Considering that one of the crucial goals for MM is to improve the final model's utility, we conclude that fine-pruning is not applicable to defend against our attacks.

\subsection{Results of Sample Filtering-based Defense}
We also evaluate the state-of-the-art sample filtering-based defense, called Scale-up~\cite{guo2023scale}. Scale-up aims to filter out test images that contain the trigger. In particular, it leverages different behaviors of the model on augmented clean images and augmented triggered images to detect the backdoor behavior. Since the sample filtering-based defense is typically utilized at the test time after the model is deployed, our experiments focus on the merged model after deployment.

Specifically, the merged model creator will first use the clean images in the small development set to query the merged model to generate a list of scores. 
Then, he/she can select a threshold such that only a small percentage of clean images have scored higher than this threshold. After that, when new test images arrive, the merged model creator can use this threshold to inspect each image. If a test image has a score higher than the threshold, it will be considered a triggered image. The corresponding results are shown in Table~\ref{table:inputdef}. We observe that even with a conservative threshold that only permits 90\% of the clean samples to be identified as benign, there are still around 43.5\% triggered images on average remaining undetected. Consequently, a large number of triggered images can bypass the detection, rendering the approach ineffective.

\subsection{Related Work: Backdoor Defenses}
\label{sec:relatedwork-defense}
To alleviate the backdoor vulnerabilities, many defense mechanisms are proposed. Based on the target and the working stage, existing defenses can be broadly categorized into four categories: (1) \emph{Detection-based defenses}~\cite{wang2019neural,wang2023mm,xiang2020detection,liu2019abs} focus on detecting whether a given model is
backdoored. (2) \emph{Model reconstruction-based defenses}~\cite{liu2018fine,wu2021adversarial,zeng2021adversarial} focus on removing the backdoor from a given model without compromising its utility. (3) \emph{Sample filtering-based defenses}~\cite{gao2019strip,guo2023scale} focus on filtering out triggered test samples during the inference time. (4) \emph{Poison suppression based defenses}~\cite{li2021anti,huang2022backdoor} focus on learning a clean model from a poisoned dataset. In this work, we focus on the first \emph{three} types of defenses because the defender (typically the merged model creator) has no access to the fine-tuning process of the adversary model.

\end{document}